\documentclass[a4paper,fleqn]{cas-sc}
\usepackage{subfigure}
\usepackage{algorithmicx,algorithm}
\usepackage[noend]{algpseudocode}
\usepackage{algorithm,algpseudocode,float}
\usepackage{amsmath}
\usepackage{textcomp}
\usepackage{setspace}
\usepackage{lineno}
\usepackage[authoryear,longnamesfirst]{natbib}

\def\tsc#1{\csdef{#1}{\textsc{\lowercase{#1}}\xspace}}
\tsc{WGM}
\tsc{QE}
\tsc{EP}
\tsc{PMS}
\tsc{BEC}
\tsc{DE}

\begin{document}
	\let\WriteBookmarks\relax
	\def\floatpagepagefraction{1}
	\def\textpagefraction{.001}
	\shorttitle{HiVision: Rapid Visualization of Large-Scale Spatial Vector Data}
	\shortauthors{M Ma et~al.} 
	
	\title [mode = title]{HiVision: Rapid Visualization of Large-Scale Spatial Vector Data}                      
	\tnotemark[1,2]
	
	\tnotetext[1]{This document is the results of the research project funded by the National Natural Science Foundation of China under Grant No. 41871284, 41971362, U19A2058 and the Natural Science Foundation of Hunan Province No.2020JJ3042.}
	
	\tnotetext[2]{Mengyu Ma designed and implemented the algorithm; Mengyu Ma, Ye Wu and Luo Chen performed the experiments and analyzed the data; Jun Li and Ning Jing contributed to the construction of experimental environment; Mengyu Ma and Xue Ouyang wrote the paper.}
	
	\author{Mengyu Ma}[orcid=0000-0002-7510-5638]
	\ead{mamengyu10@nudt.edu.cn}
	
	\author{Ye Wu}
	\ead{yewugfkd@nudt.edu.cn}
	
	\author{Xue Ouyang}
	\cormark[1]
	\ead{ouyangxue08@nudt.edu.cn}

	\author{Luo Chen}
	\ead{luochen@nudt.edu.cn}
	
	\author{Jun Li}
	\ead{junli@nudt.edu.cn}
	
	\author{Ning Jing}
	\ead{ningjing@nudt.edu.cn}
	
	\address{College of Electronic Science, National University of Defense Technology, Changsha 410073, China}

	\begin{abstract}
		Rapid visualization of large-scale spatial vector data is a long-standing challenge in Geographic Information Science. In existing methods, the computation overheads grow rapidly with data volumes, leading to the incapability of providing real-time visualization for large-scale spatial vector data, even with parallel acceleration technologies. To fill the gap, we present HiVision, a display-driven visualization model for large-scale spatial vector data. Different from traditional data-driven methods, the computing units in HiVision are pixels rather than spatial objects to achieve real-time performance, and efficient spatial-index-based strategies are introduced to estimate the topological relationships between pixels and spatial objects. HiVision can maintain exceedingly good performance regardless of the data volume due to the stable pixel number for display. In addition, an optimized parallel computing architecture is proposed in HiVision to ensure the ability of real-time visualization. Experiments show that our approach outperforms traditional methods in rendering speed and visual effects while dealing with large-scale spatial vector data, and can provide interactive visualization of datasets with billion-scale points/segments/edges in real-time with flexible rendering styles. The HiVision code is open-sourced at https://github.com/MemoryMmy/HiVision with an online demonstration.
		
	\end{abstract}
	
	\begin{graphicalabstract}
		\includegraphics[width=0.8\linewidth,trim=25 15 15 15,clip]{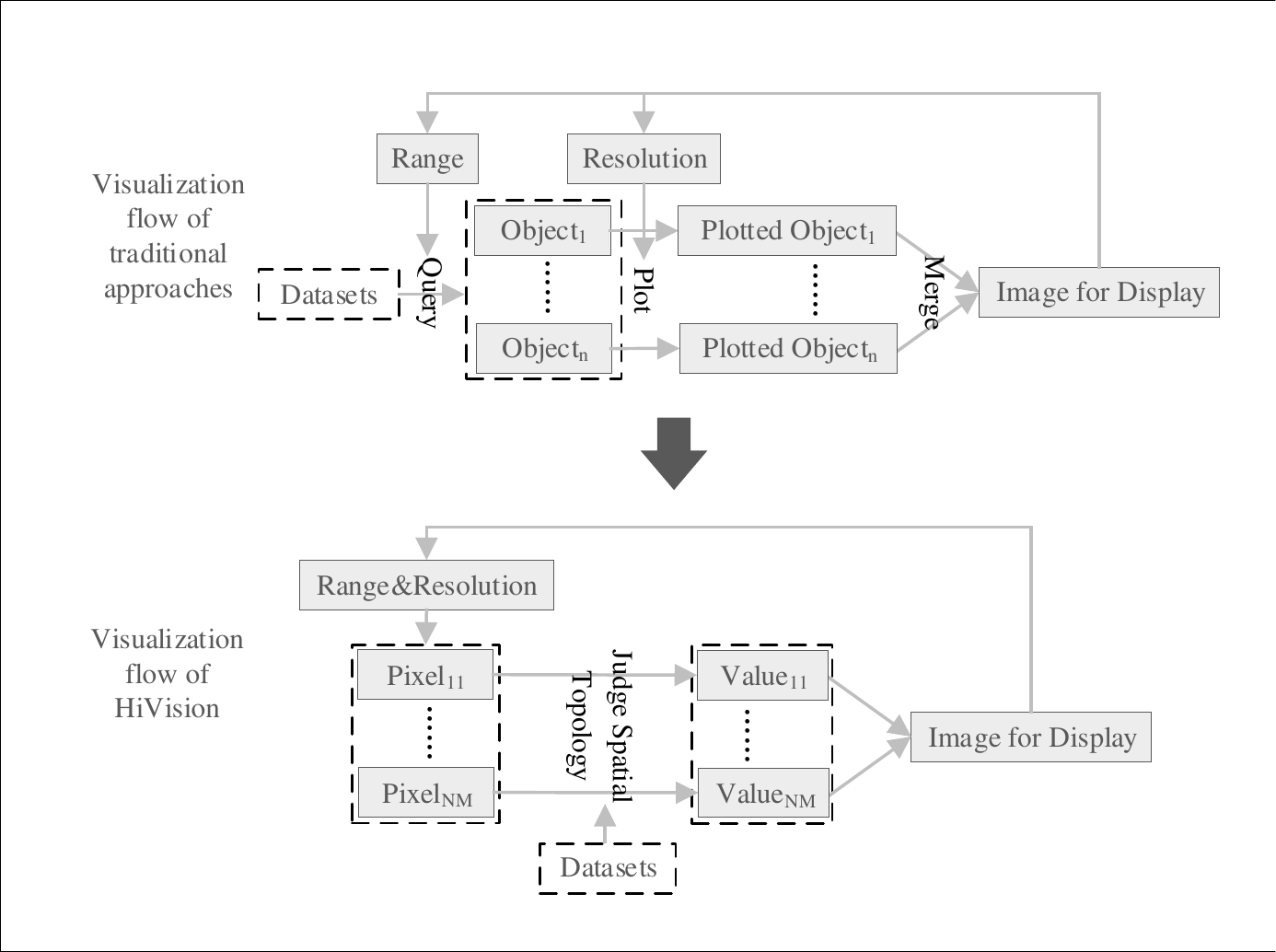}

	\end{graphicalabstract}
	
	\begin{highlights}
		\item Proposes a display-driven computing model for large-scale data visualization
		\item Designs a spatial-index-based optimization for real-time data visualization
		\item Proposes a hybrid-parallel architecture for enhanced data processing
		\item Implements an open-source tool for rapid visualization of large-scale spatial vector data
	\end{highlights}
	
	
	\maketitle
	
	\linenumbers
	
	\section{Introduction}\label{sec:introduction}
	
	There has been an explosion in the amounts of spatial data in recent years, due to the development of data acquisition technology, the prevalence of location-based services, and etc~\citep{Yao2018Big}. Visualization can make the intricate data more intuitive to human readers, thus important to discover implicit information and support further decision-making~\citep{Maceachren2004Geovisualization}. For example, effective visualization of taxi trajectories can help people better understand the urban transportation system, finding out strategies to reduce the number of accidents and traffic jams~\citep{Zuchao2013Visual}; a scatter plot of the road network nationwide can help the government to expose isolated areas, planning and constructing new roads. As an important type of spatial data, spatial vector is the abstract of real-world geographical entities, generally expressed as points, linestrings, or polygons (areas)~\citep{tong2013modeling}. In the big data era, the problem of efficient spatial vector data visualization becomes even more prominent, as visualizing spatial vector data involves the rasterizing process, which can be extremely time-consuming when the data scale is large. Rapid visualization of large-scale spatial vector data has become a severe challenge in Geographic Information Science (GIS).
	
	With the development of computer hardware, there has been an expansion in processor numbers, and parallel computing becomes increasingly important for processing large-scale spatial data. Recently, to optimize the data-intensive and computing-intensive spatial analysis using high-performance computing technologies has become a hot research topic in GIS~\citep{Yao2018Big}. Parallel computing is an effective way to accelerate the visualization process and is shown by representative works~\citep{gao2005parallel, tang2013parallel, guo2015spatially,guo2017efficient} that it can achieve highly improved performance compared with traditional serial methods. In addition, with the emergence of various parallel computing models (e.g., MPI, OpenMP, Hadoop, Storm, Spark), a series of high-performance spatial vector data visualization frameworks have been proposed and have seen some success (e.g., HadoopViz~\citep{eldawy2016hadoopviz}, GeoSparkViz~\citep{yu2018geosparkviz}, etc). 
	
\begin{figure}[htbp]
		\begin{center}
			\includegraphics[width=0.7\textwidth,trim=10 10 10 20,clip]{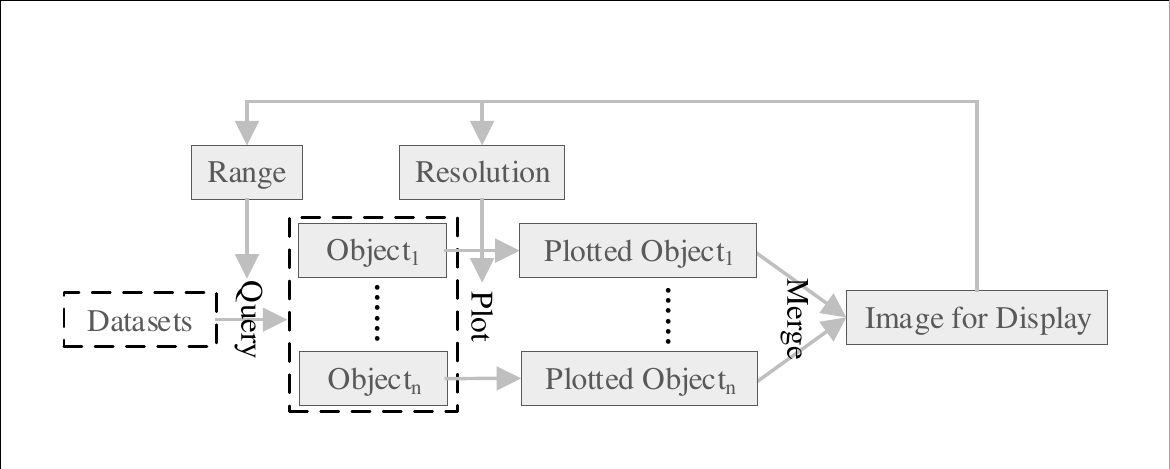}
			\caption{Processing flow of data-driven spatial vector data visualization.}
			\label{f1_1}			
		\end{center}
	\end{figure}

	However, not many existing methods can support real-time visualization of large-scale spatial vector data, even with the adopted high-performance computing technologies. Figure~\ref{f1_1} presents the general processing flow in existing visualization methods: firstly, each spatial object in the range is plotted according to the image resolution, then followed by a merge step to generate the final raster image. The computational scale of this data-driven processing flow expands rapidly with the volume of spatial objects in the image range, therefore, suffers performance drop when dealing with big data scenario and cannot meet the real-time requirements.
	
\begin{figure}[htbp]
		\begin{center}
			\includegraphics[width=0.6\textwidth,trim=10 10 10 25,clip]{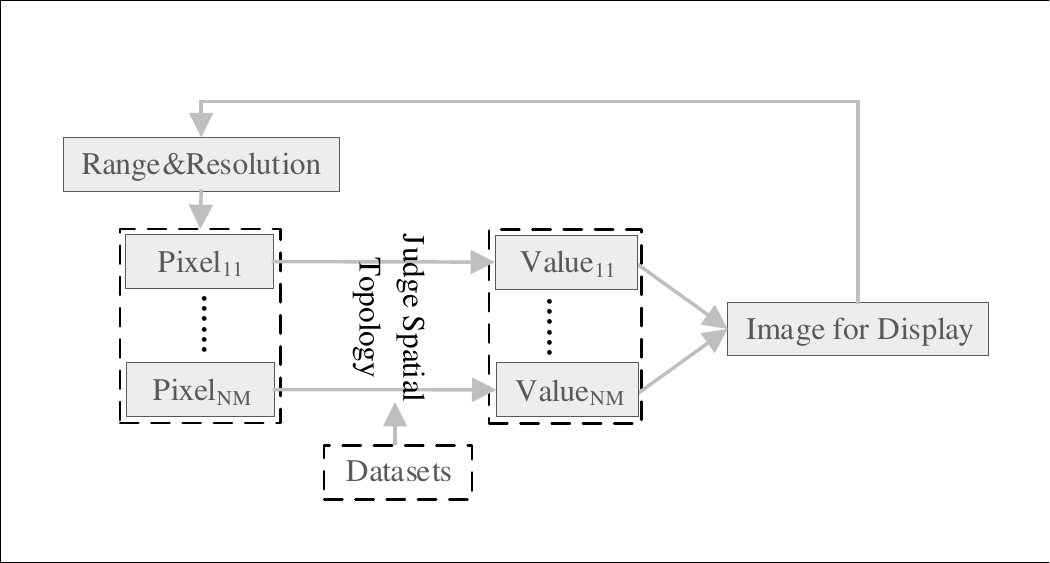}
			\caption{Spatial vector data visualization processing flow in HiVision.}
			\label{f1_2}			
		\end{center}
	\end{figure}

	To address the scale issue, we present HiVision, a display-driven vector data visualization model as shown in Figure~\ref{f1_2}. In HiVision, the computing units are pixels rather than spatial objects and the algorithms focus on determining the actual pixel value with relevant spatial objects: as the number of pixels in the image range is limited and stable, the computational complexity of HiVision remains stable while dealing with spatial data of different scales. In addition, efficient spatial-index-based strategies are introduced to estimate spatial topological relationships between pixels and spatial objects, thus determining the value of pixels. To the best of our knowledge, this is the first attempt for rapid visualization of vector data that can achieve the advantage of being less sensitive to data volumes.

	The contributions of this paper can be summarized as follows.
	
	\begin{itemize}
		\item Implements an open-source tool for rapid visualization of large-scale spatial vector data. HiVision can be used to provide an interactive exploration of massive raw spatial vector data with flexible rendering styles \textcolor{black}{and better visual effects}, so as to discover implicit information and parameter settings for further processing.
		\item Designs a display-driven vector data visualization approach and reduces the computational complexity dramatically (from O($n$) to O($log(n)$). HiVision calculates visualization results directly using a parallel per-pixel approach with efficient fine-grained spatial indexes. Our approach provides new research ideas for many related fields (e.g. map cartography~\citep{kraak2013cartography}, spatial analysis and data visualization).
		\item Carries out extensive experiment evaluations and provides an online demonstration. Experiments show that HiVision dramatically outperforms traditional data-driven methods in tile rendering speed, and it is capable of handling billion-scale spatial vector data. In addition, the demonstration verifies that a normal 4 cores CPU with 32GB Memory could perform interactive visualization of 10-million-scale spatial objects using HiVision.				
	\end{itemize}
	
	The rest of this paper proceeds as follows. Section \ref{section 2} highlights the state-of-the-art of big spatial data visualization. In Section \ref{section 3} and Section \ref{section 4}, the techniques of HiVision are described in details. The experimental results are presented and analyzed in Section \ref{section 5} with an online demonstration of HiVision introduced in Section \ref{section 6}. And the conclusions are drawn in Section \ref{section 7}.

	\section{Related Work}\label{section 2}

	\subsection{Big Spatial Data}
	
	There are many studies on big spatial data that discuss the challenges brought by data volume and diverse application requirements. Most of the studies focus on solving the problem of spatial query that emerged when processing large-scale spatial data~\citep{ bellur2014parallelizing,fries2014phidj, aly2015aqwa, zhu2015novel, Eldawy2013CG, scitovski2018density}. Typically, \cite{aly2015aqwa} presented a workload-aware big spatial data partitioning strategies for range and k-nearest-neighbor queries which achieved an order of magnitude enhancement compared with the state-of-the-art systems; \cite{Eldawy2013CG} proposed a set of efficient MapReduce algorithms for some basic spatial analysis operations, including polygon union, farthest/closest pair, skyline query and convex hull. Besides, many high-performance frameworks/systems have been proposed to process or analyze big spatial data, among them are ScalaGiST~\citep{lu2014scalagist}, Sphinx~\citep{eldawy2015sphinx}, Hadoop-GIS~\citep{aji2013hadoop},  SpatialHadoop~\citep{ eldawy2015spatialhadoop}, GeoSpark~\citep{yu2015geospark}, Simba~\citep{xie2016simba}, and etc. However, none of them is capable of providing visualization of large-scale spatial data in real-time.
	
	\subsection{Big Spatial Data Visualization}
	
	Spatial data visualization, as an important means of spatial analysis, is a core issue in map cartography. To visualize large-scale spatial data, the Open Geospatial Consortium (OGC) has provided a standard Web Map Tile Service (WMTS)~\citep{wmts}, in which pre-rendered or run-time computed georeferenced map images are organized into the tile-pyramid structure and transferred as map tiles over the Internet. Tile-pyramid is a multi-resolution data structure model widely used for map browsing on the web. At the lowest level of tile-pyramid (level 0), a single tile summarizes the whole map. On each higher level, there are up to \({4^z}\) tiles, in which \(z\) is the zoom level. Each tile has the same size of \(n \times n\) pixels and corresponds to a same geographic range. However, existing solutions to large-scale map data visualization are not ideal due to the following problems: 1) (\textbf{long generating time}) on the one hand, it will take a long time to render one tile which intersects large amounts of spatial objects, on the other hand, it may take dozens of hours or even more to slice all the tiles to provide free exploration of one spatial dataset; 2) (\textbf{massive tiles}) a tile-pyramid of world-scale with zoom from 0 to 16 contains billions of tiles, requiring TB level in storage; 3) (\textbf{inflexible styles}) the style of rendered tiles can not be changed. In other words, a new set of tiles has to be re-generated if one wants to change the style.
	
	Several studies focus on improving tile rendering performance, a typical yet important benchmark of spatial data processing. In the field of map cartography, there are many mature tools for spatial data visualization, such as Mapnik~\citep{Mapnik}, GeoServer~\citep{GeoServer} and MapServer~\citep{MapServer}. These tools are widely used for generating maps due to their efficient rendering algorithms and rich rendering styles. In order to further improve the rendering performance of large-scale spatial data, researchers have provided several parallel methods, and various acceleration technologies are adopted. For example, \cite{gao2005parallel} presented a parallel multi-resolution volume rendering algorithm for visualizing large data sets, in which the raw data is converted to a wavelet tree to achieve load-balanced rendering. \cite{tang2013parallel} proposed a parallel construction method of large circular cartograms based on graphics processing units (GPUs) and achieved significant acceleration performance. In order to achieve load-balance, \cite{guo2015spatially} developed a spatially adaptive decomposition approach for polyline and polygon visualization to divide the visualization domain into unequally sized sub-domains, such that they entail approximately the same amount of computational intensity. \cite{guo2017efficient} proposed an approach of vector data rendering by using the parallel computing capability of many-core GPUs, which involves a balancing allocation strategy to take full advantage of all processing cores of the GPU. OmniSci~\citep{omniSci} is a GPU-based analytics platform which allows users to interactively query and visualize large-scale spatial dataset. It can process billions of points and millions of polygons and generate custom pointmaps, heatmaps, choropleths, scatterplots, and other visualizations, enabling zero-latency visual interaction at any scale. \cite{eldawy2016hadoopviz} proposed a MapReduce-based framework, HadoopViz, for visualizing big spatial data, and experiments showed that HadoopViz can efficiently produce giga-pixel images for billions of input records. \cite{yu2018geosparkviz} proposed a big spatial data visualization framework, GeoSparkViz, which takes advantage of the in-memory architecture of Spark, and experiments verified that GeoSparkViz can generate a gigapixel image of 1.3 billion taxi trips in 5 minutes on a four-node commodity cluster. 
	
	In addition, researchers have proposed several other approaches to improve data visualization effects. \cite{yang2005performance} listed several possible techniques to improve the performance of data exploration on Web-based GIS, including pyramids and hash indices for large images, multi-threading, data catching and binary compression. To manage the massive map tiles, \cite{wan2016effective} developed a tile storage approach based on the NoSQL database. To provide interactive exploration of large-scale spatial data while avoiding generating all the image tiles, \cite{ ghosh2019aid} proposed an adaptive image data index, which pre-generates image tiles for the regions where spatial objects are dense; other typical methods generate tile-like intermediate variables through precomputing thus to compute requested tiles on the fly~\citep{liu2013immens, pahins2016hashedcubes, lins2013nanocubes}. The vector tile technology~\citep{Vectortiles} has been a popular approach over the recent years to visualize large-scale spatial vector data; it transfers packets of geographic data rather than images to clients and can change the map styles without generating new tiles. However, as it involves the complex cartographic generalization operations, it is more time-consuming to generate vector tiles than image tiles.
	
	To summarize, the existing solutions to rapid visualization of large-scale vector data are normally data-driven, with the computational scales expanding rapidly with the volume of spatial objects, leading to the result that it is difficult to provide visualization of large-scale vector data in real-time.

	\subsection{Display-driven Computing}
	
	Display-driven computing (DisDC) is a computing model that is especially suitable for data-intensive problems in GIS. In DisDC, the computing units are pixels rather than the spatial objects. The core issue in DisDC is to identify spatial topological relationships between pixels and spatial objects, thus determining the value of pixels for display. DisDC has a broad prospect of applications and researches in big data analysis.

	In our previous works~\citep{ijgi7120467, ijgi8010021}, the primary idea of DisDC was first proposed and applied to solve some basic analysis problems in GIS. We have successively brought forward HiBuffer and HiBO to provide interactive buffer and overlay analysis of large-scale spatial data. In \citep{ijgi7120467}, we conducted an experiment on a high-performance server, in which the optimized parallel buffer analysis methods proposed in recent years and the popular GIS software programs are discussed and compared (key results are summarized in Table~\ref{t2_1}), and the display-driven buffer analysis method, HiBuffer, is deployed and tested in the same hardware environment. Experiments verified that HiBuffer reduced computation time by up to orders of magnitude while dealing with large-scale spatial vector data, and DisDC has significant advantages compared with data-driven computing (DataDC). In this paper, we have applied DisDC to the field of rapid vision for large-scale spatial vector data to explore its effects. 
	
	\begin{table}[htbp]
		\caption{Performance of data-driven methods and HiBuffer~\citep{ijgi7120467}.}
		\centering
		\begin{tabular}{|c|c|c|c|c|}
			\hline
			\textbf{Algorithm} & \textbf{40,927 linestrings} & \textbf{208,067 linestrings}&\textbf{597,829 linestrings} &\textbf{21,898,508 linestrings}\\
			\hline
			$\text{Parallel Method}$$_{\text{1}}$$^{\mathrm{a}}$		& 9.0 s			& 38.8 s	 &332.3 s 	& Failed		\\
			$\text{Parallel Method}$$_{\text{2}}$$^{\mathrm{b}}$		& 15.4 s	 &128.2 s 		&936.9 s	 & Failed			\\
			$\text{Parallel Method}$$_{\text{3}}$$^{\mathrm{c}}$	  & 12.3 s	 &75.5 s		   		&661.9 s & Failed	 		\\
			$\text{Parallel Method}$$_{\text{4}}$$^{\mathrm{d}}$			& 17.2 s	    &220.8 s	 		&2813.4 s	  	& Failed	\\
			$\text{PostGIS}$	& 34.9 s     &295.8 s		&2380.2	s  	& Failed		\\
			$\text{QGIS}$		&129 s   	&2788 s			&Failed 	& Failed	\\
			$\text{ArcGIS}$		& 139 s	 	&2365 s					&Failed	& Failed 	\\
			$\text{HiBuffer}$		&\textless 1 s		&\textless 1 s		&\textless 1 s &\textless 1 s	\\
			\hline
		\end{tabular}
		
		{\scriptsize $^{\mathrm{a}}$\citep{shen2018buffer}
			$^{\mathrm{b}}$\citep{fan2014optimization}
			$^{\mathrm{c}}$\citep{Huang2013Parallel}
			$^{\mathrm{d}}$\citep{wang2016parallel}}
		\label{t2_1}
	\end{table}

	\section{Methodology}\label{section 3}
	
\begin{figure}[htbp]
		\begin{center}
			\subfigure[Point objects]{
				\includegraphics[width=0.4\textwidth,trim=10 20 20 15,clip]{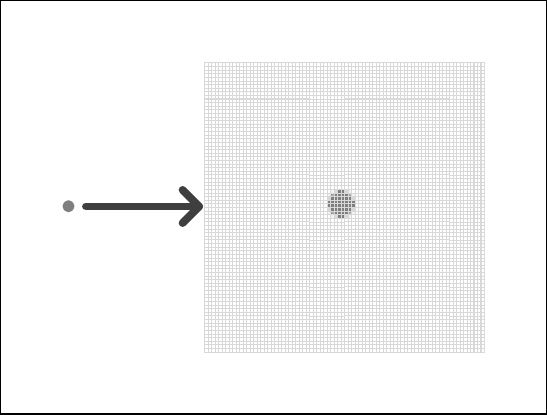}}%
			\subfigure[Linestring objects]{
				\includegraphics[width=0.4\textwidth,trim=10 20 20 15,clip]{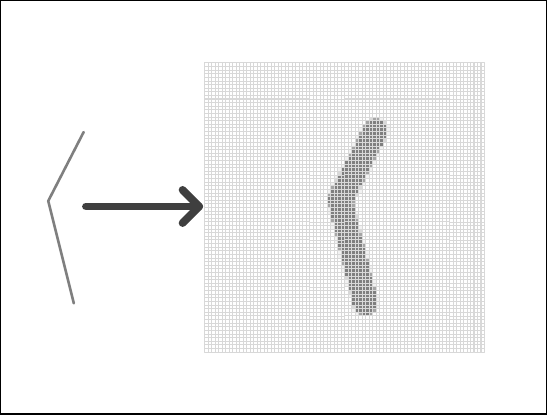}}%
			\vfill
			\subfigure[Polygon objects]{
				\includegraphics[width=0.55\textwidth,trim=10 20 20 15,clip]{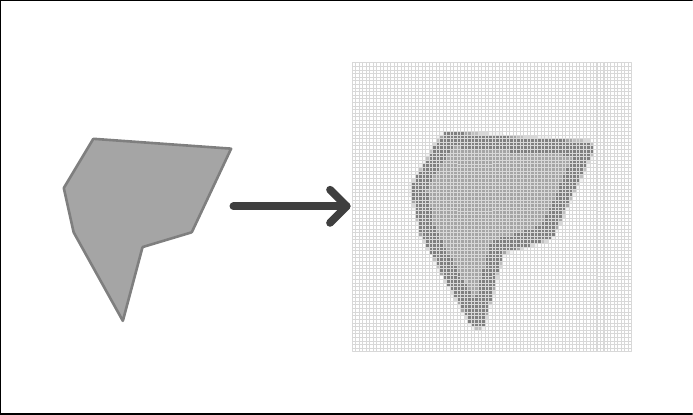}}%
			\caption{Plot spatial objects for visualization.}%
			\label{f3_1}
		\end{center}
	\end{figure}

	In this section, the key ideas for spatial vector data visualization in HiVision are introduced. Given the fact that the core task of visualizing spatial vector is to rasterize the spatial objects and render the final raster images for display, we have applied DisDC to scatter plot the spatial point, the linestring and the polygon objects in HiVision. And to provide better visualization effects, points, linestrings and boundaries of polygons are generally plotted with widths, and the anti-aliasing process is needed (see Figure~\ref{f3_1}). In HiVision, we process each pixel of the final raster image as an independent computing unit; and spatial indexes are utilized to identify the spatial topological relationships between pixels and spatial objects, thus determining the value of pixels for display. We design a DisDC oriented vector data organization structure for data visualization. Specifically, for point, linestring and polygon edges, we propose a visualization method named Spatial-Index-Based Visualization (SIBV); and for the filling problem in polygon visualization, we present Spatial-Index-Based Filling (SIBF) algorithm.
	
	\subsection{Data Organization}
	
	The core issue in DisDC is to identify spatial topological relationships between pixels and spatial objects, so as to calculate the value of pixels for display. To support the rapid visualization of large-scale spatial vector data using DisDC, we design a specialized data organization structure in HiVision. Spatial indexes are widely used to organize spatial data so that efficient spatial object accessing can be guaranteed. R-tree, as an efficient tree data structure widely used for indexing and querying spatial data, can be built efficiently by grouping nearby objects and representing them with their Minimum Bounding Rectangle (MBR) in the next higher level of the tree~\citep{choubey1999gbi}. In HiVision, an R-tree based design is proposed to organize spatial objects. As spatial objects, in reality, may have complex structures and different shapes that are difficult to identify accurately by the MBRs, use the MBR of each spatial object directly as R-tree record nodes can cause low query performance in the display-driven analyzing process. Accordingly, linestring or polygon objects are separated to segments or edges to be stored in the R-tree indexes in HiVision.

\begin{figure}[htbp]
		\begin{center}
			\includegraphics[width=0.8 \textwidth,trim=30 5 15 20,clip]{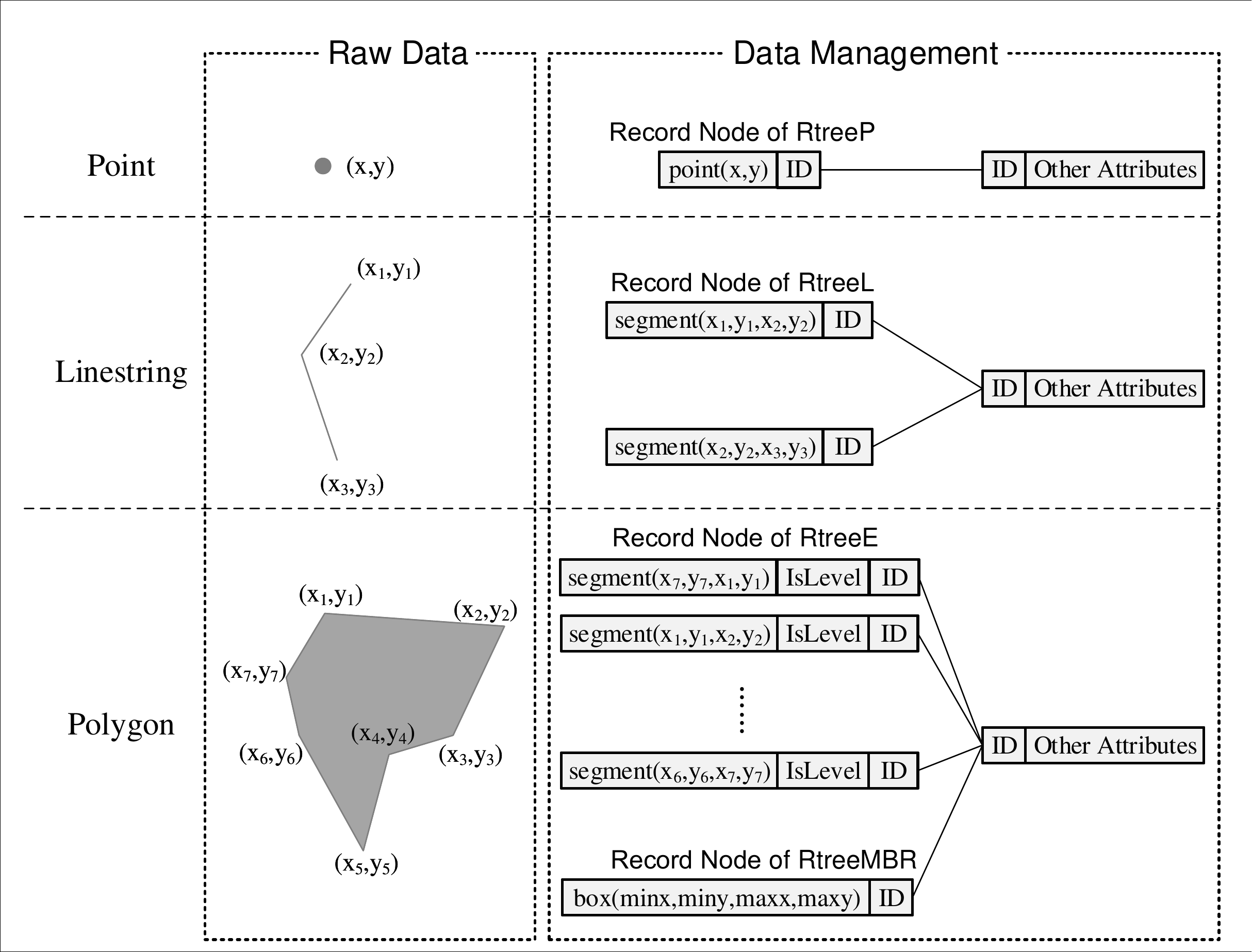}
			\caption{Vector data organization in HiVision.}
			\label{f3_2}			
		\end{center}
	\end{figure}

\begin{figure}[htbp]
		\begin{center}
			\subfigure[Edges reverse direction at endpoint (cutting process not required)]{
				\includegraphics[width=0.35\textwidth,trim=25 25 25 25,clip]{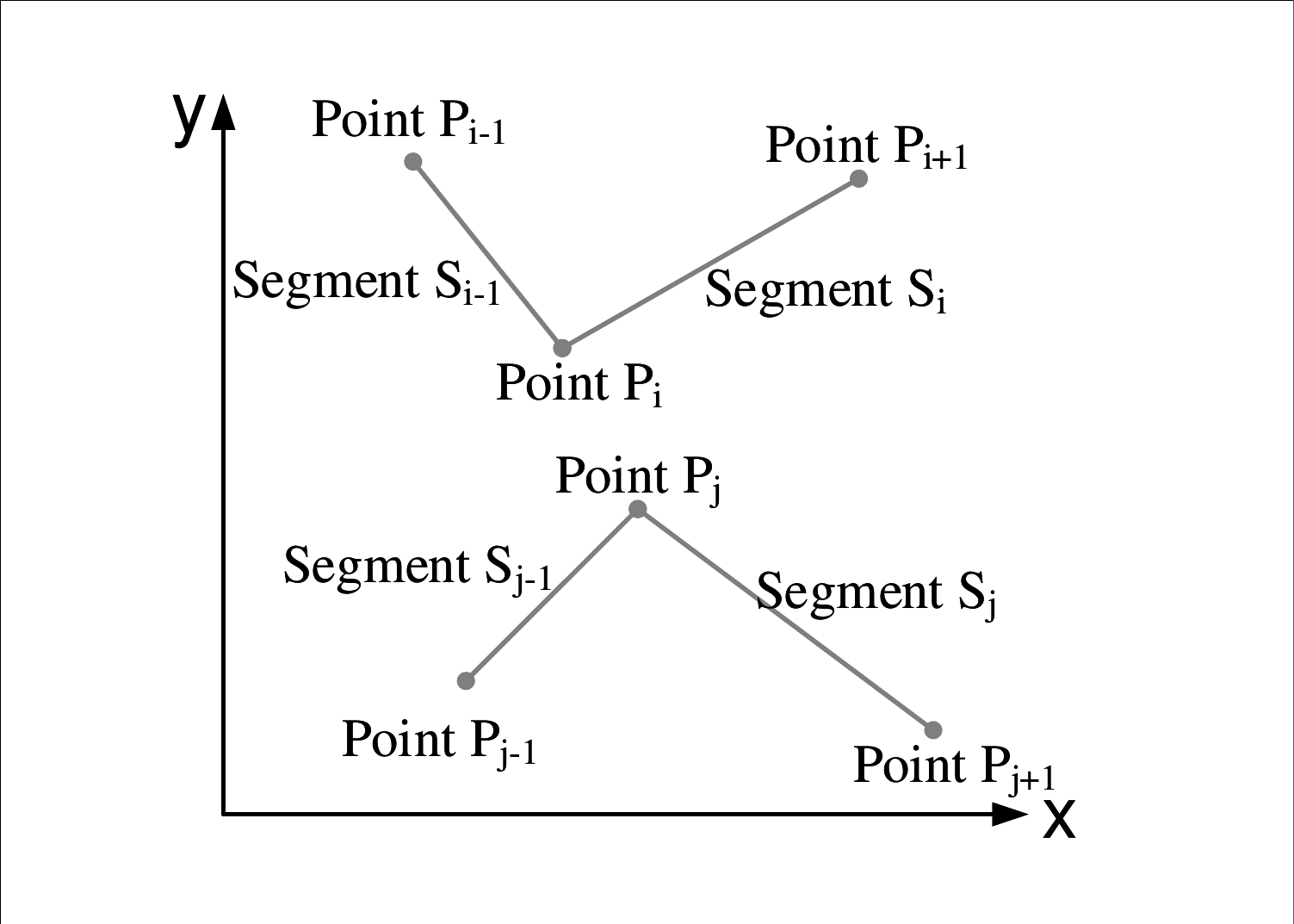}}%
			\subfigure[Edges monotonically increase or decrease (cutting process required)]{
				\includegraphics[width=0.65\textwidth,trim=25 25 25 25,clip]{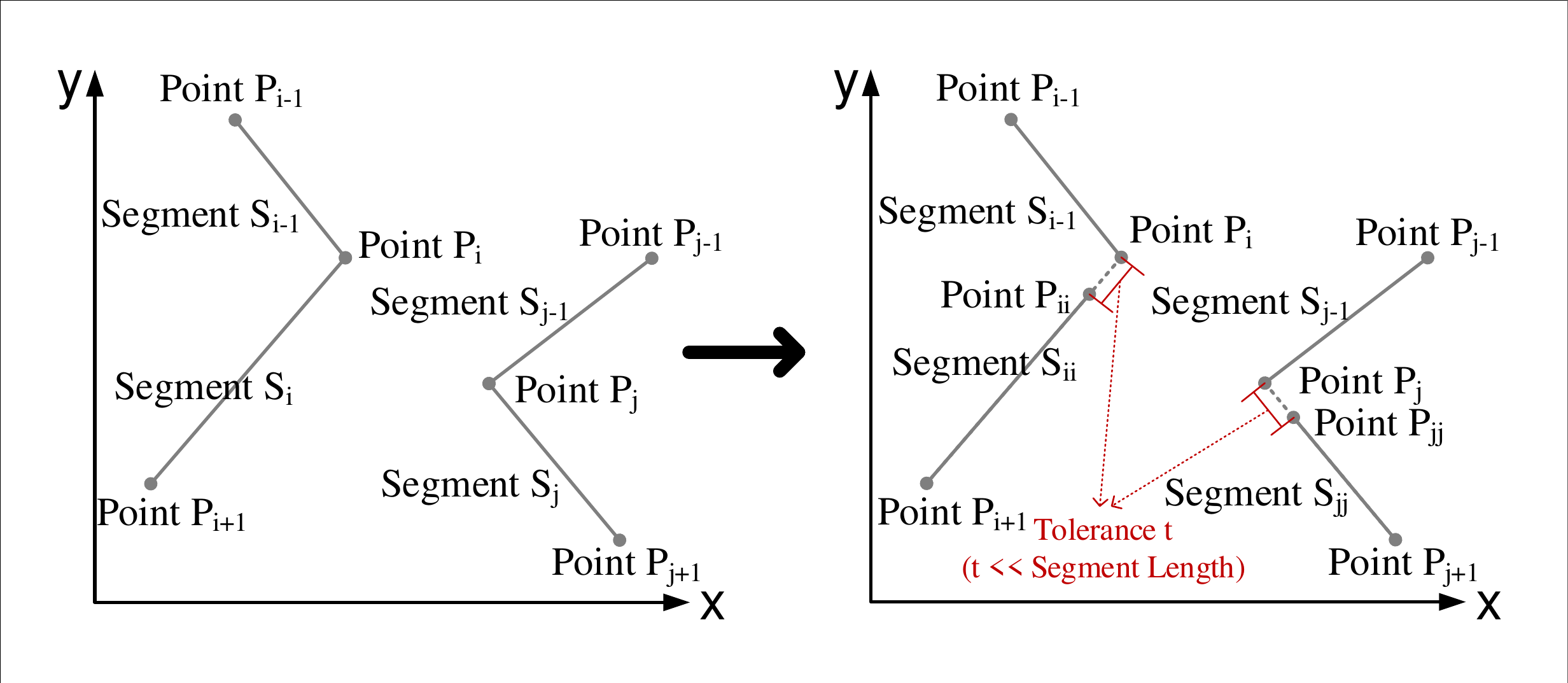}}%
			\caption{Segment cutting for polygon edges.}%
			\label{f3_3}
		\end{center}
	\end{figure}

	As shown in Figure~\ref{f3_2}, for point and linestring objects, we create R-tree indexes with point and segment as nodes types; for polygon objects which involve a filling step, we design a multi-level index architecture (MLIA). In MLIA, each edge of the polygon objects is stored as a segment in $RtreeE$ and the polygon MBRs are stored as boxes in $RtreeMBR$. In particular, to support spatial judging in SIBF, two operations are executed: 1) node information ($IsLevel$) is included in $RtreeE$ to identify whether the edge is parallel to the x-axis; 2) for the edges which monotonically increase or decrease, the segment cutting process is adopted (see Figure~\ref{f3_3}).
	
	\subsection{Data Visualization}
	
	\subsubsection{SIBV for Point, Linestring and Polygon Edges}
	
\begin{figure}[htbp]
		\begin{center}
			\includegraphics[width=0.18 \textwidth,trim=15 15 15 15,clip]{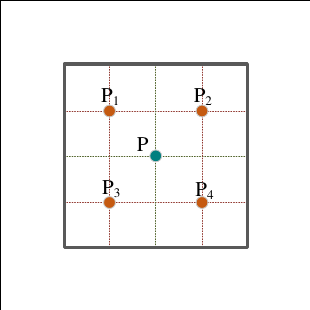}
			\caption{Super-sampling of pixel $P$ for anti-aliasing in SIBV.}
			\label{f3_4}			
		\end{center}
	\end{figure}

\begin{figure}[htbp]
		\begin{center}
			\subfigure[Before anti-aliasing]{
				\includegraphics[width=0.45 \textwidth,trim=60 40 40 120,clip]{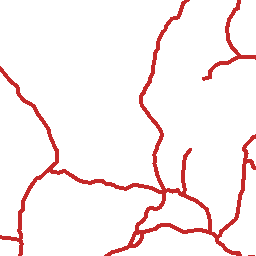}}%
			\subfigure[After anti-aliasing]{
				\includegraphics[width=0.45 \textwidth,trim=60 40 40 120,clip]{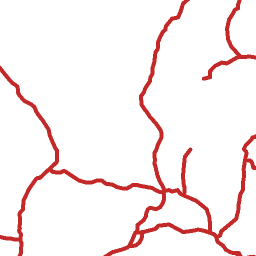}}%
			\caption{Improvement of visual effects with anti-aliasing in SIBV.}%
			\label{f3_5}
		\end{center}
	\end{figure}

	As points, linestrings and polygon edges are visualized with widths, it can be regarded as generating spatial buffers~\citep{Sommer2006A} of the objects. Different from general spatial buffer analysis which identifies areas by surrounding geographic features with a given spatial distance, for visualization, the widths of spatial objects are measured by the number of pixels. In SIBV, we extend the buffer generation method in HiBuffer~\citep{ijgi7120467} to visualize spatial point, linestring and the boundaries of polygon objects; moreover, we design a super-sampling approach for anti-aliasing: as shown in Figure~\ref{f3_4}, the pixel $P$ is split into four sub-pixels and a sample is taken from each sub-center. The value of P is generated by weighting the values of sub-pixels (Figure~\ref{f3_5} shows the improvement of visual effects in SIBV with the anti-aliasing approach).
	
	The details of SIBV are as shown in Algorithm 1, and the query boxes used in SIBV is illustrated in Figure~\ref{f3_6}. Two main factors are considered to optimize the algorithm: 1) the super-sampling process should only be used in the color transition regions, as it will surely increase the calculation amount; 2) when R-tree is used, intersect operators work well for queries using bounding-box rather than other shapes, and nearest-neighbor search has much higher computation complexity than the bounding-box query. We introduce $R_1$ (= $R - \sqrt{2}/4 \times R_z$) and $R_2$ (= $R + \sqrt{2}/4 \times R_z$). If the distance from $P$ to the nearest spatial object, defined as $D$, is less than $R_1$, it means that all the sub-pixels of $P$ are in the zones of rasterized spatial objects; if $D$ is between $R_1$ and $R_2$, $P$ belongs to the color transition regions; otherwise, $P$ belongs to the background. The query process in SIBV can be divided into two steps: 
	
	\begin{figure}[htbp]
		\begin{center}
			\includegraphics[width=0.4\textwidth,trim=15 15 15 15,clip]{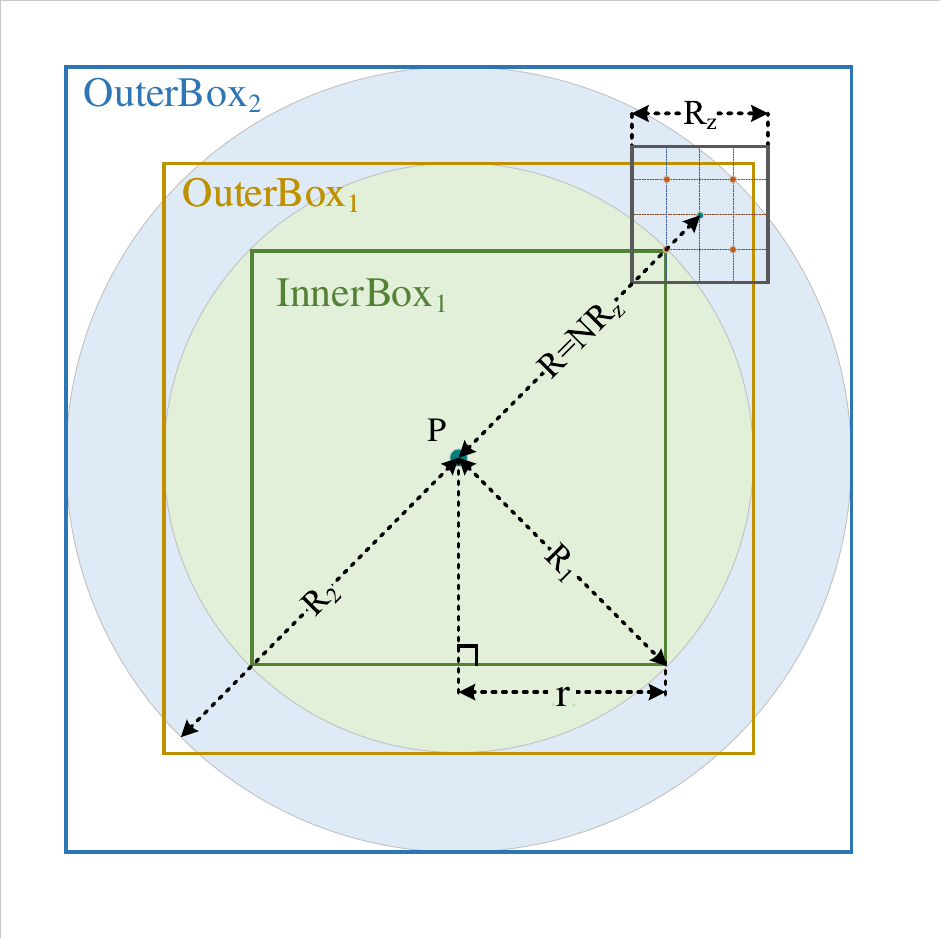}
			\caption{Query boxes for calculating pixel $P$ with $N$ as radius in SIBV ($R_z$: resolution at zoom level $Z$).}
			\label{f3_6}			
		\end{center}
	\end{figure}
	
	\textbf{Step 1} To determine whether $P$ is in the buffer area of spatial objects with $R_1$ as radius. We introduce $InnerBox_1$ and $OuterBox_1$ to deal with different situations in this step (if there are lots of spatial objects within the distance $R_1$ from $P$, we query the spatial objects intersects the $InnerBox_1$, as high density of spatial objects in the neighbor is very likely to intersect the inner box; and if there are few spatial objects in the neighbor of $P$, we use the $OuterBox_1$ to filter out the spatial objects which are far from $P$).
	
	\textbf{Step 2} $OuterBox_2$ is used as a filter to determine whether $P$ belongs to the color transition regions. If so, we calculate the number of sub-pixels that are in the plotting region. Specially, the number of sub-pixels in the plotting region is used as an indicator to identify the degree which $P$ belongs to the zones of rasterized spatial objects.
	
	\begin{algorithm}[!ht]
		\textbf {Algorithm 1:} Spatial-Index-Based Visualization\\
		\rule[5pt]{16.4 cm}{0.05em}
		\hspace*{0.02in} {\bf Input:} Pixel $P$, zoom level $Z$, radius $N$ (pixels) and spatial index $Rtree$ ($RtreeP$, $RtreeL$ or $RtreeE$).\\
		\hspace*{0.02in}{\bf Output:} 0 - 4 (0: $P$ belongs to the background region; 1 - 3: $P$ belongs to color transition regions; 4: $P$ totally belongs to the zones of rasterized spatial objects).
		\begin{spacing}{1}
			\begin{algorithmic}[1]
				\scriptsize
				\State $R_z \gets$ \Call{resolution}{$Z$} 
				\State $R \gets N \times R_z$ 
				\State $R_1 \gets R - \sqrt{2}/4 \times R_z$
				\State $R_2 \gets R + \sqrt{2}/4  \times R_z$
				\State $r \gets \sqrt{2}/2 \times R_1 $
				\State $InnerBox_1 \gets$ \Call{box}{$P.x-r$, $P.y-r$, $P.x+r$, $P.y+r$}
				\State $Tmp \gets$ satisfying $Rtree$.\Call{intersect}{$InnerBox_1$}.\Call{limit}{$1$}
				\If {$Tmp$ is not $null$}
				\Return 4
				\Else
				\State $OuterBox_1 \gets$ \Call{box}{$P.x-R_1$, $P.y-R_1$, $P.x+R_1$, $P.y+R_1$}
				\State $Tmp \gets$ satisfying $Rtree$.\Call{intersect}{$OuterBox_1$} and $Rtree$.\Call{nearest}{$P$}
				\If {$Tmp$ is not $null$ \&\& \Call{distance}{$Tmp$,$P$} $\leq R_1$}
				\Return 4
				\Else 
				\State $OuterBox_2 \gets$ \Call{box}{$P.x-R_2$, $P.y-R_2$, $P.x+R_2$, $P.y+R_2$}
				\State $Tmp \gets$ satisfying $Rtree$.\Call{intersect}{$OuterBox_2$} and $Rtree$.\Call{nearest}{$P$}
				\If {$Tmp$ is not $null$ \&\& \Call{distance}{$Tmp$,$P$} $\leq R_2$} 
				\State $P_1 \gets$ \Call{point}{$P.x-1/4 \times R_z$, $P.y+1/4\times R_z$}	\Comment{Super-sampling for anti-aliasing.}
				\State $P_2 \gets$ \Call{point}{$P.x+1/4 \times R_z$, $P.y+1/4 \times R_z$}	
				\State $P_3 \gets$ \Call{point}{$P.x-1/4 \times R_z$, $P.y-1/4\times R_z$}	
				\State $P_4 \gets$ \Call{point}{$P.x+1/4\times R_z$, $P.y-1/4\times R_z$}
				\State \Return $\sum_{i=1}^4\Call{distance}{Tmp,P_i} \leq R$ $?$ $1$ $:$ $0$	
				\EndIf
				\EndIf
				\EndIf
				\State \Return 0
			\end{algorithmic}
		\end{spacing}
	\end{algorithm}

	\subsubsection{SIBF for Polygon Filling}
	
\begin{figure}[htbp]
		\begin{center}
			\makeatletter
			\def
			\@captype{figure}
			\makeatother
			\subfigure[Monochromatic colors filling]{
				\includegraphics[width=0.3 \textwidth,trim=0 150 0 0,clip]{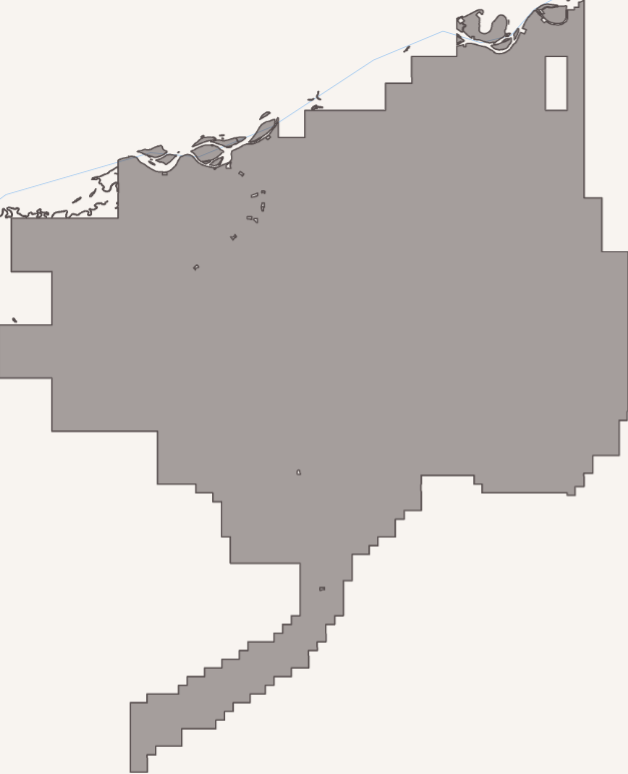}}%
			\hspace{1.4cm}
			\subfigure[Patterns filling]{
				\includegraphics[width=0.3 \textwidth,trim=2 152 0 0,clip]{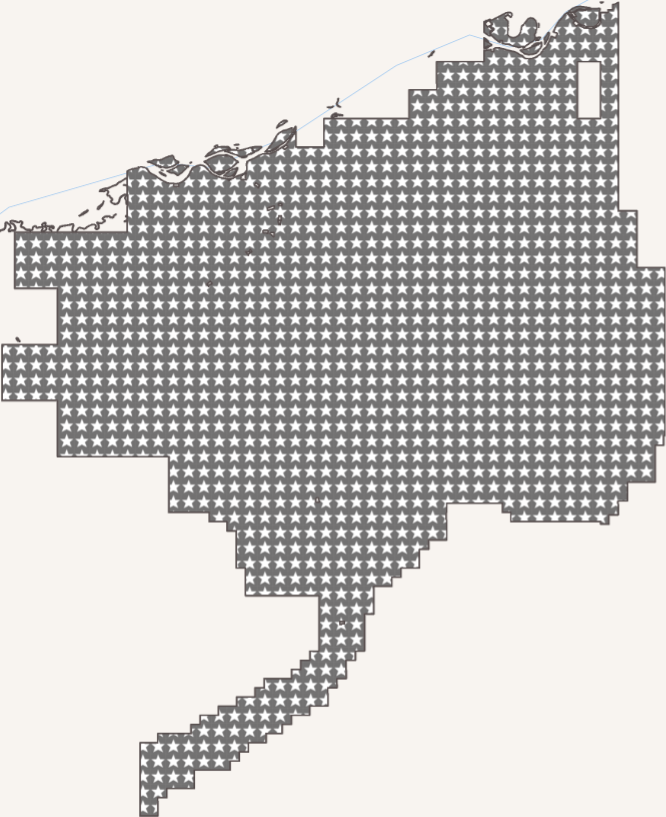}}%
			\caption{Visualization of polygon objects in HiVision.}%
			\vspace{-0.5cm}
			\label{f3_7}
		\end{center}
	\end{figure}

	We design the SIBF to determine whether the pixel $P$ is inside the polygon objects, so as to visualize the zones inside polygon objects. The details of SIBF are shown in Algorithm 2, we use the $RtreeMBR$ to find the candidate polygons and then measure the spatial relationship between the pixel and each candidate polygon one by one until the polygon which contains the pixel is found. We apply the ray casting algorithm~\citep{shimrat1962algorithm} to determine whether a pixel is inside a polygon. To be more specific, given a pixel and a polygon, a segment ($QuerySegment$) is drawn from the MBR boundaries of the polygon to the pixel which is parallel to the x-axis, then $RtreeE$ is used to calculate how many times the segment intersects the edges of the polygon (the edges in parallel with the x-axis are processed as invalid edges). The pixel is classified as 'inside the polygon' if the number of crossings is odd, or 'outside' if it is an even number. The result holds for polygons with inner rings. Moreover, as longer $QuerySegment$ may intersect large amounts of edges which belong to other polygons and thus cause performance degradation, two optimizations have been made to minimize the length of the $QuerySegment$: 1) the polygons with smaller x spans are used for spatial judging preferentially (in line 2 in Algorithm 2); 2) the vertical segment from the pixel to the closer edge of the polygon MBR is used as $QuerySegment$ (details are given in line 6-9). When the spatial relationships are determined, we can then render the pixels inside polygon objects according to the given styles. In the current implementation, Monochromatic colors and patterns filling are both supported. Figure~\ref{f3_7} shows the visual effects of polygon objects in HiVision.

	\begin{algorithm}[!htbp]
		\textbf {Algorithm 2:} Spatial-Index-Based Filling\\
		\rule[5pt]{16.4 cm}{0.05em}
		\hspace*{0.02in} {\bf Input:} Pixel P, RtreeE and RtreeMBR.\\
		\hspace*{0.02in}{\bf Output:} True or False (whether P is in polygons).
		\begin{spacing}{1}
			\begin{algorithmic}[1]
				\scriptsize
				\State $TmpMBR \gets$ satisfying $RtreeMBR$.\Call{intersect}{$P$}
				\State \Call{sort}{$TmpMBR$} \Comment{Polygon with smaller x span has higher priority.}
				\For{$v \in TmpMBR$}
				\State $EdgeCount \gets 0$
				\State $vMinx \gets v.Box.minx$, $vMaxx \gets v.Box.maxx$
				\If {$P.x-vMinx < vMaxx-P.x$}
				\State $QuerySegment \gets$ \Call{segment}{$vMinx$, $P.y$, $P.x$, $P.y$}
				\Else
				\State $QuerySegment \gets$ \Call{segment}{$P.x$, $P.y$, $vMaxx$, $P.y$}
				\EndIf
				\State $TmpS \gets$ satisfy $RtreeE$.\Call{intersect}{$QuerySegment$}
				\For{$s \in TmpS$}
				\If {(not $s.IsLevel)\&\& s.ID==v.ID$} 
				\State $EdgeCount ++$
				\EndIf
				\EndFor
				\If {$EdgeCount$ is odd}
				\Return True
				\EndIf
				\EndFor
				\State \Return False
			\end{algorithmic}
		\end{spacing}
	\end{algorithm}
	
	\vspace{-0.5cm}
	
	\vspace{-0.8cm}
	
	\subsection{Superiority analysis}
	
	HiVision outperforms traditional data-driven solutions in the following two aspects:
	
	\begin{itemize}
		\item (\textbf{low computational complexity}) Assume $n$ to be the number of spatial objects for visualization. \textcolor{black}{As the observation resolution limit of human eyes\citep{eyes} and the processing capacity limit of the browsers, the number of calculated pixels for display in a screen has an upper limit and can be regarded as a constant factor.} In data-driven solutions, as each object will be computed and analyzed successively, the total \textcolor{black}{time complexity} is O($n$). In contrast, the computing units in HiVision are pixels and we introduce R-tree indexes to accelerate the process of finding the objects to determine the value of each pixel; as a result, the \textcolor{black}{time complexity} is reduced to O($log(n)$).
		
		\item (\textbf{easy to parallel}) Real-world spatial datasets have the spatial unbalanced distribution property, which creates challenges with respect to efficient parallel processing. Using DataDC, in general, complex partitioning and merging strategies need to be designed; however, few strategies are capable of handling all kinds of spatial distributions with good load balancing. In contrast, as the \textcolor{black}{time complexity} of the algorithm is O($log(n)$), it indicates that our approach is less sensitive to spatial distributions and simply partitioning the analysis task by dividing the pixels equally can achieve good load balancing.
		
	\end{itemize}

	\section{Architecture}\label{section 4}

\begin{figure}[htbp]
		\begin{center}
			\makeatletter
			\def
			\@captype{figure}
			\makeatother
			\includegraphics[width=0.93\linewidth,trim=25 15 15 15,clip]{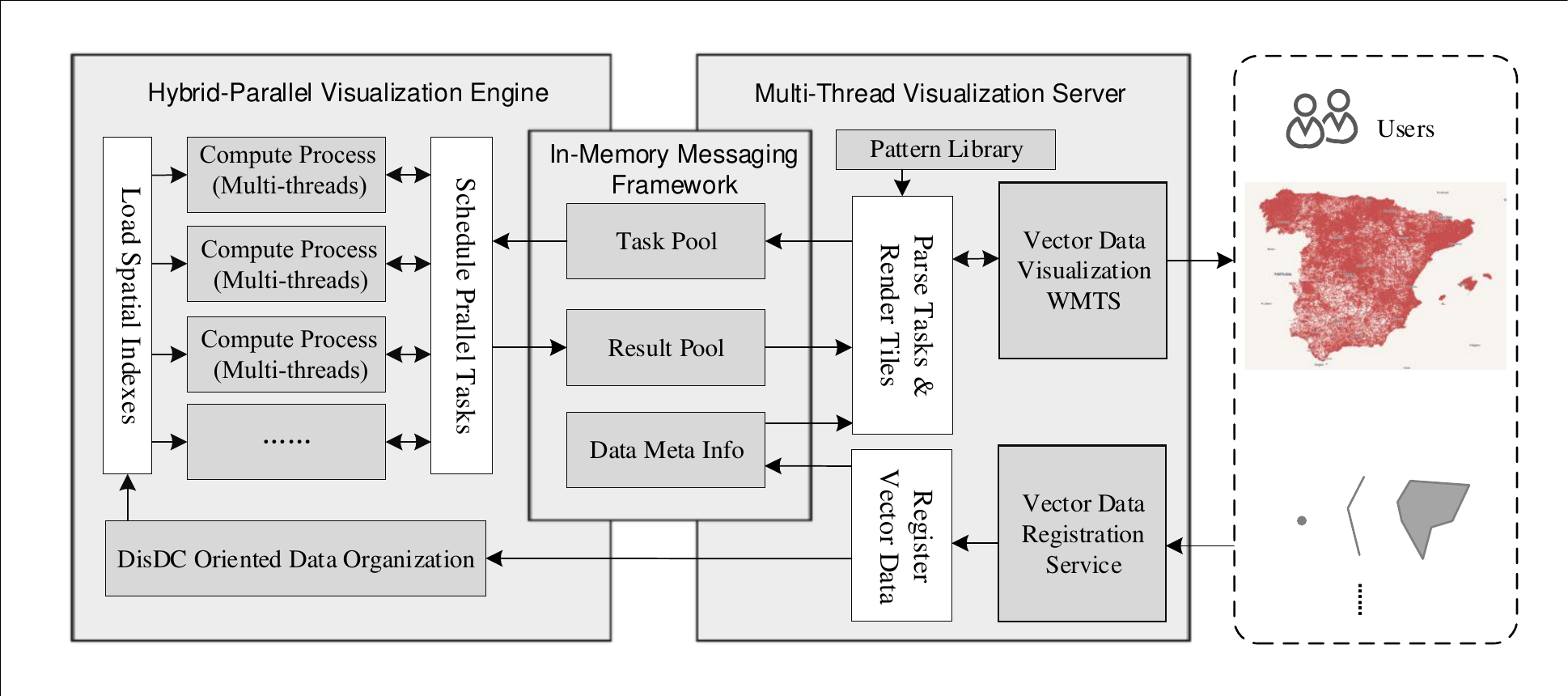}
			\caption{Architecture of HiVision.}
			\label{f4_1}			
		\end{center}
	\end{figure}

	To provide an interactive exploration of large-scale spatial vector data, we design a high-performance parallel processing architecture as illustrated in Figure~\ref{f4_1}. The DisDC oriented vector data organization structure is stored as memory-mapped files~\citep{memorymapped}, which do not need to be totally loaded into memory while accessing the files. The architecture of HiVision adopts the browser-server application model. In HiVision, visualization results are organized into the tile-pyramid structure with \(256 \times 256\) pixels as the tile size. When a user browses the spatial datasets, tiles in the display range will be rendered on the fly according to the rendering styles. The server side of the architecture consists of three parts: Multi-Thread Visualization Server (MTVS), In-Memory Messaging Framework (IMMF) and Hybrid-Parallel Visualization Engine (HPVE).

	\subsection{Multi-Thread Visualization Server}
	
	In MTVS, the spatial data visualization service is encapsulated as a WMTS, which can be easily added to web maps as a raster layer. The visualizing process of each tile is treated as independent tasks. The \textbf{Parse Tasks} process analyzes tile requests and generates visualization tasks in \textbf{Task Pool}; in particular, the following types of tiles will not lead to new tasks: 1) tiles that are not in the spatial scope of data MBRs; 2) tiles that are previously processed and the visualization results that are still preserved in the \textbf{Result Pool}; 3) tiles with wrong request expressions. The \textbf{Render Tiles} process gets visualization results from the \textbf{Result Pool} once the visualizing process is done in HPVE, and renders tiles according to the style provided by users; The \textbf{Pattern Library} stores the patterns used for pattern filling style of polygon objects. MTVS provides a data registration interface, and in the \textbf{Register Vector Data} process, MTVS creates spatial indexes and writes dataset meta-data (e.g., MBRs) to \textbf{Data Meta Info}. In addition, multi-thread technology is adopted in MTVS to improve concurrency.
	
	\subsection{In-Memory Messaging Framework}
	
	To reduce message transmission time, tasks, results, control messages are delivered in memory without disk I/O in IMMF. IMMF is implemented based on Redis, a distributed, in-memory key-value database. \textbf{Task Pool} is a first-in-first-out queue that stores the requested tile tasks; the tasks are pushed to the list by MTVS and popped to the task processors in HPVE, and the operations are executed in blocking mode to avoid repeated allocation of tasks. \textbf{Result Pool} stores the visualization results in key-value structure: the key identifies the tile request expression, while the value stores the visualization results (in the form of a two-dimensional array indicating different zones for the rendering process). Once a task is finished in MPVE, the visualization result will be written to the \textbf{Result Pool}, and then a task completion message will be sent to MTVS through subscribing/publishing in Redis. To avoid overwhelming memory consumption, visualization results are set with an expire time window and expired results will be cleaned up if memory usage reaches the upper limit.		
	
	\subsection{Hybrid-Parallel Visualization Engine}
	
	HPVE adopts a hybrid MPI-OpenMP parallel processing model and a dynamic task partitioning strategy to achieve real-time exploration of large-scale spatial vector data. In HiVision, each tile is partitioned by lines and processed with multiple OpenMP threads in one MPI process. When a user browses the spatial datasets, the tasks are generated in a streaming way and handled at a first-in/first-served basis. An MPI process will be suspended if no tasks are assigned or the assigned tasks are finished. Tasks are dynamically allocated to a suspended MPI process, and if there are no available free MPI processes, the extra task will be temporarily stored in the \textbf{Task Pool} waiting for idle processes.

	\section{Experimental Evaluation}\label{section 5}
	
	In this section, we conduct several experiments to evaluate the performance of HiVision. Firstly, we compare HiVision with the typical data-driven methods which are popular in recent years; then, we test the ability of HiVision to support interactive exploration of large-scale spatial vector data; moreover, we carry out an experiment to analyze the influence of request rates while providing interactive visualization in HiVision; finally, the parallel scalability of HiVision is tested through running with varying numbers of MPI processes and OpenMP threads.
	
	All the experiments are conducted on a cluster with four nodes (Table~\ref{t5_1}). The computer code of HiVision is implemented in C++, based on Boost C++ 1.64, MPICH 3.04, GDAL 2.1.2 and Redis 3.2.12. The data-driven methods are based on Hadoop 2.9.2, Spark 2.3.3, SpatialHadoop 2.4.2, GeoSpark 1.2.0 and Mapnik 3.0.22. In HiVision, the spatial indexes could be constructed quickly~\citep{BoostGeometry} and the experiments are conducted based on the pre-built DisDC oriented vector data organization structure. \textcolor{black}{The spatial indexes can be used to provide interactive buffer and overlay analysis as well\citep{ijgi7120467,ijgi8010021}}. To deploy HiVision on the cluster, we keep a copy of index files in each cluster node so that all the processes can access the spatial vector data efficiently. Table~\ref{t5_2} shows the datasets used in the experiments, and the datasets are all on the planet level. P$_{1}$ is from OpenCellID\footnote{ https://opencellid.org}, which is the world's largest collaborative community project that collects GPS positions of base stations. Other datasets are from OpenStreetMap, which is a digital map database built through crowdsourced volunteered geographic information. L$_{7}$, P$_{2}$ and A$_{2}$ respectively contain all the linestrings, points and polygons on the planet from OpenStreetMap; and there are more than 1 billion segment/point/edge items in each of the dataset.
	
	\begin{table}[htbp]
		\caption{Experimental environment.}
		\centering
		{
			\begin{tabular}{ll}
				\hline
				\textbf{Item}&\textbf{Description}\\
				\hline
				\ CPU& $4 \times 32$ cores, Intel(R) Xeon(R) E5-4620@2.60 GHz\\
				\ Memory& $4 \times 256$ GB\\
				\ Operating System& Centos 7.1\\
				\hline
			\end{tabular}	
		}
		\label{t5_1}
	\end{table}
	
	\begin{table}[htbp]
		\caption{Datasets used in the experiments.}
		\centering
		{
			\begin{tabular}{@{}lllll}
				\hline
				\textbf{Dataset}& \textbf{Type}&\textbf{Records}& \textbf{Size}\\
				\hline
				L$_{1}$: OSM postal code areas boundaries&Linestring&171,226& 65,334,342 segments\\
				L$_{2}$: OSM boundaries of cemetery areas&Linestring&193,076&1,800,980 segments\\
				L$_{3}$: OSM sporting areas boundaries&Linestring&1,767,137&18,969,047 segments\\
				L$_{4}$: OSM water areas boundaries&Linestring&8,419,324& 376,208,235 segments\\
				L$_{5}$: OSM parks green areas boundaries&Linestring&9,961,896& 454,636,308 segments\\
				L$_{6}$: OSM roads and streets&Linestring&72,339,945&717,048,198 segments\\
				L$_{7}$: OSM all linestrings on the planet&Linestring&106,269,321&1,578,947,752 segments\\
				P$_{1}$: OpenCelliD cell tower locations&Point&40,719,479&40,719,478 points\\
				P$_{2}$: OSM all points on the planet&Point&2,682,401,763&2,682,401,763 points\\
				A$_{1}$: OSM buildings&Polygon&114,796,734& 689,197,342 edges\\
				A$_{2}$: OSM all polygons on the planet&Polygon&177,662,806&2,077,524,465 edges\\
				\hline
			\end{tabular}
		}
		\label{t5_2}
	\end{table}
	
	\vspace{-0.3cm}
	
	\subsection{Experiment 1. Outperforming Data-driven Methods}
	
	In order to highlight the superiority of HiVision, we compare HiVision with three typical data-driven methods, namely, HadoopViz, GeoSparkViz and Mapnik. All the methods are deployed on the cluster with four nodes. HadoopViz and GeoSparkViz are respectively implemented based on the Hadoop and the in-memory Spark with well load-balance task partition strategies; given a spatial dataset, the methods can generate all the tiles of given zoom levels. Mapnik is an open-source toolkit for rendering maps. The inputs of Mapnik are the spatial objects in the tile range and the rendering styles, and the output is rendered map tile. In the experiment, the tile rendering tasks are stored in a queue and multiple Mapnik rendering processes are launched to process the tasks successively. We have totally started 128 Mapnik rendering processes in the cluster. HiVision is set to run with 128 MPI processes and 2 OpenMP threads in each process. For each dataset, we generate tiles of zoom levels 1, 3, 5, 7 and 9 with different methods; the numbers of tiles in each level are 4, 64, 1024, 16384 and 262144.
	
\begin{figure}[htbp]
		\centering
		\includegraphics[width=0.9\textwidth,trim=80 115 70 120,clip]{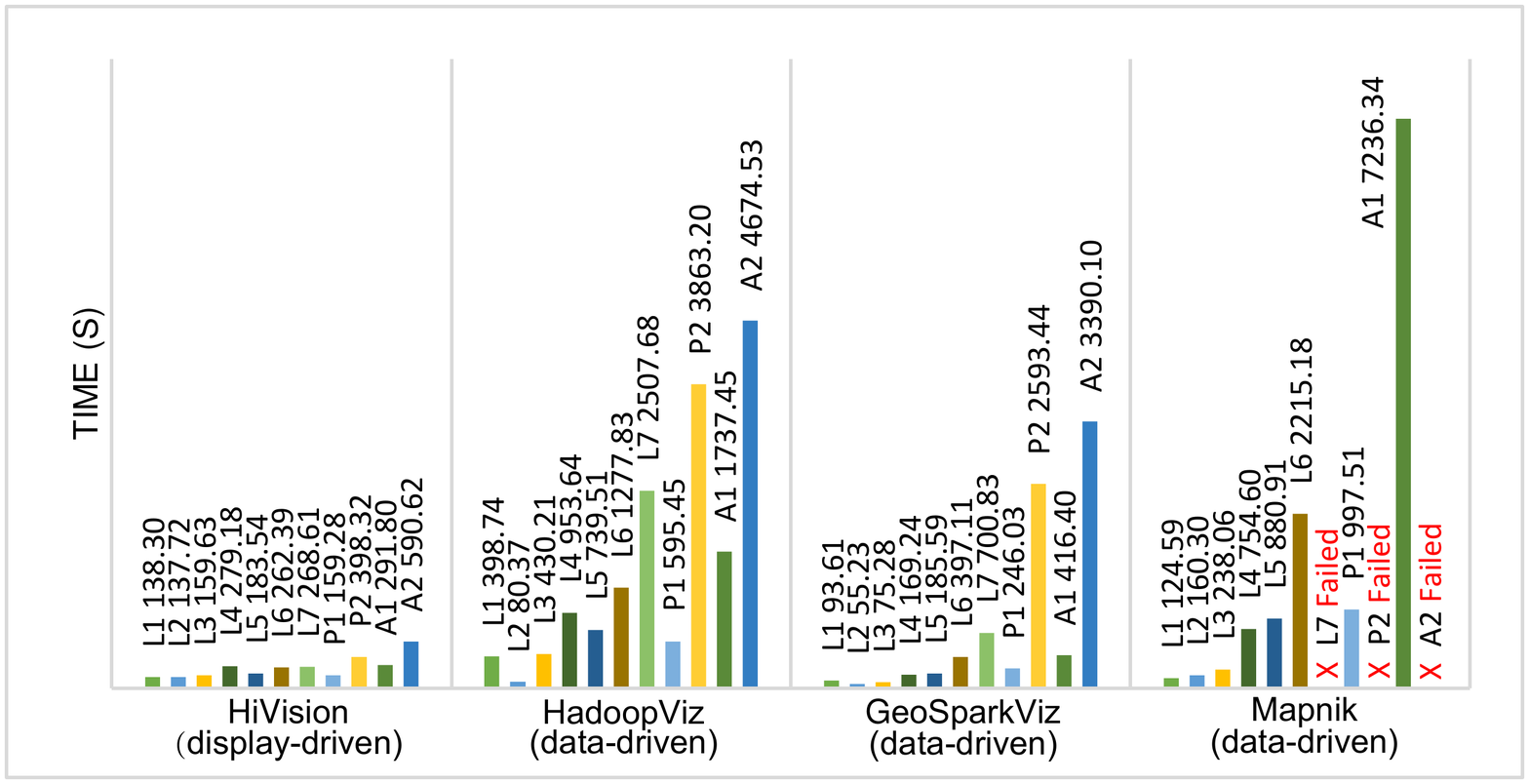}
		\caption{Total rendering time of generating all the tiles in zoom levels 1, 3, 5, 7 and 9.}
		\label{f5_1}	
	\end{figure}

	Figure~\ref{f5_1} shows the total rendering time of all the tiles in zoom levels 1, 3, 5, 7 and 9 with different methods. GeoSparkViz shows high performance among the data-driven methods: for all the datasets, GeoSparkViz takes the shortest time than other data-driven methods. Taking HiVision and GeoSparkViz for comparison, GeoSparkViz shows higher performance while the dataset scale is small (L$_{1-4}$), and HiVision outperforms GeoSparkViz on larger datasets (L$_{5-7}$, P$_{1-2}$, A$_{1-2}$). For the billion-scale datasets L$_{7}$, P$_{2}$ and A$_{2}$, HiVision shows the high performance and it respectively takes about 38.33\% ($=$268.61s $\div$ 700.83s), 15.36\% ($=$398.32s $\div$ 2593.44s), 17.42\% ($=$590.62s $\div$ 3390.10s) of the rendering time using GeoSparkViz on each dataset. From L$_{1}$ to L$_{7}$, the data size increases sequentially, there is no significant uptrend in the tile rendering time using HiVision; in contrast, the tile rendering time of data-driven methods expands rapidly with the increase of data scales. Surprisingly, using HiVision, L$_{7}$, the largest linestring dataset with more than 1 billion segments, produces better performance than L$_{4}$, which has a much smaller scale. Experiment results show that HiVision is less sensitive to data volumes. In addition, HiVision produces better visual effects. Neither HadoopViz nor GeoSparkViz contains the anti-aliasing and polygon filling processes, and polygon objects are treated as linestring objects in the methods. Mapnik, as a mature map cartography tool, can visualize spatial objects with various styles; however, Mapnik failed to process the billion-scale datasets L$_{7}$, P$_{2}$ and A$_{2}$. In conclusion, compared with traditional data-driven methods, HiVision produces higher performance with better visual effects while dealing with large-scale spatial vector data.
	
\begin{figure}[htbp]
		\begin{center}
			\makeatletter
			\def
			\@captype{figure}
			\makeatother
			\subfigure[L$_{1}$]{
				\includegraphics[width=0.33\textwidth,trim=80 110 60 160,clip]{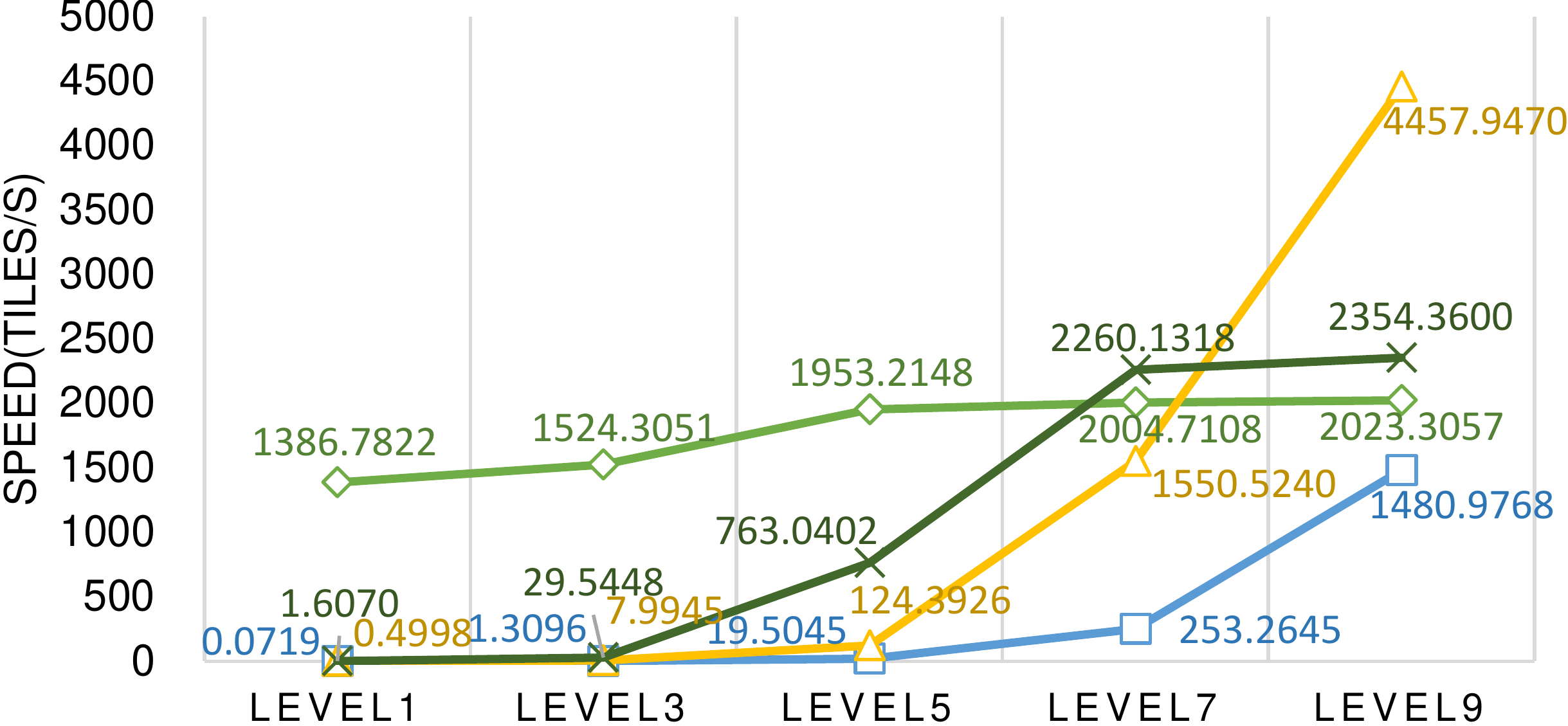}}%
			\subfigure[L$_{2}$]{
				\includegraphics[width=0.33\textwidth,trim=80 110 60 160,clip]{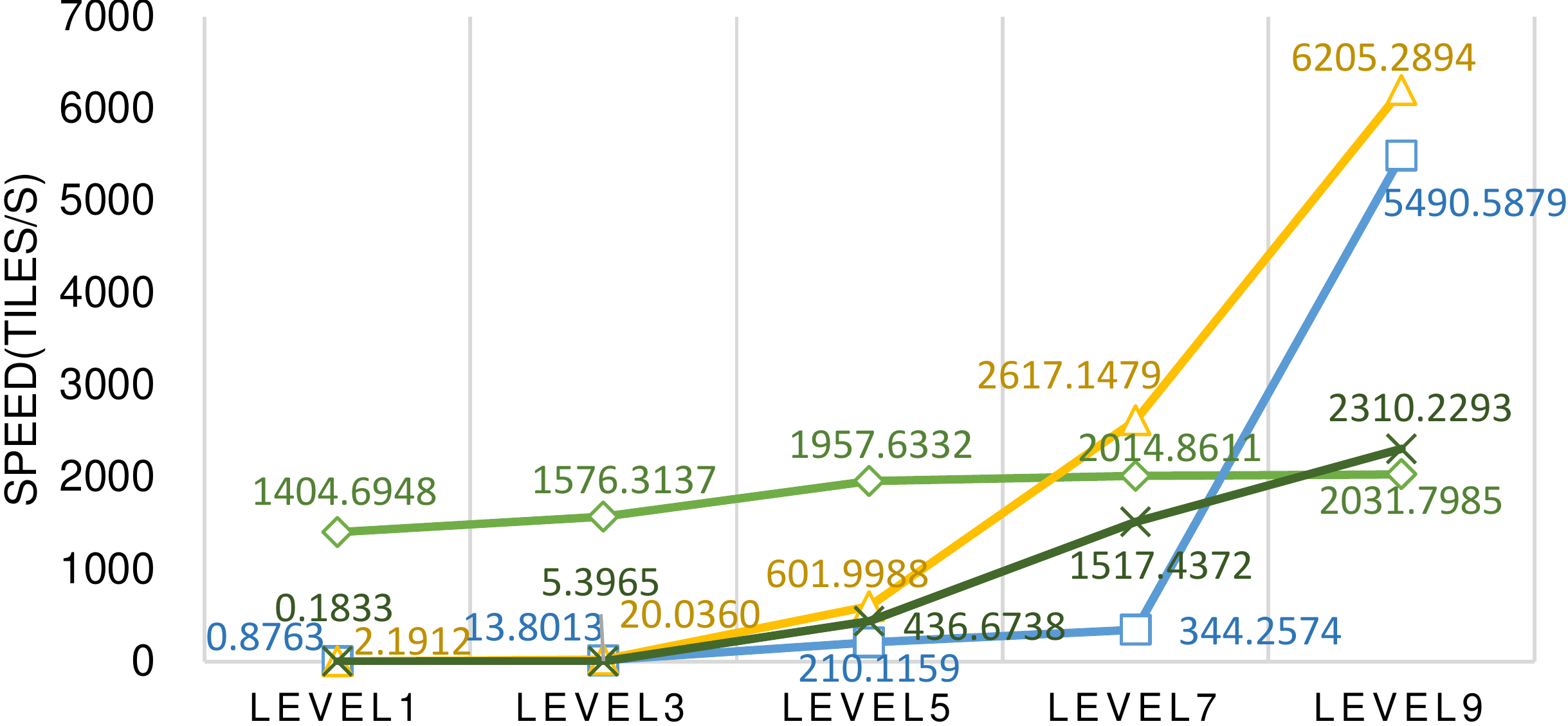}}%
			\subfigure[L$_{3}$]{
				\includegraphics[width=0.33\textwidth,trim=80 110 60 160,clip]{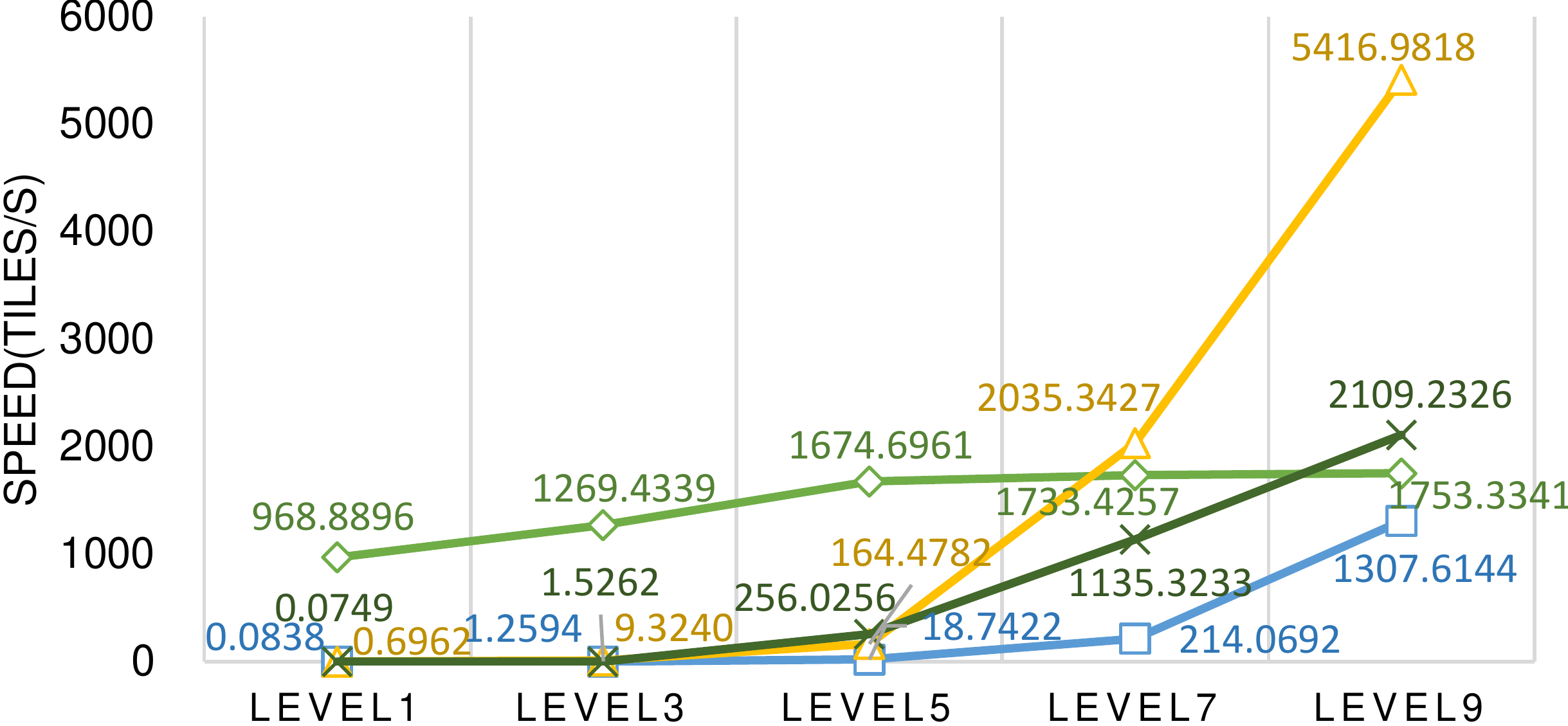}}%
			\vfill
			\subfigure[L$_{4}$]{
				\includegraphics[width=0.33\textwidth,trim=80 110 60 160,clip]{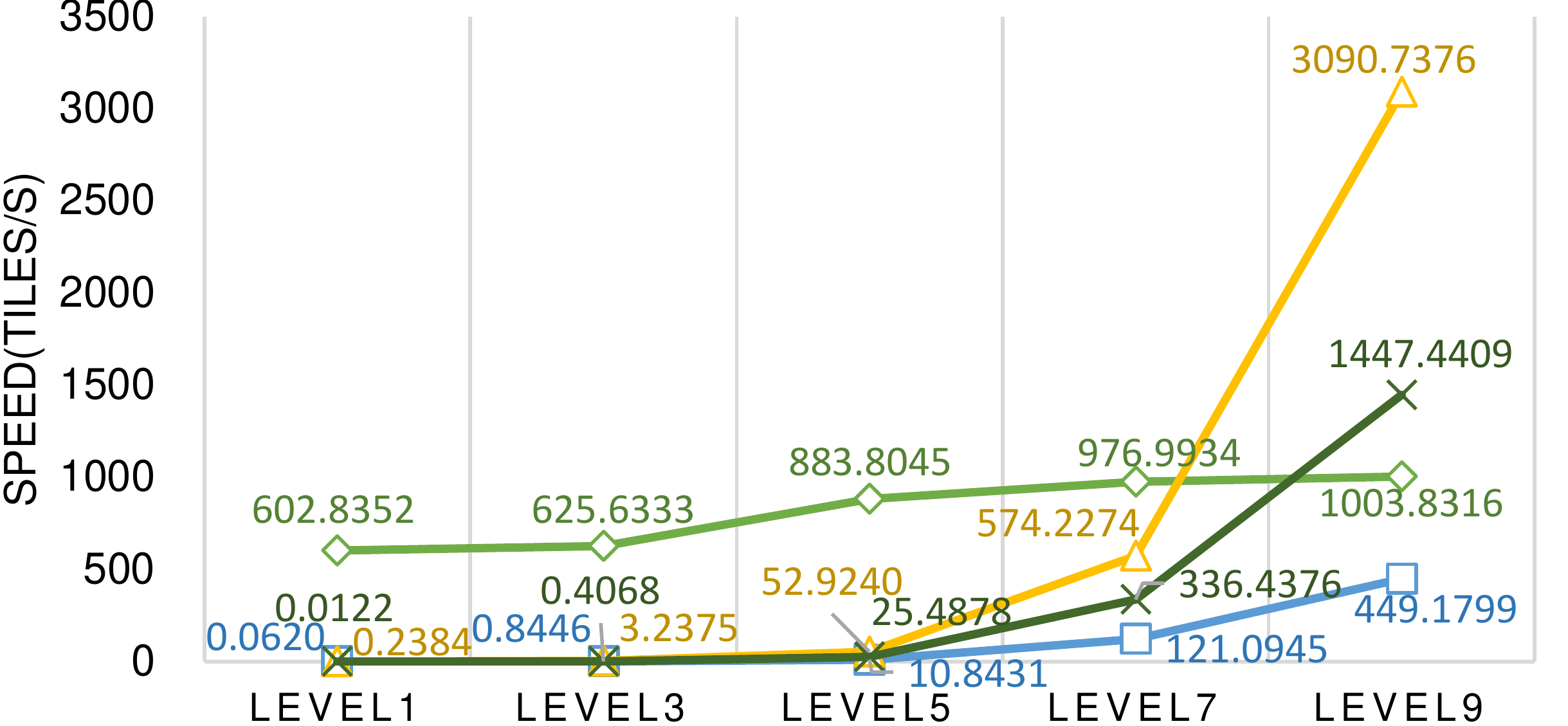}}%
			\subfigure[L$_{5}$]{
				\includegraphics[width=0.33\textwidth,trim=80 110 60 160,clip]{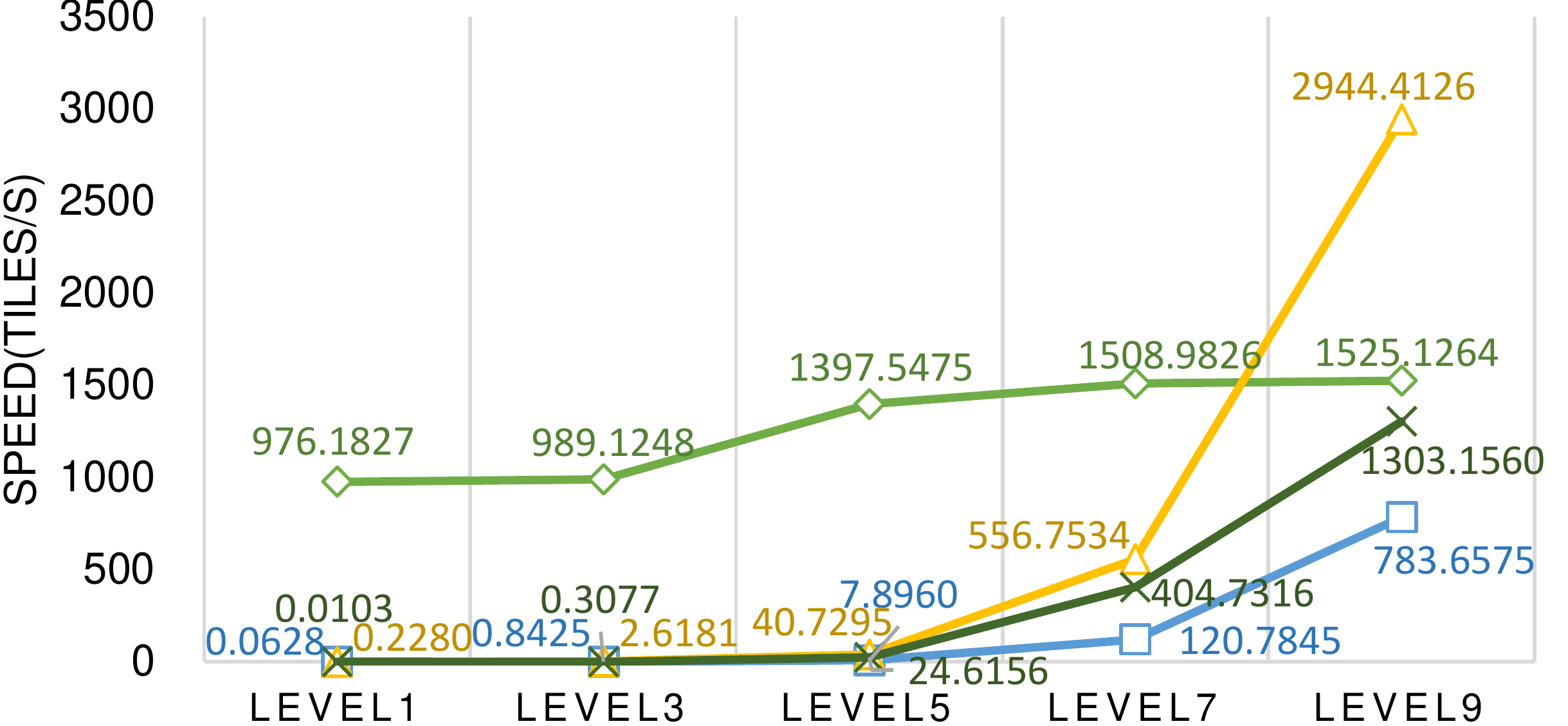}}%
			\subfigure[L$_{6}$]{
				\includegraphics[width=0.33\textwidth,trim=80 110 60 160,clip]{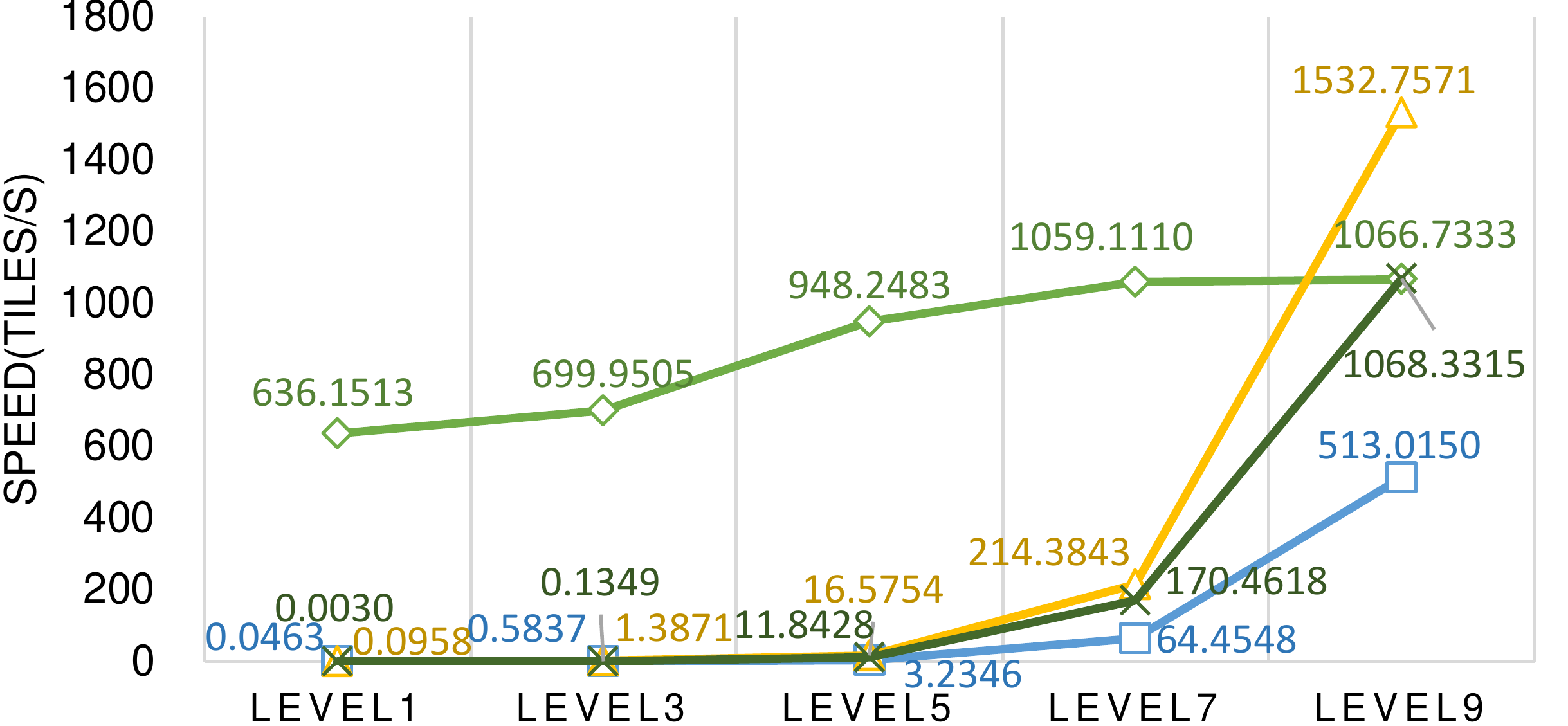}}%
			\vfill
			\subfigure[L$_{7}$ (Mapnik failed)]{
				\includegraphics[width=0.33\textwidth,trim=75 100 60 170,clip]{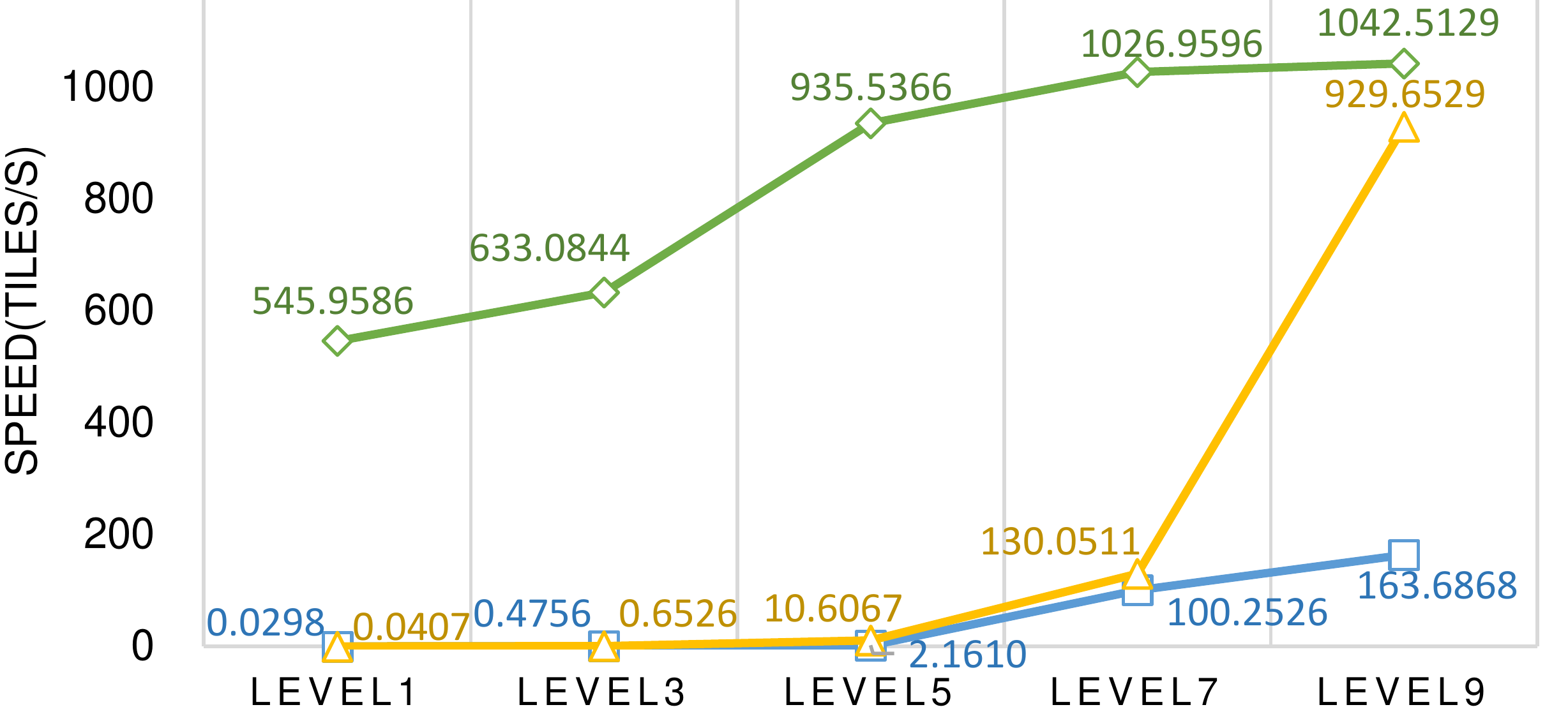}}%
			\subfigure[P$_{1}$]{
				\includegraphics[width=0.33\textwidth,trim=80 110 60 160,clip]{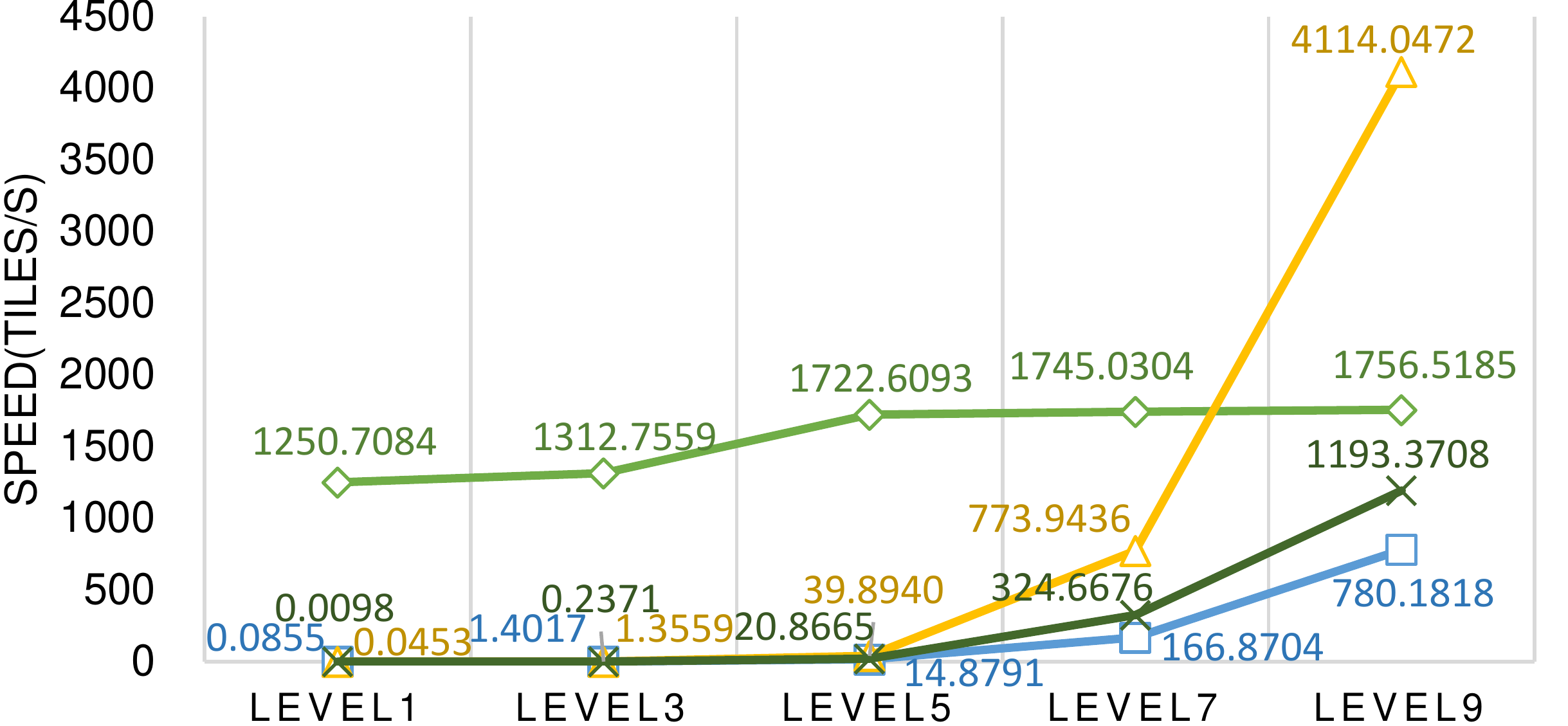}}%
			\subfigure[P$_{2}$ (Mapnik failed)]{
				\includegraphics[width=0.33\textwidth,trim=80 100 60 170,clip]{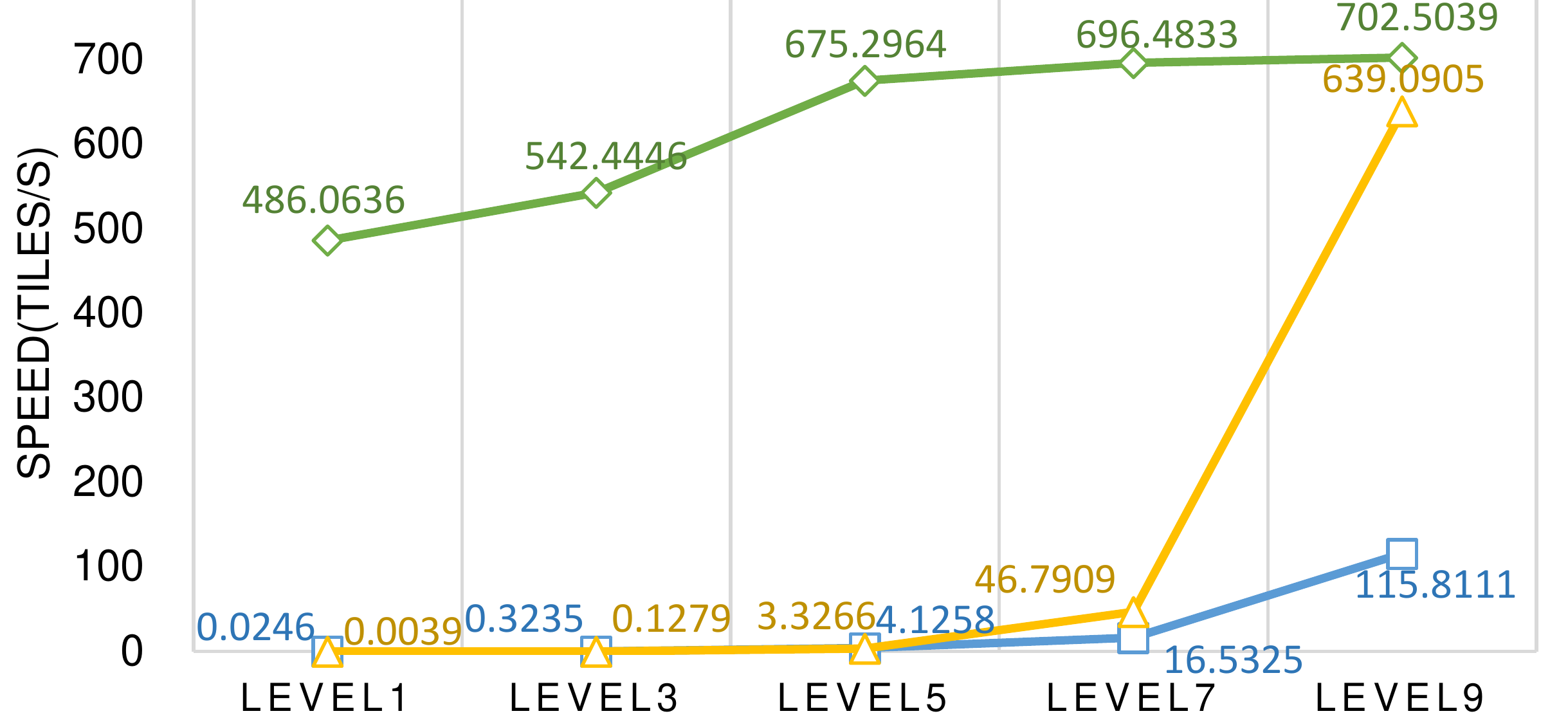}}%
			\vfill
			\subfigure[A$_{1}$]{
				\includegraphics[width=0.33\textwidth,trim=80 110 60 160,clip]{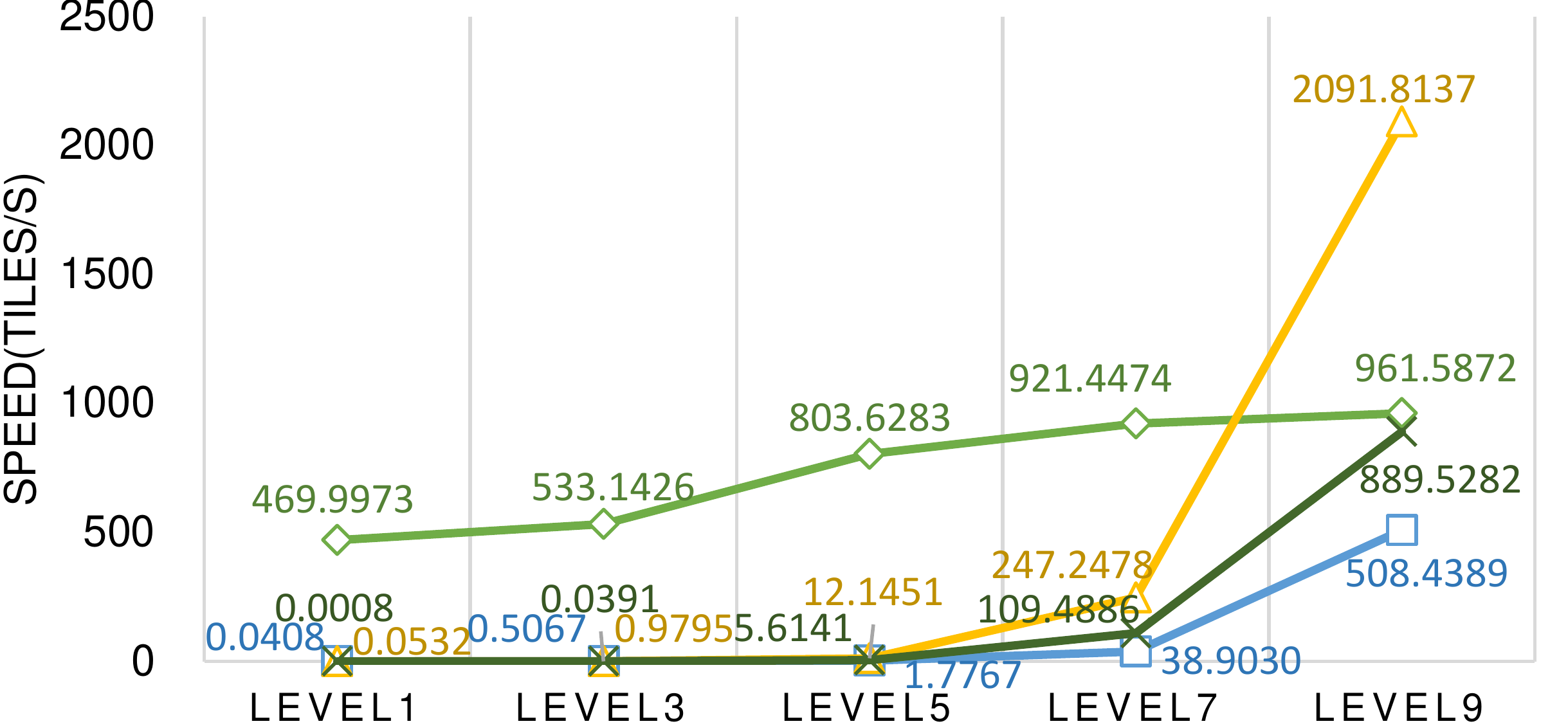}}%
			\subfigure[A$_{2}$ (Mapnik failed)]{
				\includegraphics[width=0.33\textwidth,trim=80 100 60 170,clip]{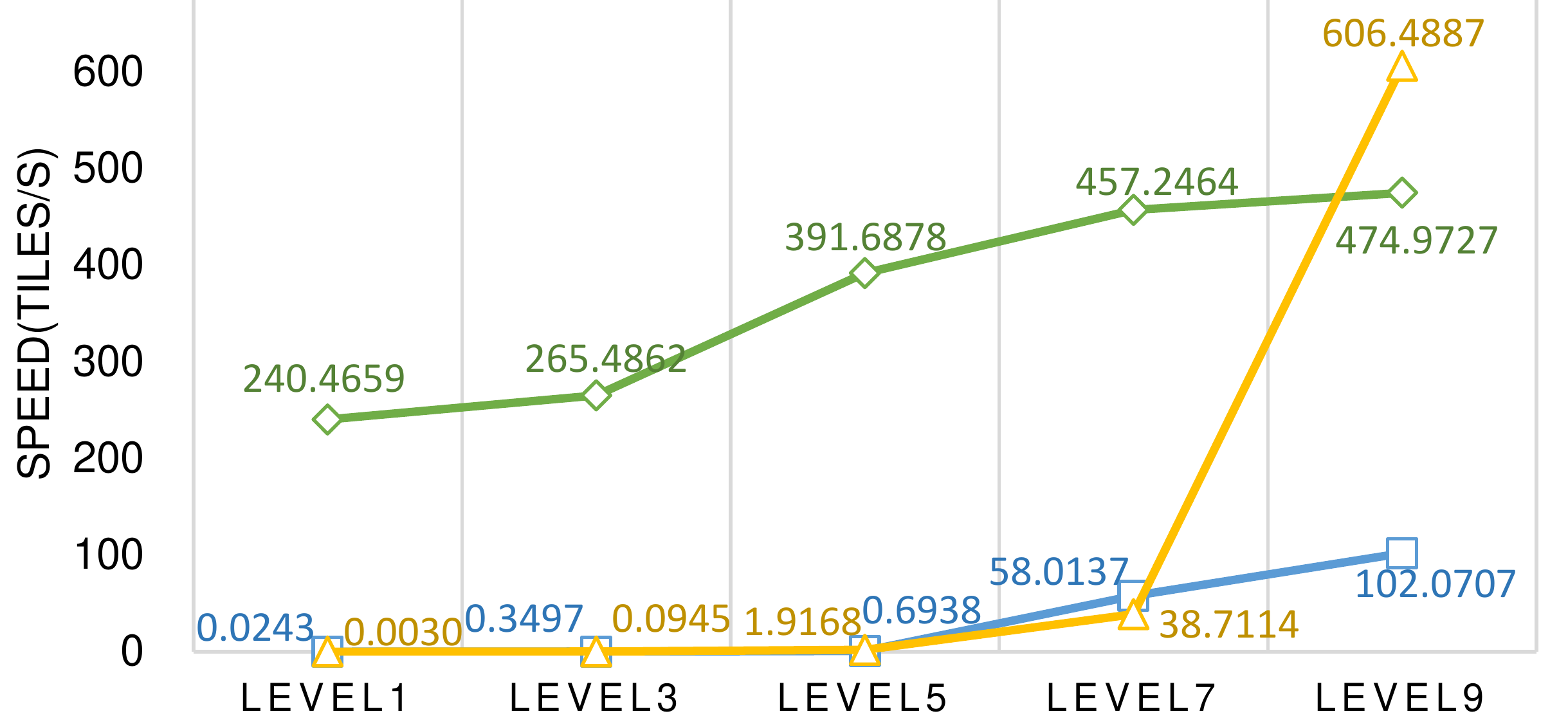}}%
			\subfigure{
				\includegraphics[width=0.2\textwidth,trim=80 310 510 110,clip]{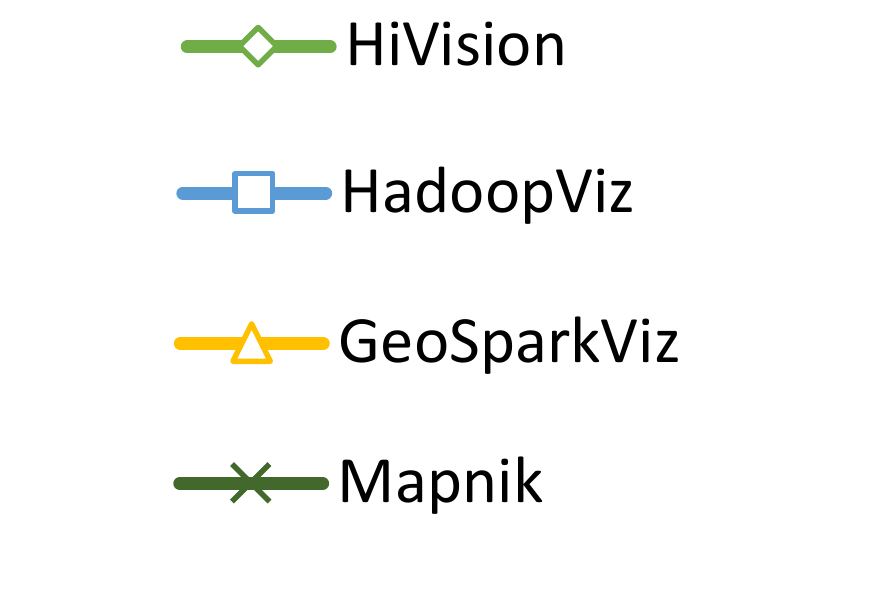}}%
			\caption{Tile rendering speed of different zoom levels.}%
			\label{f5_2}
		\end{center}
	\end{figure}

	Figure~\ref{f5_2} shows the tile rendering speed of different zoom levels. As illustrated by the results, data-driven methods show high tile rendering speed while zoom levels are high, however, the speed decreases rapidly while the zoom level is low. It is because the spatial objects in a tile range could be extremely large if the zoom level is low, and using DataDC, each object needs to be plotted and merged successively to generate the final visual effects. Given a large-scale spatial dataset, as the amounts of spatial objects in the views are unpredictable, it is difficult to provide efficient visualization of the dataset on all zoom levels using data-driven methods; by contrast, using the display-driven HiVision, as the tile rendering speed remains stable with the different zoom levels, it is easy to provide an interactive exploration of the dataset in all the zoom levels. Compared with data-driven methods, HiVision shows obvious advantages while the density of spatial objects is high but tends to be slower when density is low. In our future works, we will consider combining both display-driven and data-driven computing to provide interactive spatial visual analysis with better performance.
	
	\subsection{Experiment 2. Visualizing Large-Scale Vector Data in Real Time}
	
	In this experiment, we test the ability of HiVision to support the real-time exploration of large-scale spatial vector data. HiVision is set to run with 64 MPI processes and 4 OpenMP threads in each process. For each dataset, we generate 10000 tile rendering tasks through a test program, which randomly requests tiles from zoom levels 3 to 15. HiVision is set with no cache preserved in \textbf{Result Pool} to ensure that each requested tile will lead to a new task in \textbf{Task Pool}, thus evaluating the performance of the visualization engine more precisely. 
	
\begin{figure}[htbp]
		\begin{center}
			\makeatletter
			\def
			\@captype{figure}
			\makeatother
			\subfigure[Rendering time of 10000 tiles]{
				\includegraphics[width=0.54\textwidth,trim=95 140 60 130,clip]{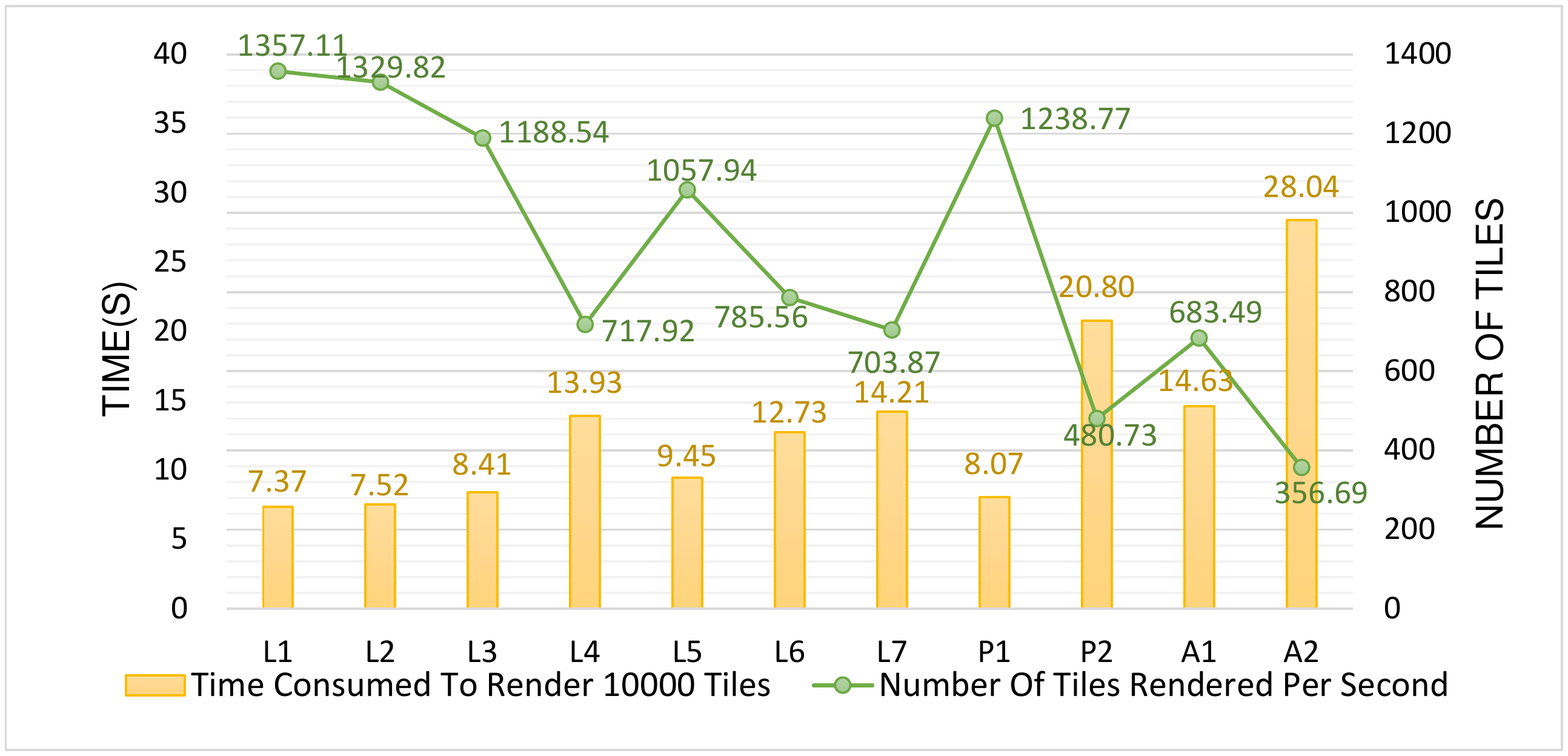}}%
			\subfigure[Rendering time of each tile]{
				\includegraphics[width=0.45\textwidth,trim=60 100 60 100,clip]{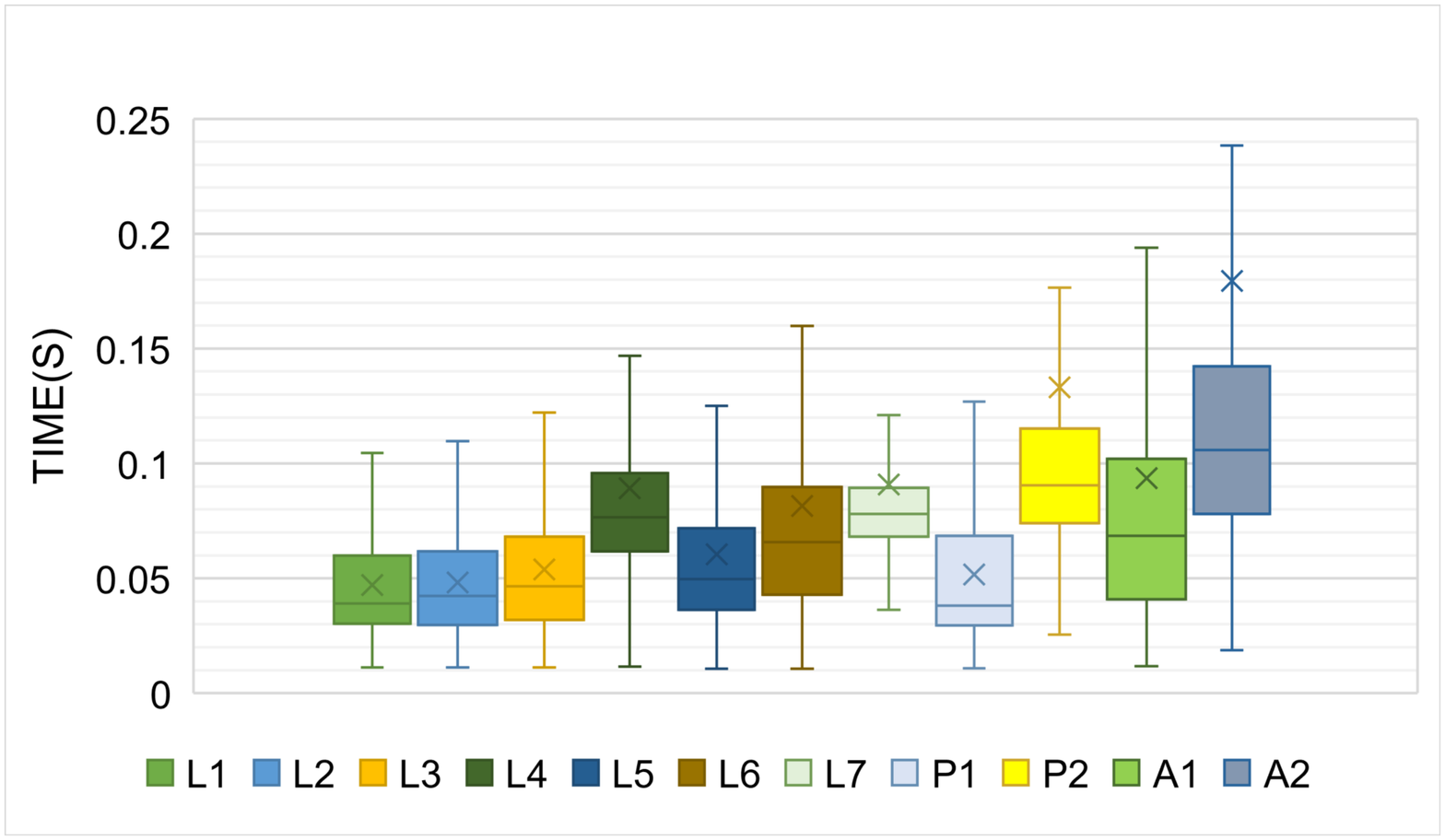}}%
			\caption{Tile rendering time of HiVision on different datasets.}
			\label{f5_3}			
		\end{center}
	\end{figure}

	Figure~\ref{f5_3} (a) shows the total rendering time of 10000 tiles on different datasets using HiVision. For all the datasets, A$_{2}$ produces the poorest performance with the speed of 356.69 tiles/s; as the number of tiles for display in a screen is generally no more than 100 (much fewer than the number of tiles generated per second of A$_{2}$), it is possible to perform real-time visualization with HiVision on all the datasets. As shown in Figure~\ref{f5_3} (b), the rendering time distributions of each tile on different datasets are visualized through box plots ('$\times$' represents the average rendering time of each tile). For A$_{2}$ which produces the poorest performance, most of the tiles are rendered in 0.24s. Assume that a browser requests 100 different tiles of A$_{2}$. As there are totally 64 running MPI processes, all the 100 tasks will be processed in two rounds with 28 (${=\rm{64_{processes}}} \times {\rm{2 - 100_{tasks}}}$) MPI processes suspended in the second round, and all the tasks will be most likely completed in less than 0.48s (${=\rm{0.24s}} \times {\rm{2}}$). In conclusion, HiVision is able to provide an interactive exploration of large-scale spatial vector data.

	\subsection{Experiment 3. Impact of Request Rate}
	
	In the experiments other than this experiment, all the tile requests are dispatched to HiVision simultaneously, which means that the request rate is set to infinity, and HiVision keeps running at full load until all the tasks have been finished. In practical applications, the tile requests are generally generated at much lower rates. In this experiment, HiVision is set to run with 64 MPI processes and 4 OpenMP threads in each process. It means that there are 64 tile rendering processes that can render 64 tiles at the same time (the redundant tasks are stored in the Task Pool waiting for idle processes). The number of tiles to request per second in the experiment is set to multiple of 64 and the request rate is respectively set to 128, 256, 512, 1024, 2048, 4096 and infinity (INF) tiles/s. For each rate, we generate 10000 random tile tasks from zoom levels 3 to 15.
	
\begin{figure}[htbp]
		\begin{center}
			\makeatletter
			\def
			\@captype{figure}
			\makeatother
			\subfigure[L$_{1}$ (Performance limit: 1357.11 tiles/s)]{
				\includegraphics[width=0.33\textwidth,trim=60 120 60 145,clip]{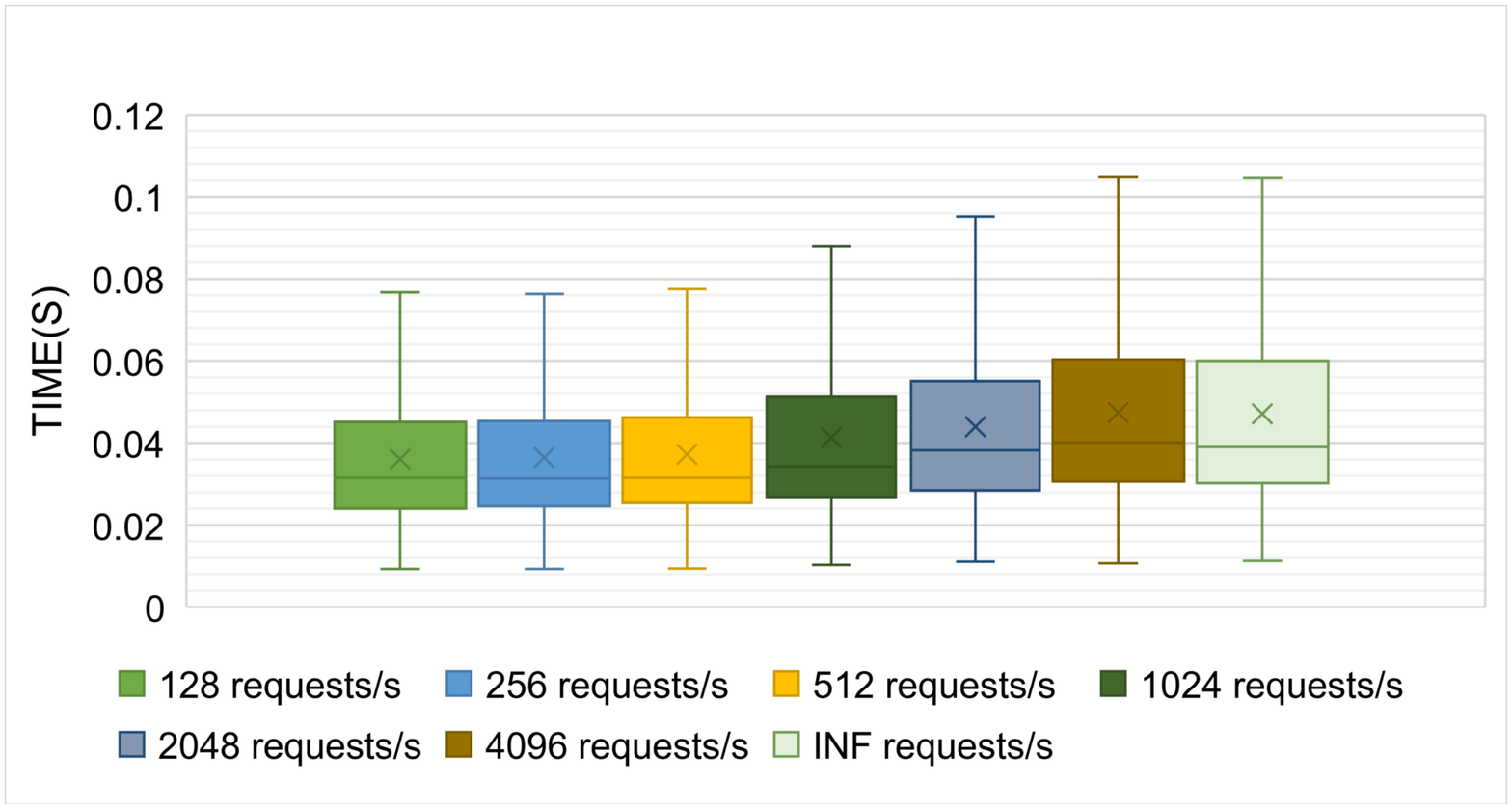}}%
			\subfigure[L$_{2}$ (Performance limit: 1329.82 tiles/s)]{
				\includegraphics[width=0.33\textwidth,trim=60 120 60 145,clip]{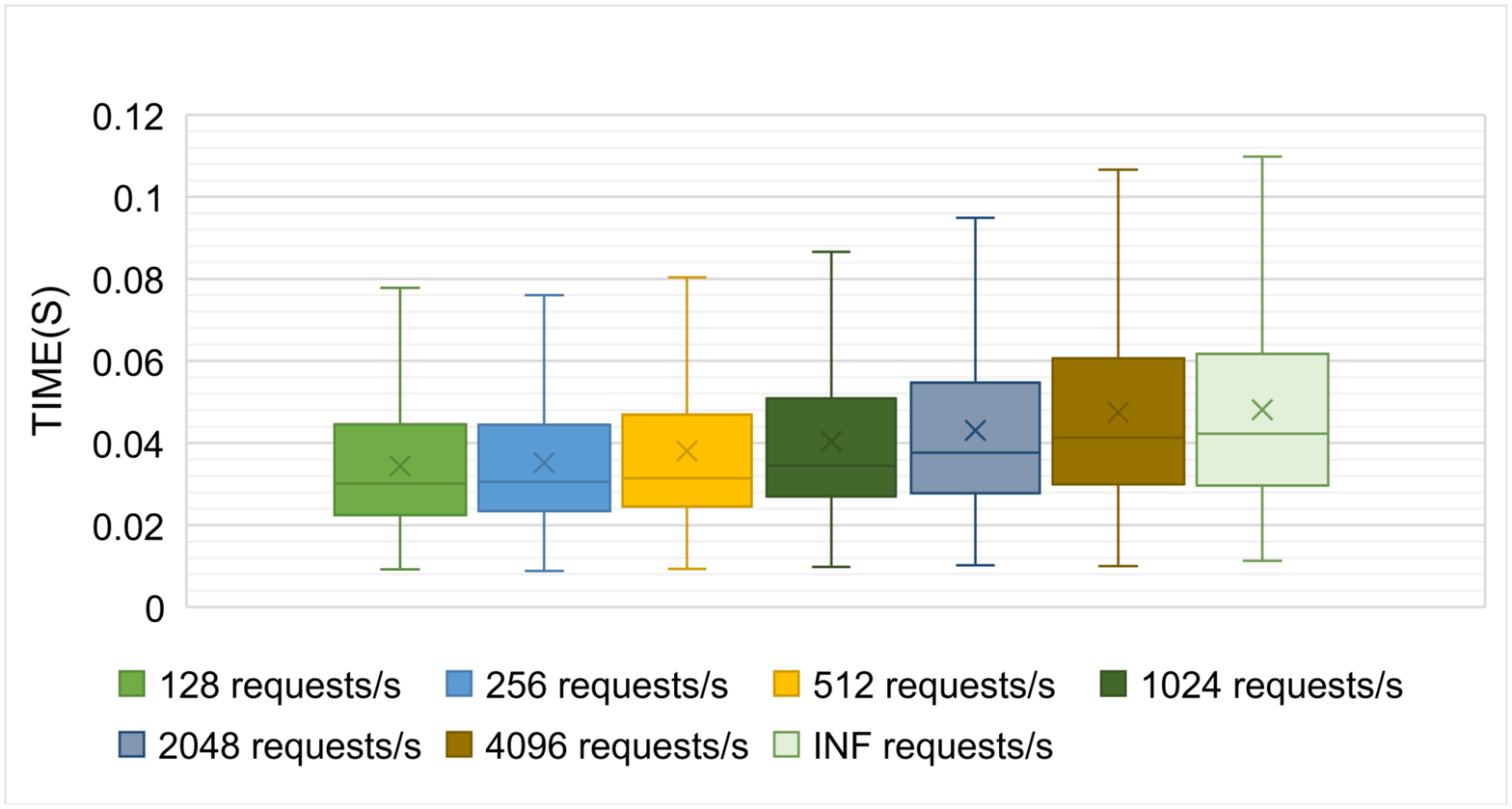}}%
			\subfigure[L$_{3}$ (Performance limit: 1188.54 tiles/s)]{
				\includegraphics[width=0.33\textwidth,trim=60 120 60 145,clip]{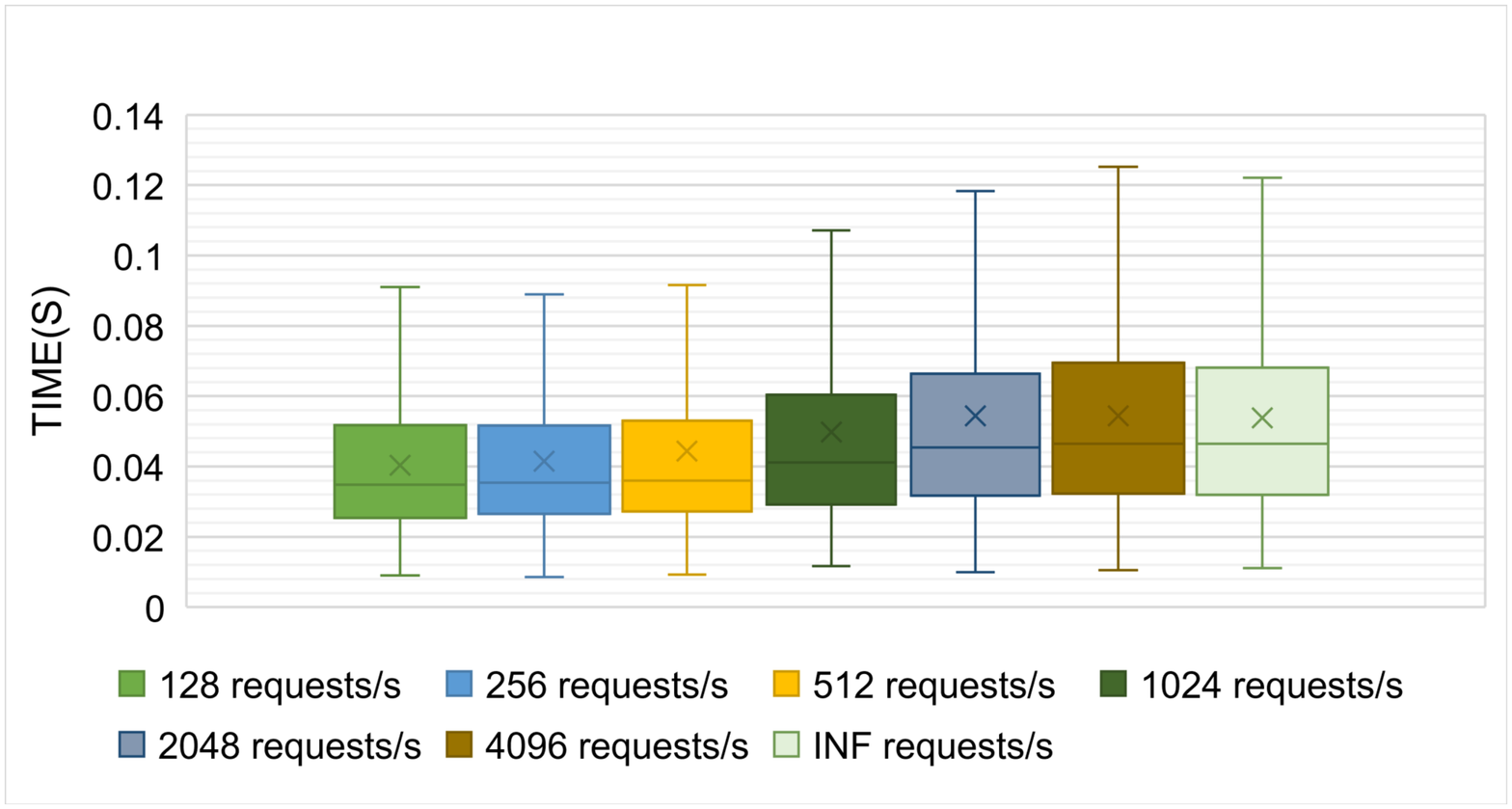}}%
			\vfill
			\subfigure[L$_{4}$ (Performance limit: 717.92 tiles/s)]{
				\includegraphics[width=0.33\textwidth,trim=60 120 60 145,clip]{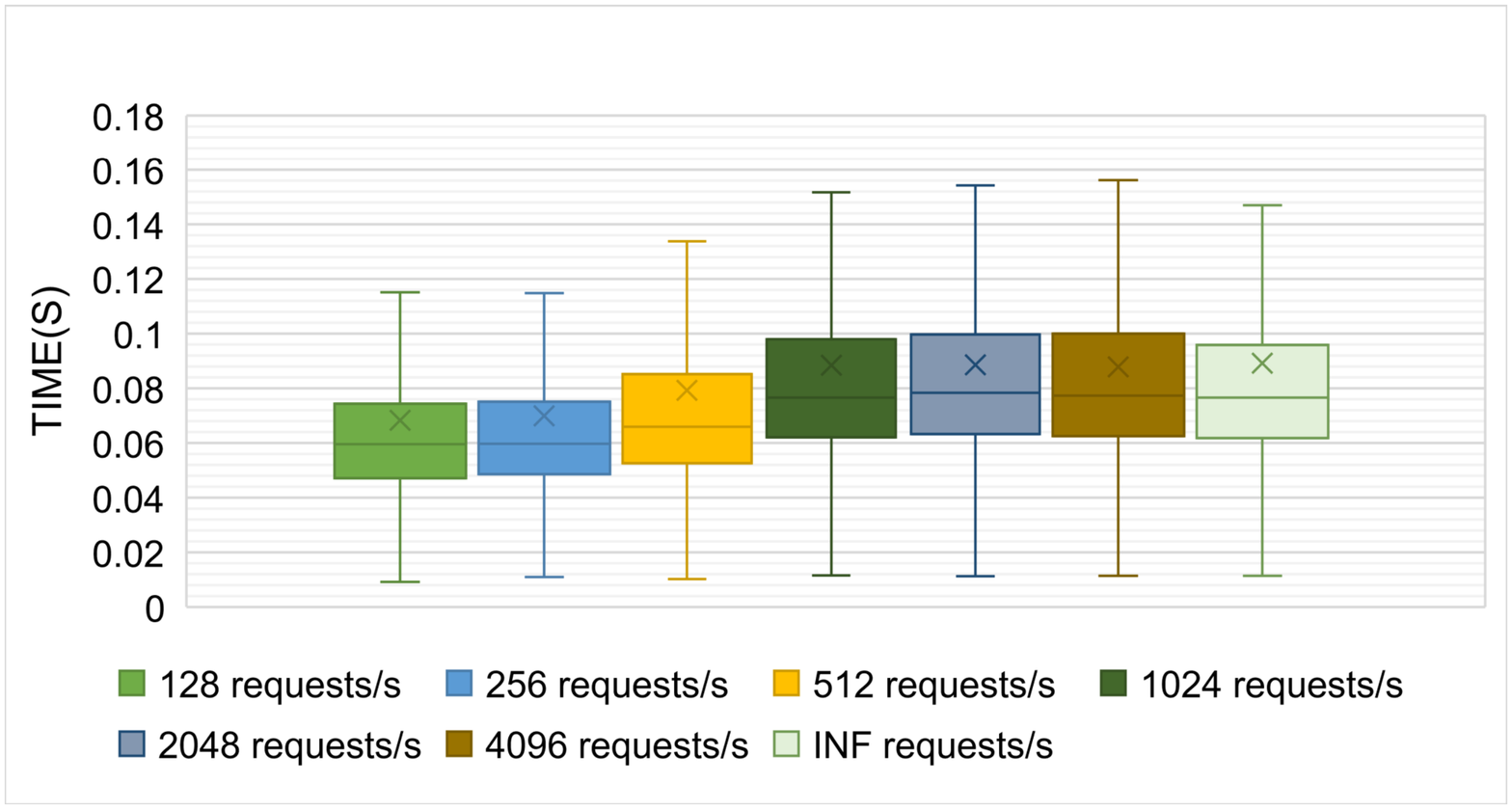}}%
			\subfigure[L$_{5}$ (Performance limit: 1057.94 tiles/s)]{
				\includegraphics[width=0.33\textwidth,trim=60 120 60 145,clip]{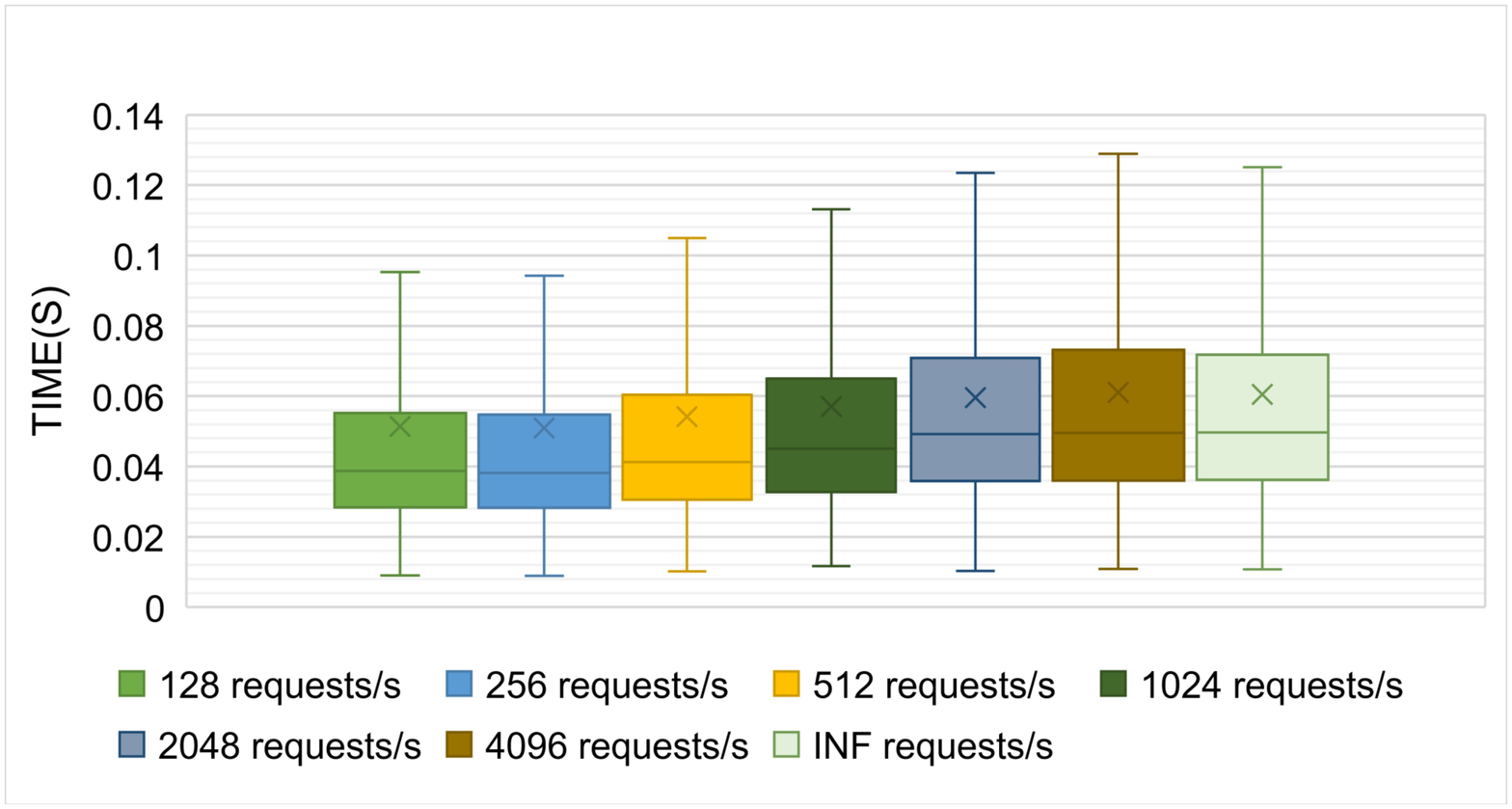}}%
			\subfigure[L$_{6}$ (Performance limit: 785.56 tiles/s)]{
				\includegraphics[width=0.33\textwidth,trim=60 120 60 145,clip]{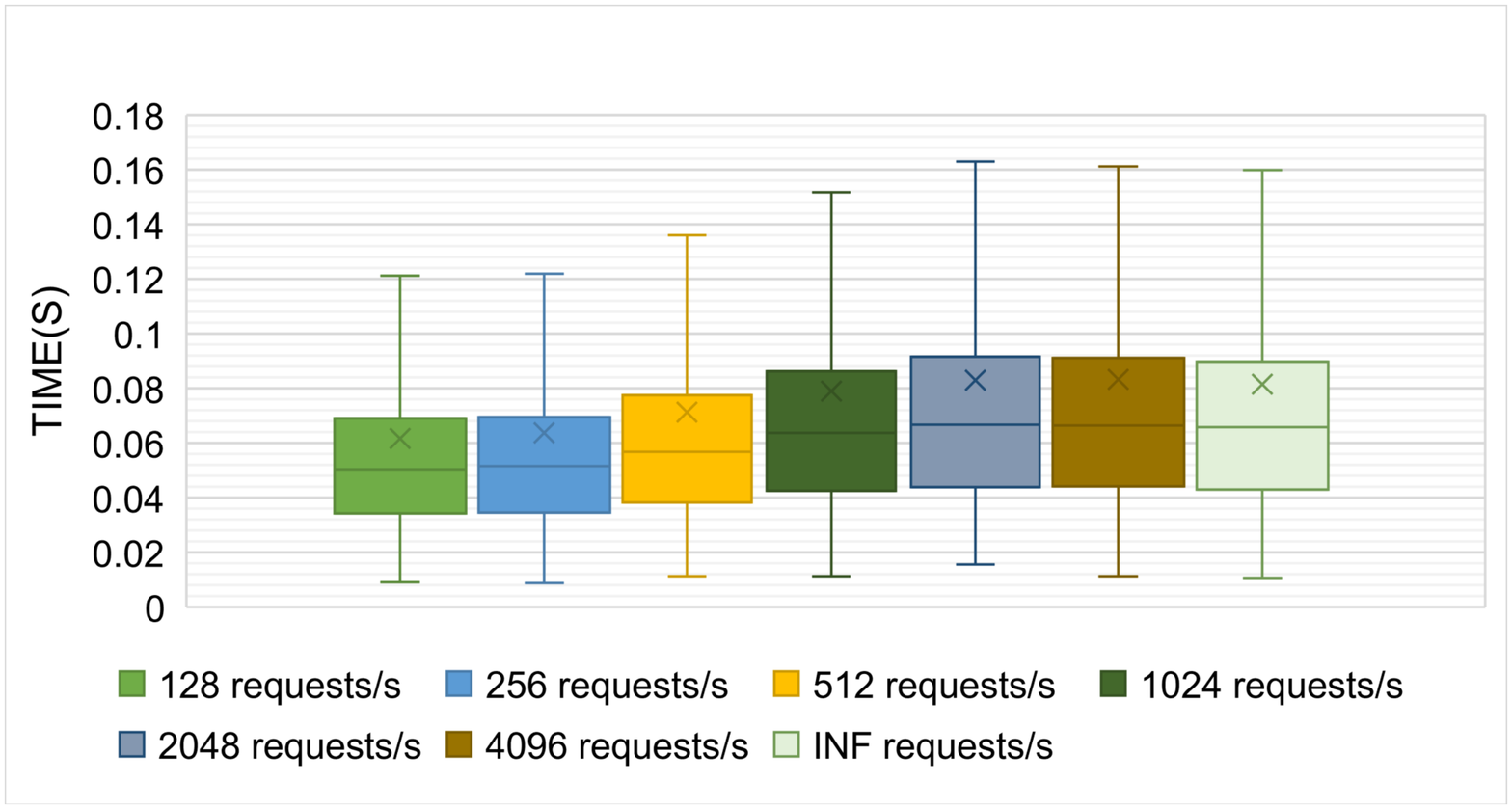}}%
			\vfill
			\subfigure[L$_{7}$ (Performance limit: 703.87 tiles/s)]{
				\includegraphics[width=0.33\textwidth,trim=60 120 60 145,clip]{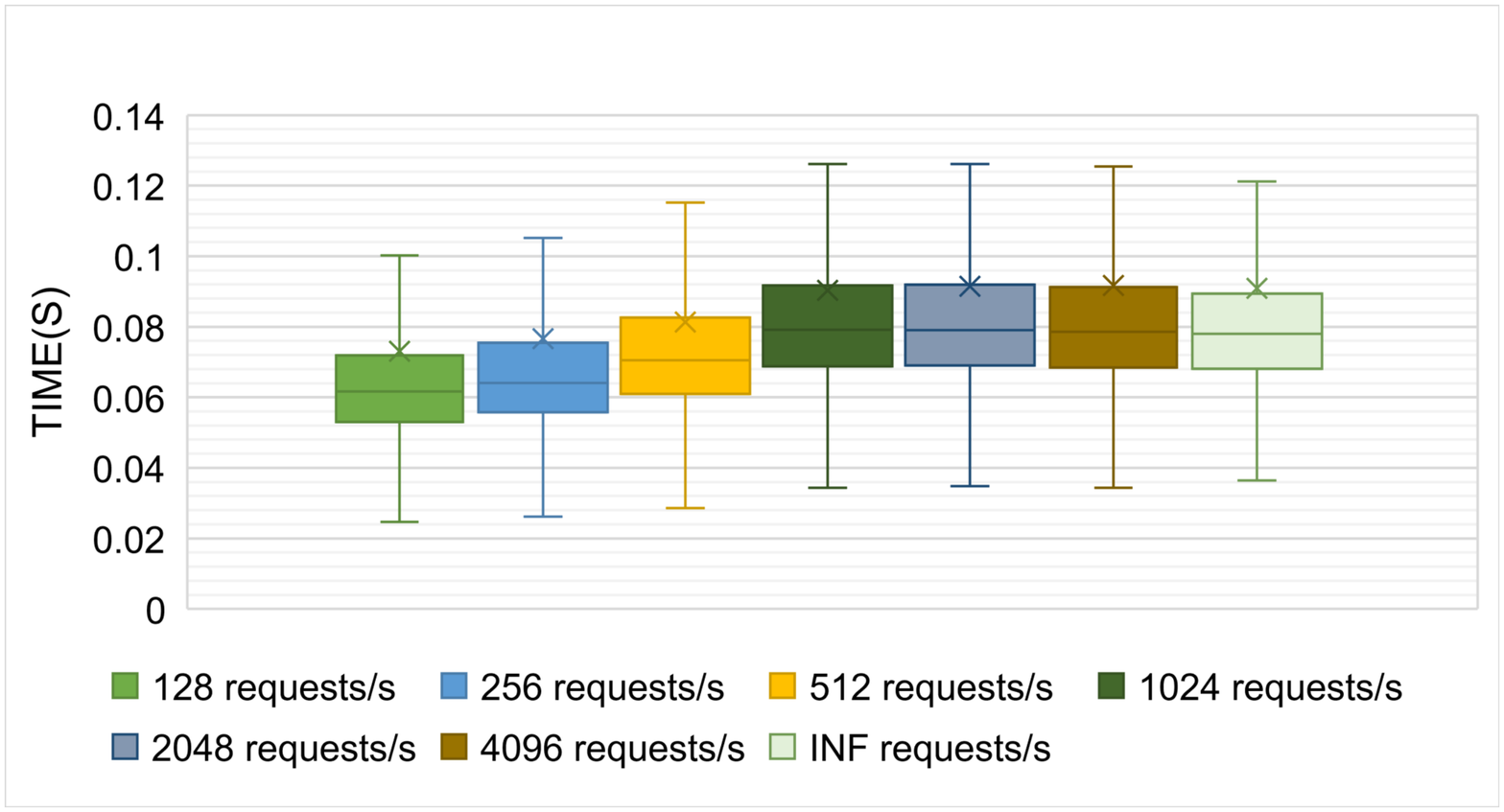}}%
			\subfigure[P$_{1}$ (Performance limit: 1238.77 tiles/s)]{
				\includegraphics[width=0.33\textwidth,trim=60 120 60 145,clip]{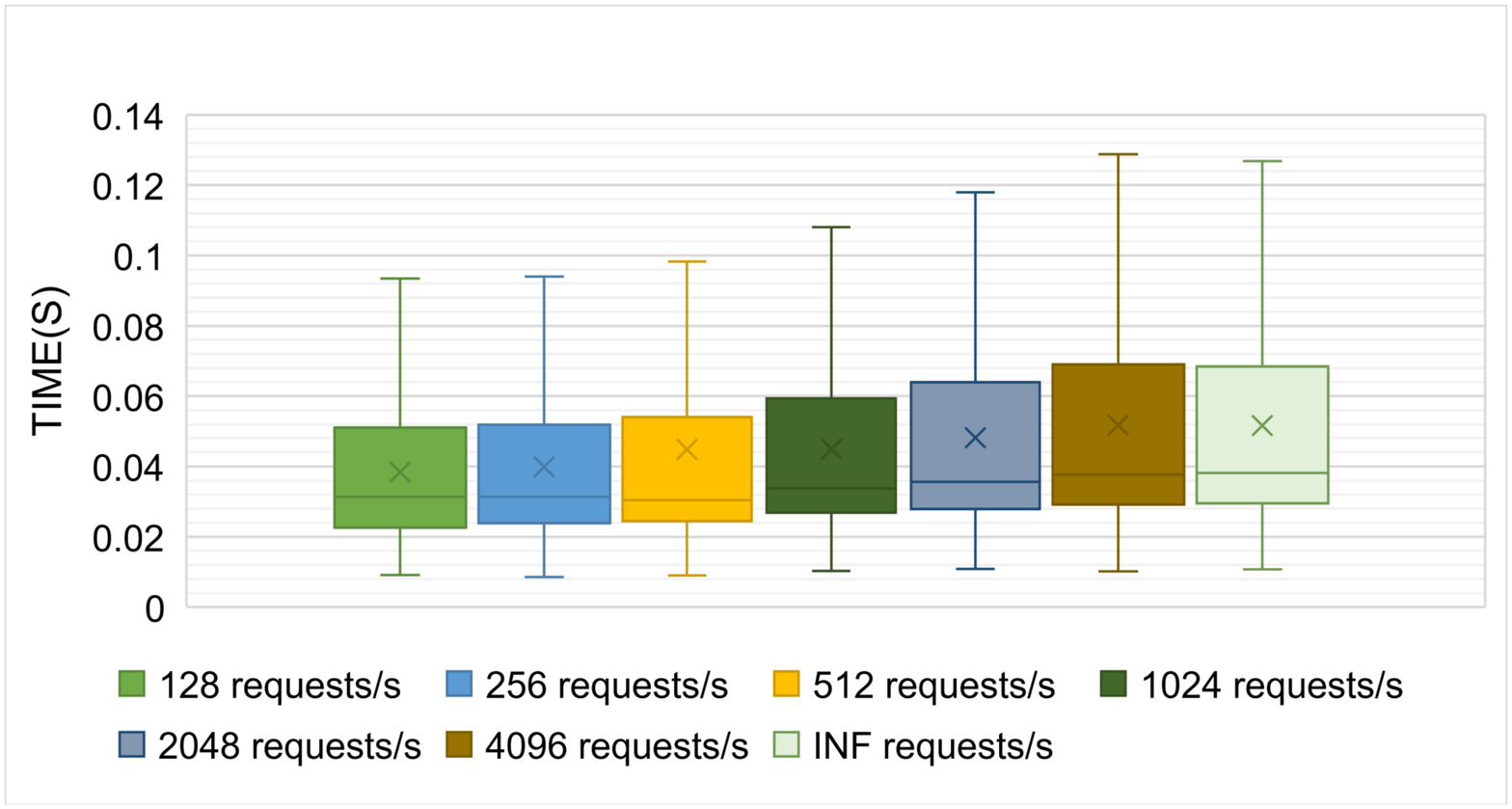}}%
			\subfigure[P$_{2}$ (Performance limit: 480.73 tiles/s)]{
				\includegraphics[width=0.33\textwidth,trim=60 120 60 145,clip]{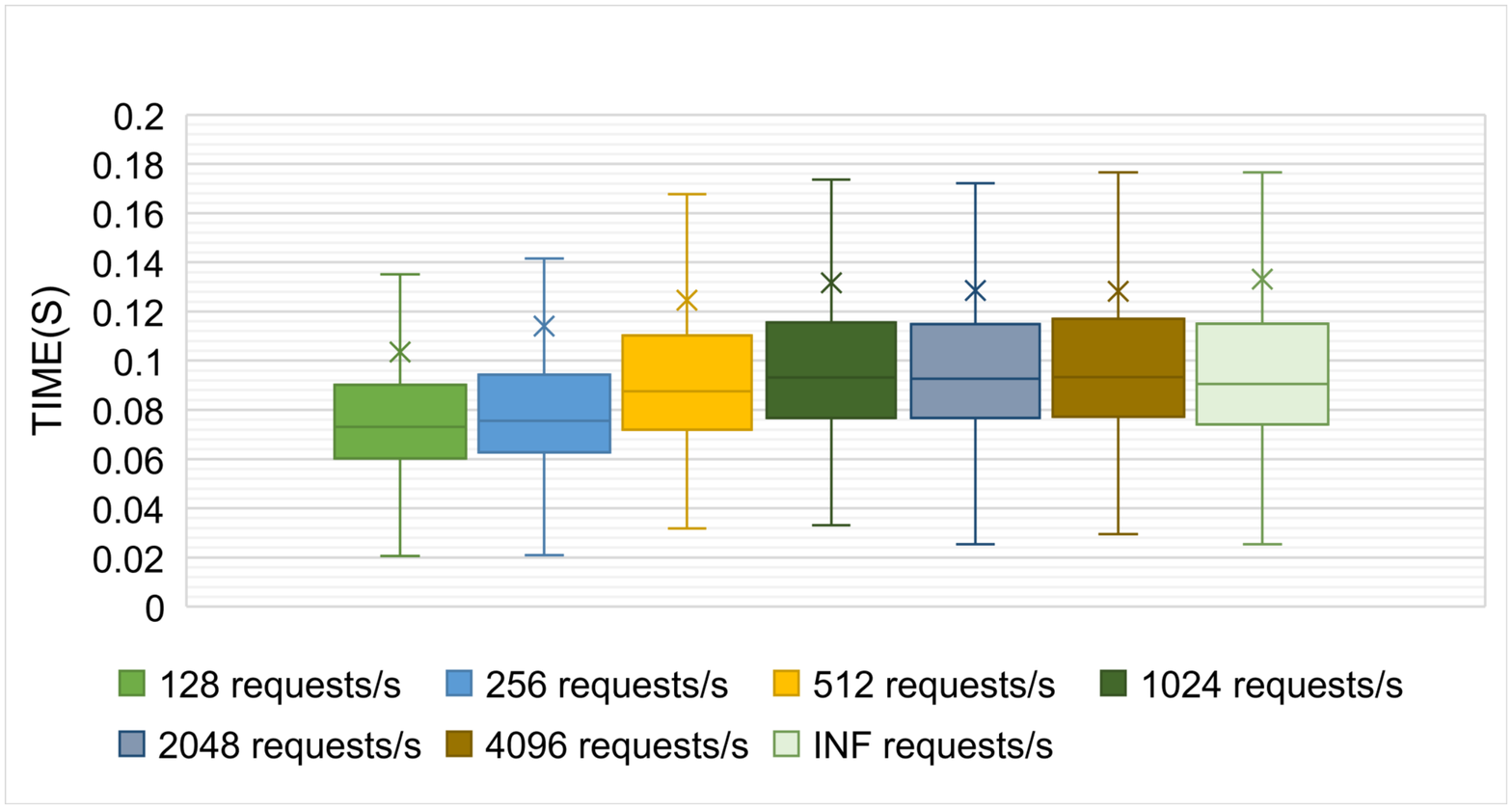}}%
			\vfill
			\subfigure[A$_{1}$ (Performance limit: 683.49 tiles/s)]{
				\includegraphics[width=0.33\textwidth,trim=60 120 60 145,clip]{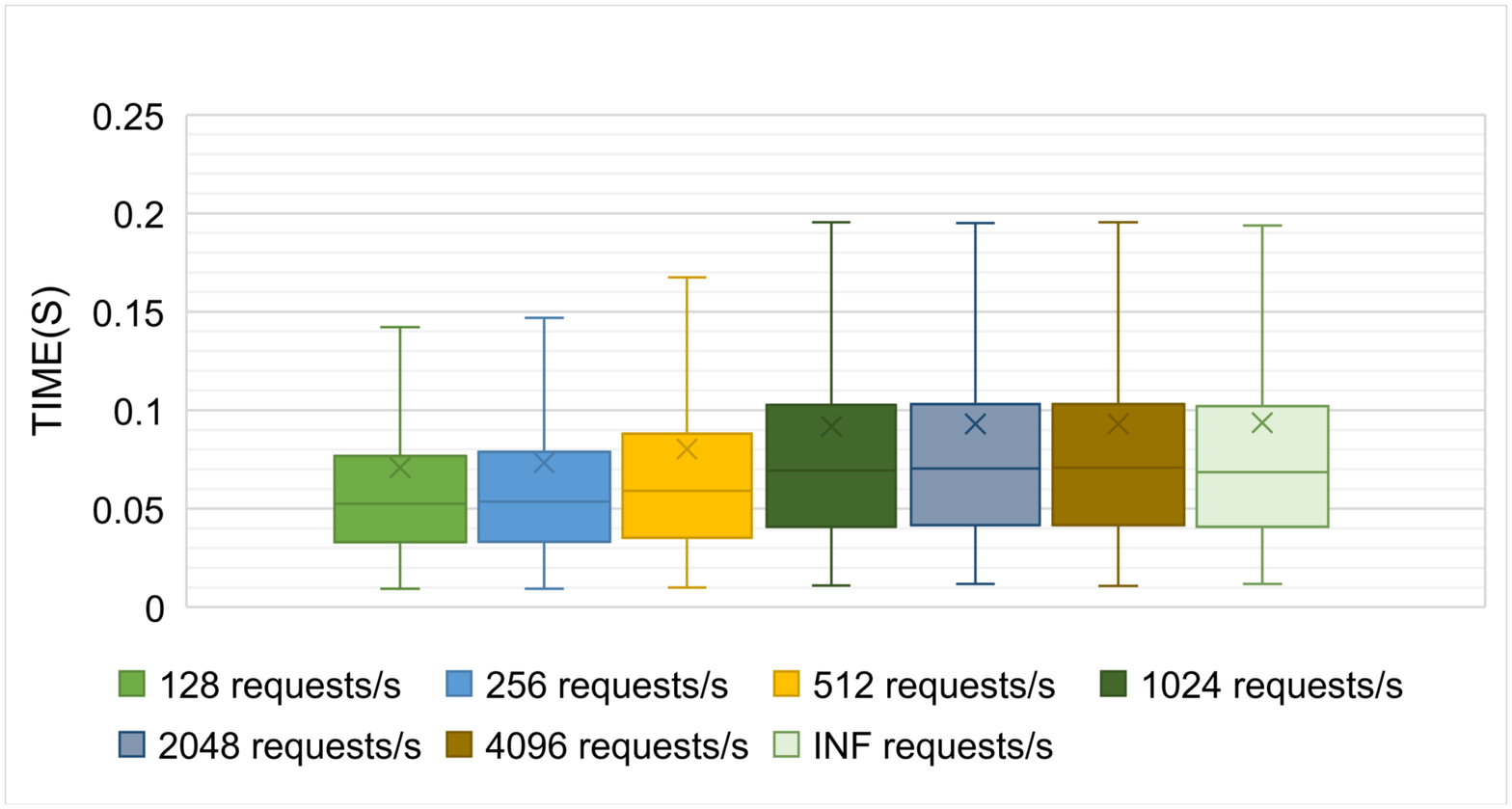}}%
			\subfigure[A$_{2}$ (Performance limit: 356.69 tiles/s)]{
				\includegraphics[width=0.33\textwidth,trim=60 120 60 145,clip]{figures/f5-4a.pdf}}%
			\caption{Rendering time of each tile with different request rates in HiVision.}%
			\label{f5_4}
		\end{center}
	\end{figure}

	The rendering time distributions of each tile with different request rates are shown in Figure~\ref{f5_4}. The performance limit is the number of tiles rendered per second at the rate of INF tiles/s (Figure~\ref{f5_3} (a)). For all the datasets, the rendering time of tiles increases obviously with the request rates when the rate is less than the performance limit. It is because of the increasing resources competition between processes. And when the request rate exceeds the performance limit, the rendering time of tiles does not change significantly. It is because that the visualization engine is running at full load and the increase of request rates will not cause obvious effects on the performance; in particular, as the increase of extra tasks in {\bf Task Pool}, the response time of HiVision will increase rapidly if the request rate exceeds the performance limit. In conclusion, compared with the results in other experiments, higher performance can be achieved in practical applications as a result of the lower request rates.
	
	\subsection{Experiment 4. Parallel Scalability}
	
\begin{figure}[htbp]
		\begin{center}
			\makeatletter
			\def
			\@captype{figure}
			\makeatother
			\subfigure[L$_{1}$]{
				\includegraphics[width=0.28\textwidth,trim=215 65 190 90,clip]{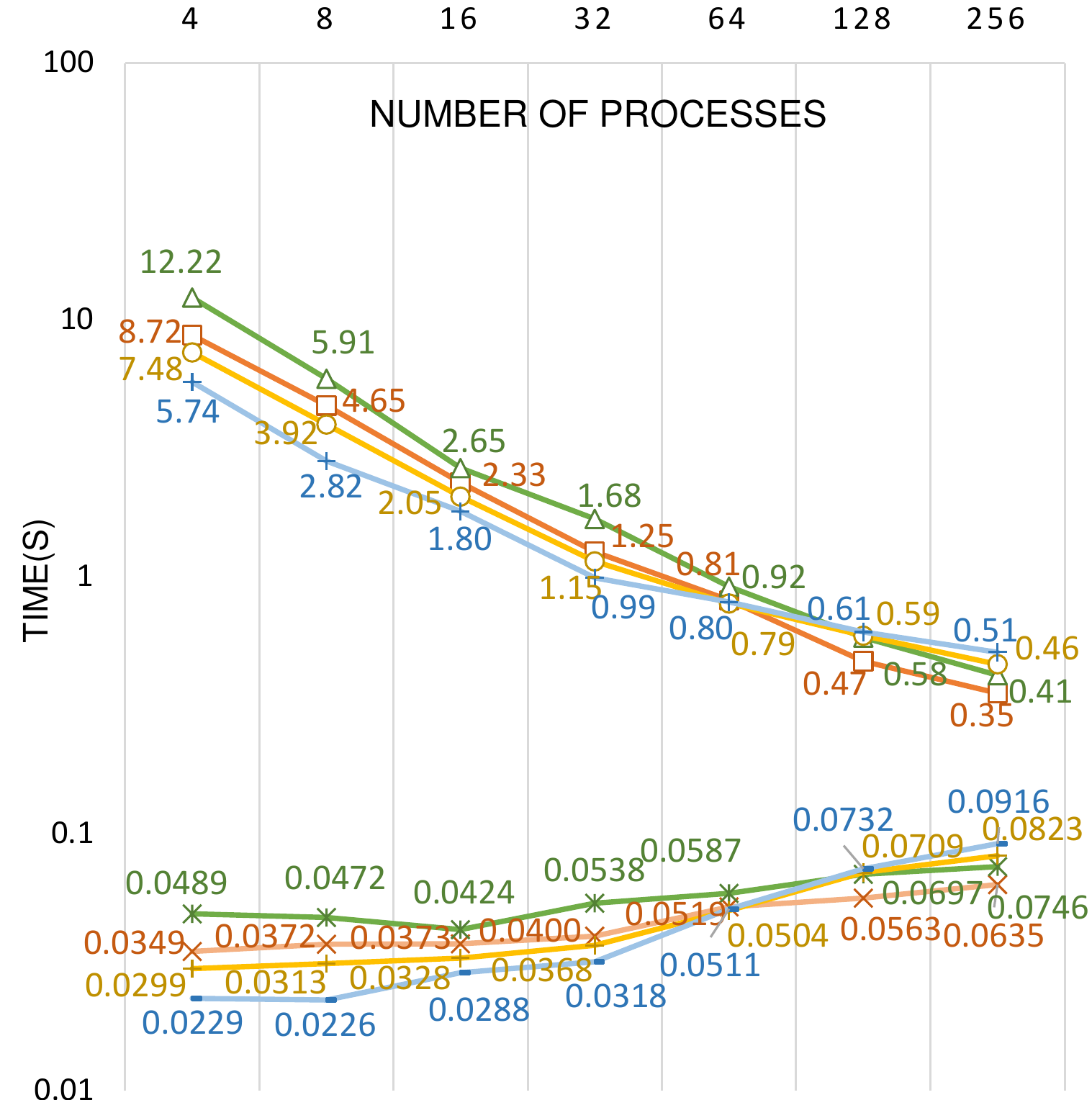}}%
			\subfigure[L$_{2}$]{
				\includegraphics[width=0.28\textwidth,trim=215 65 190 90,clip]{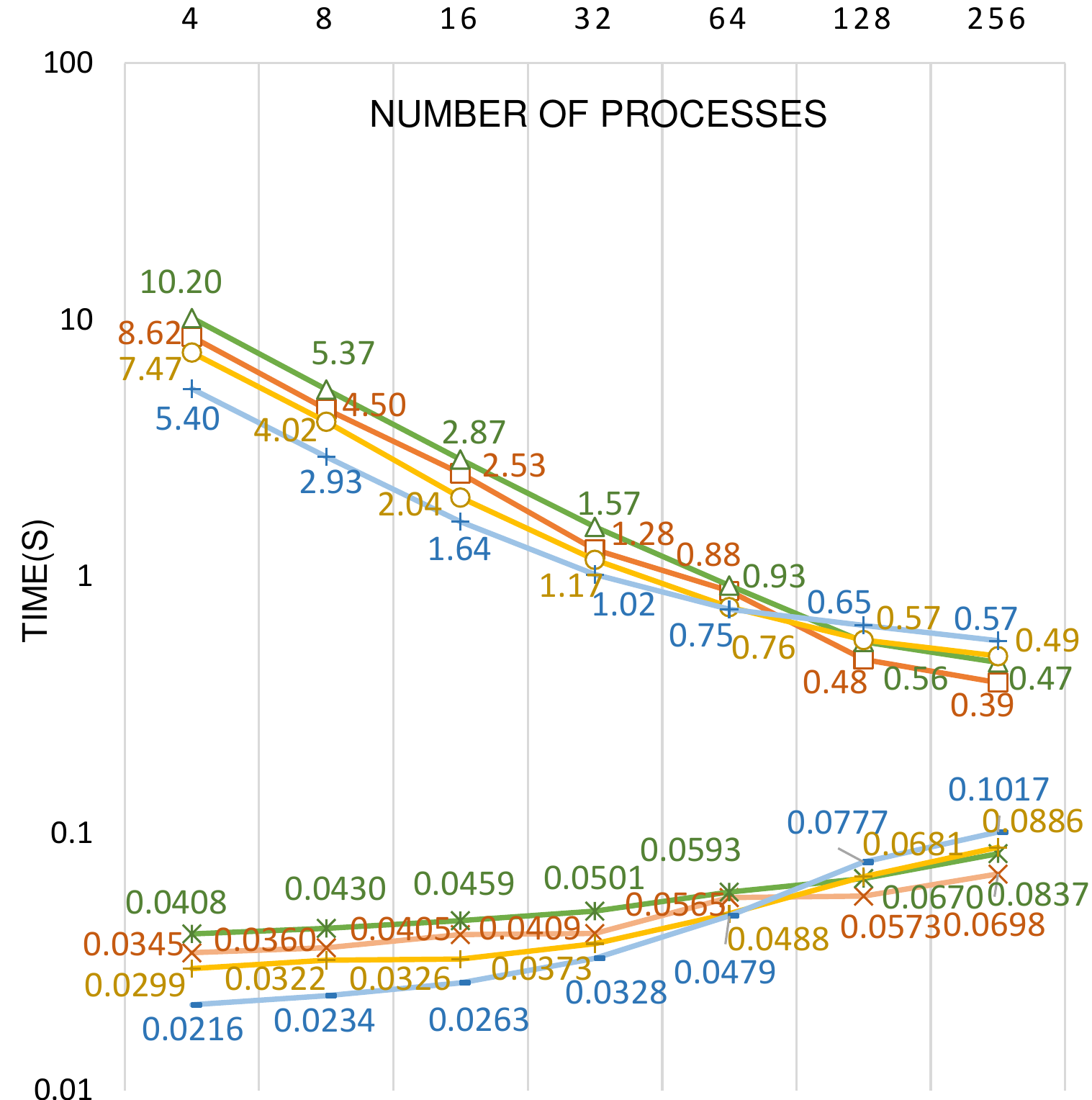}}%
			\subfigure[L$_{3}$]{
				\includegraphics[width=0.28\textwidth,trim=215 65 190 90,clip]{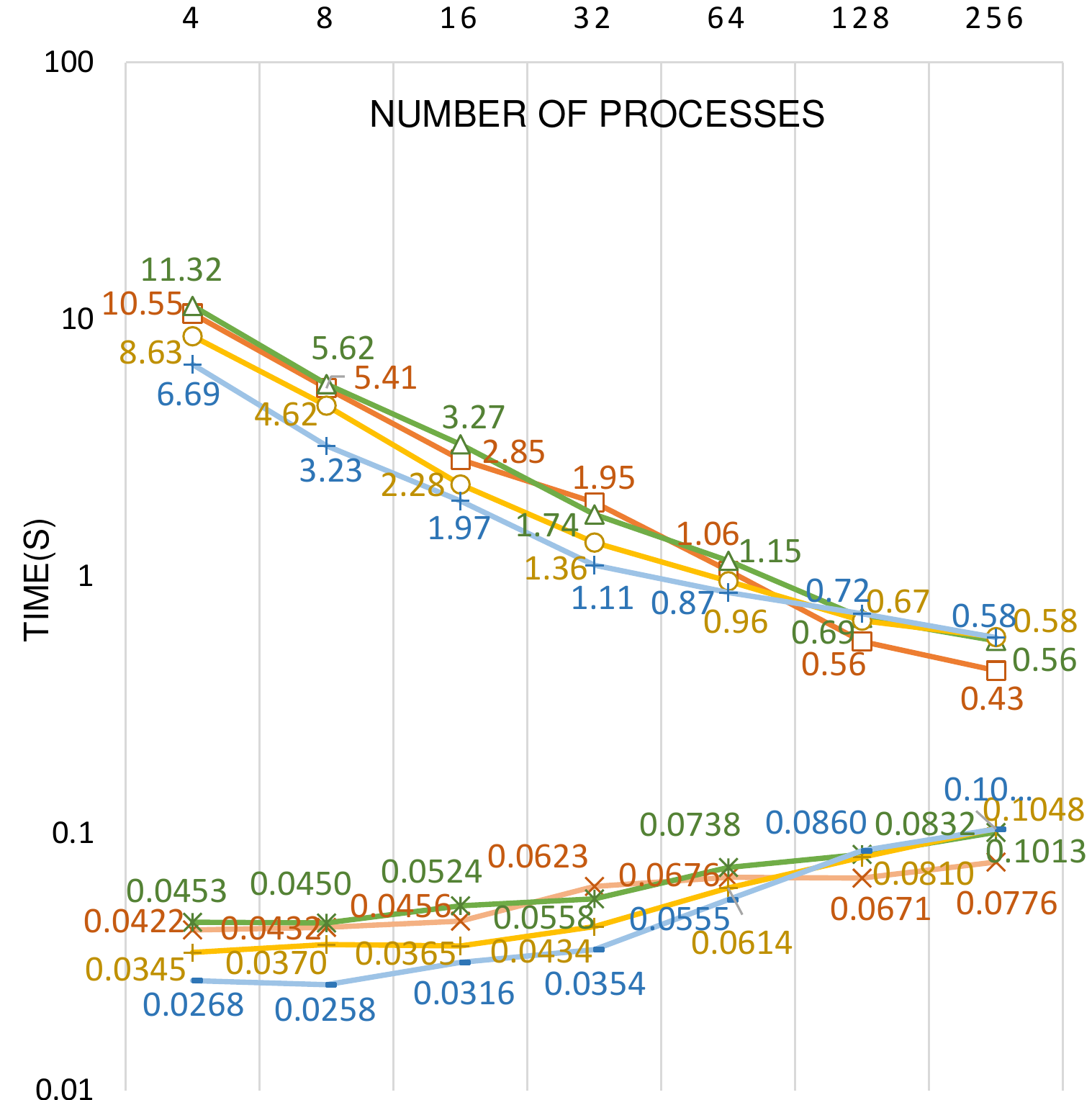}}%
			\vfill
			\subfigure[L$_{4}$]{
				\includegraphics[width=0.29\textwidth,trim=215 65 190 90,clip]{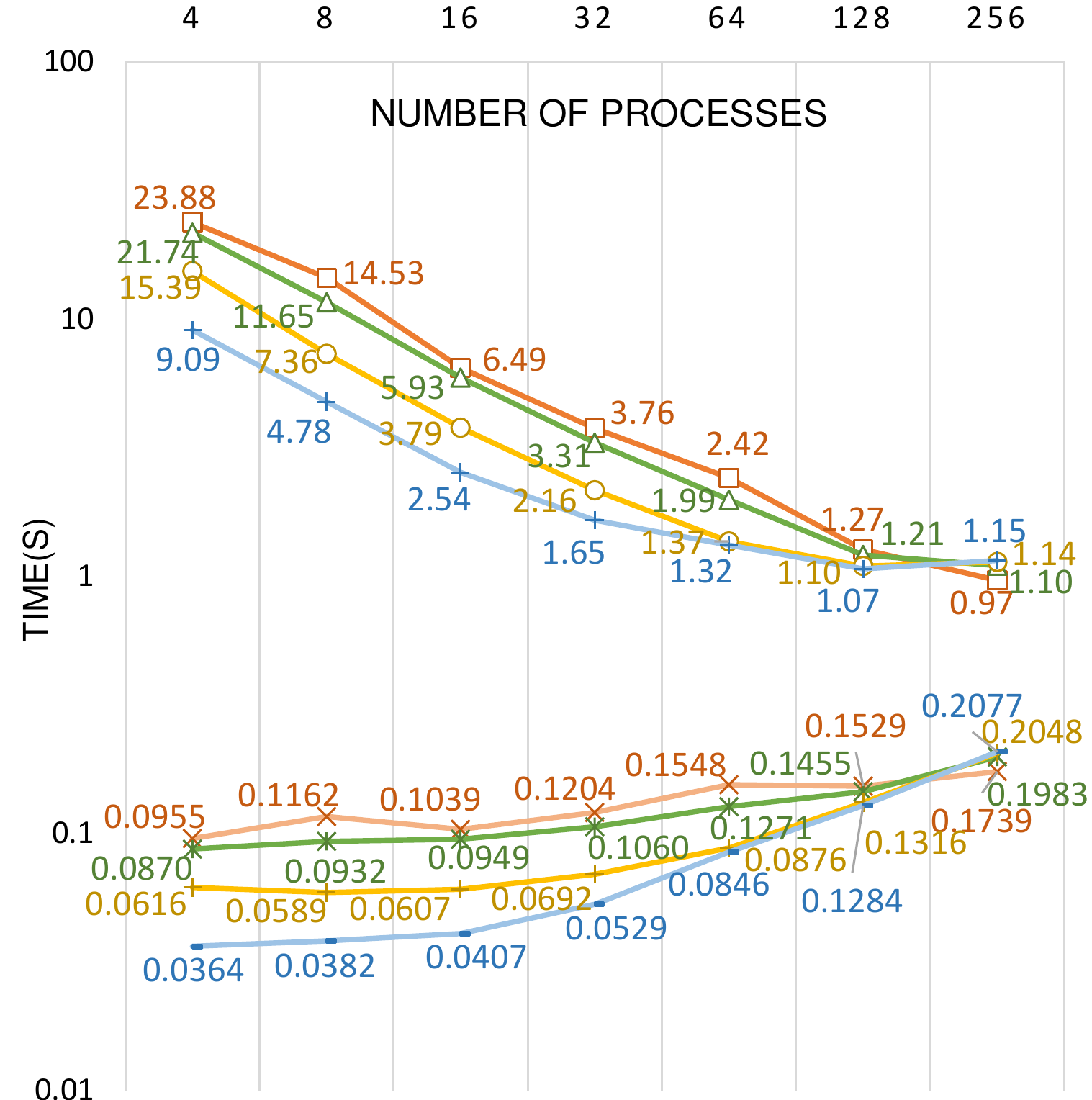}}%
			\subfigure[L$_{5}$]{
				\includegraphics[width=0.29\textwidth,trim=215 65 190 90,clip]{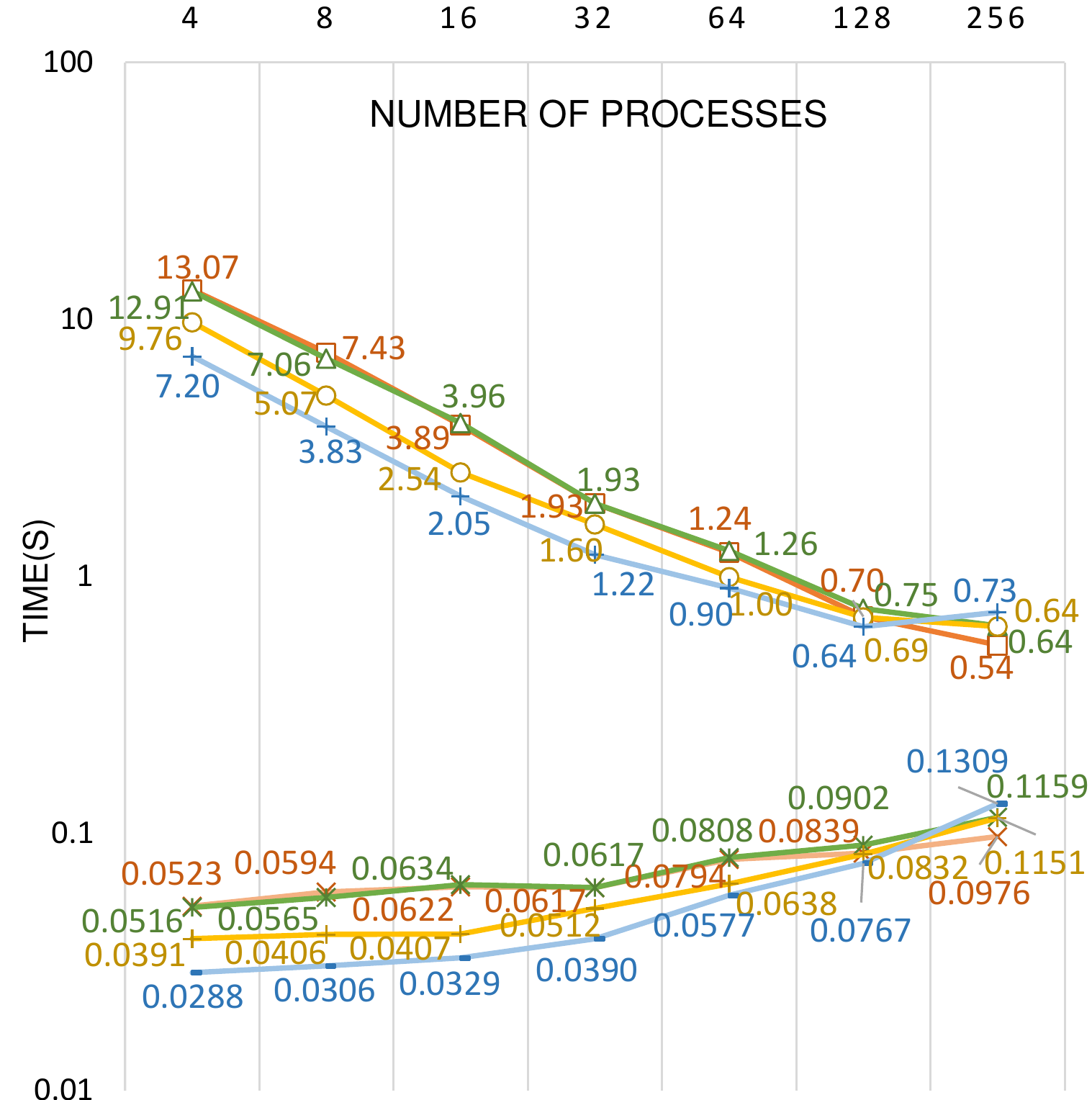}}%
			\subfigure[L$_{6}$]{
				\includegraphics[width=0.29\textwidth,trim=215 65 190 90,clip]{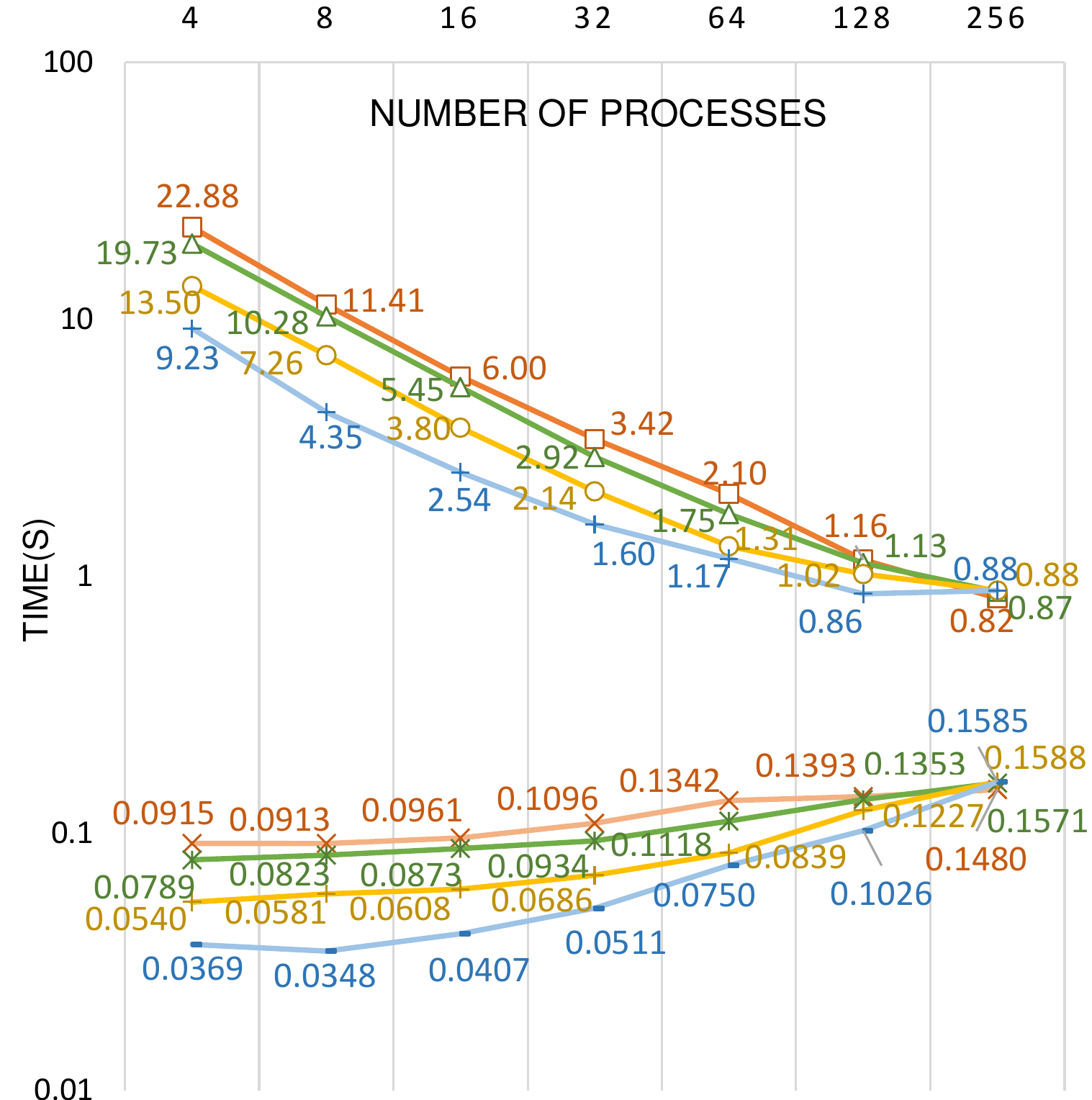}}%
			\vfill
			\subfigure[L$_{7}$]{
				\includegraphics[width=0.29\textwidth,trim=215 65 190 90,clip]{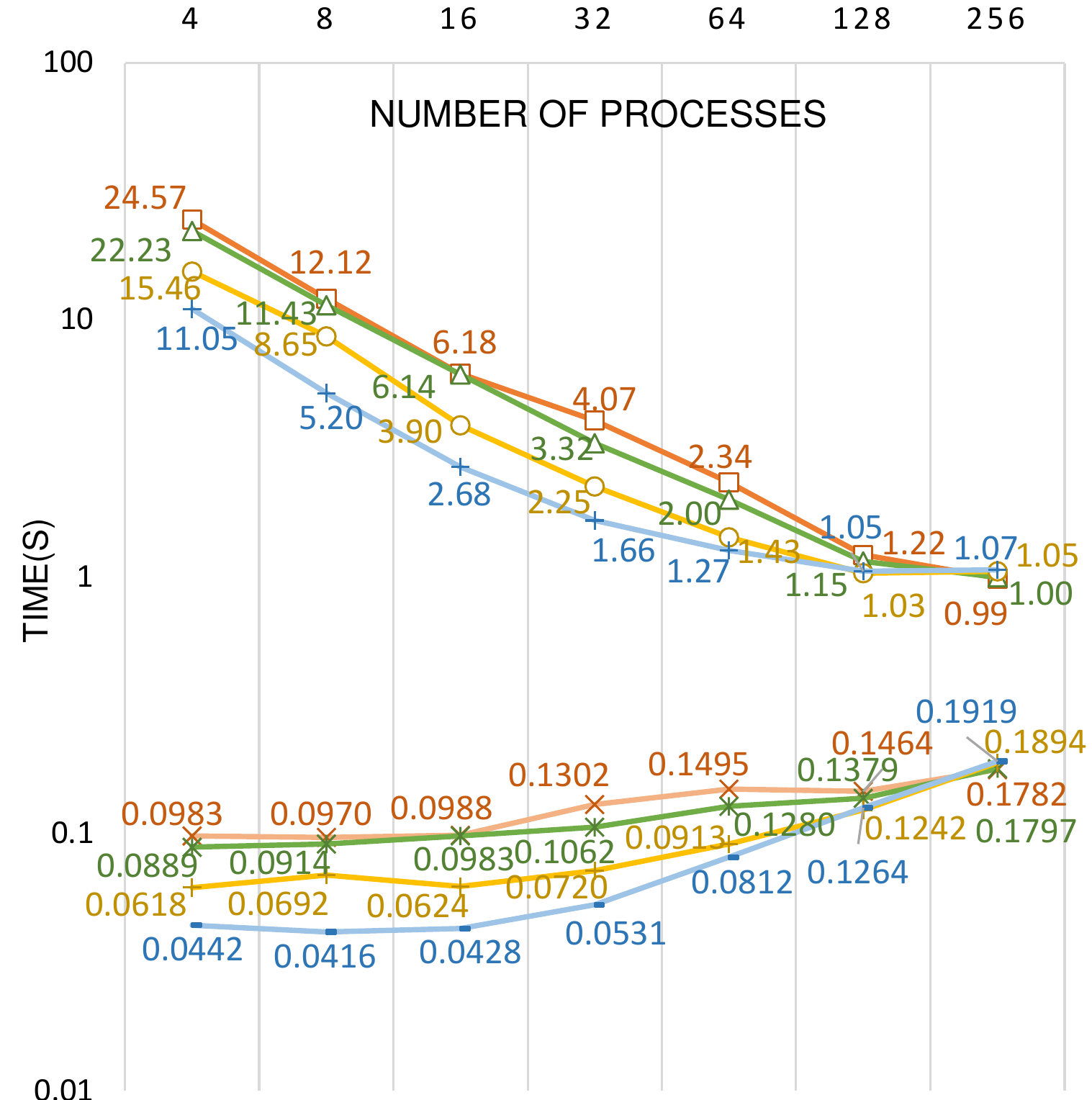}}%
			\subfigure[P$_{1}$]{
				\includegraphics[width=0.29\textwidth,trim=215 65 190 90,clip]{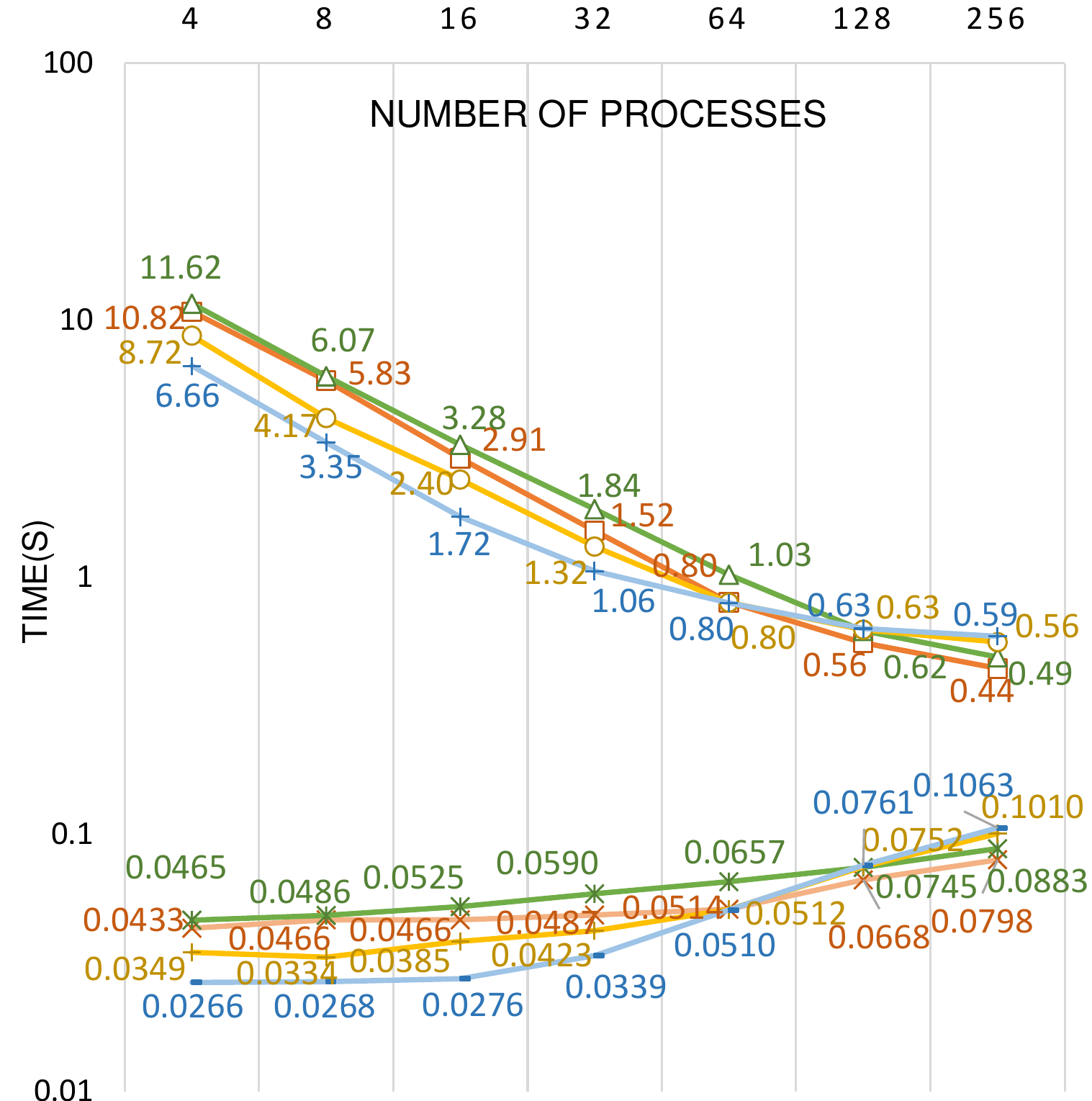}}%
			\subfigure[P$_{2}$]{
				\includegraphics[width=0.29\textwidth,trim=215 65 190 90,clip]{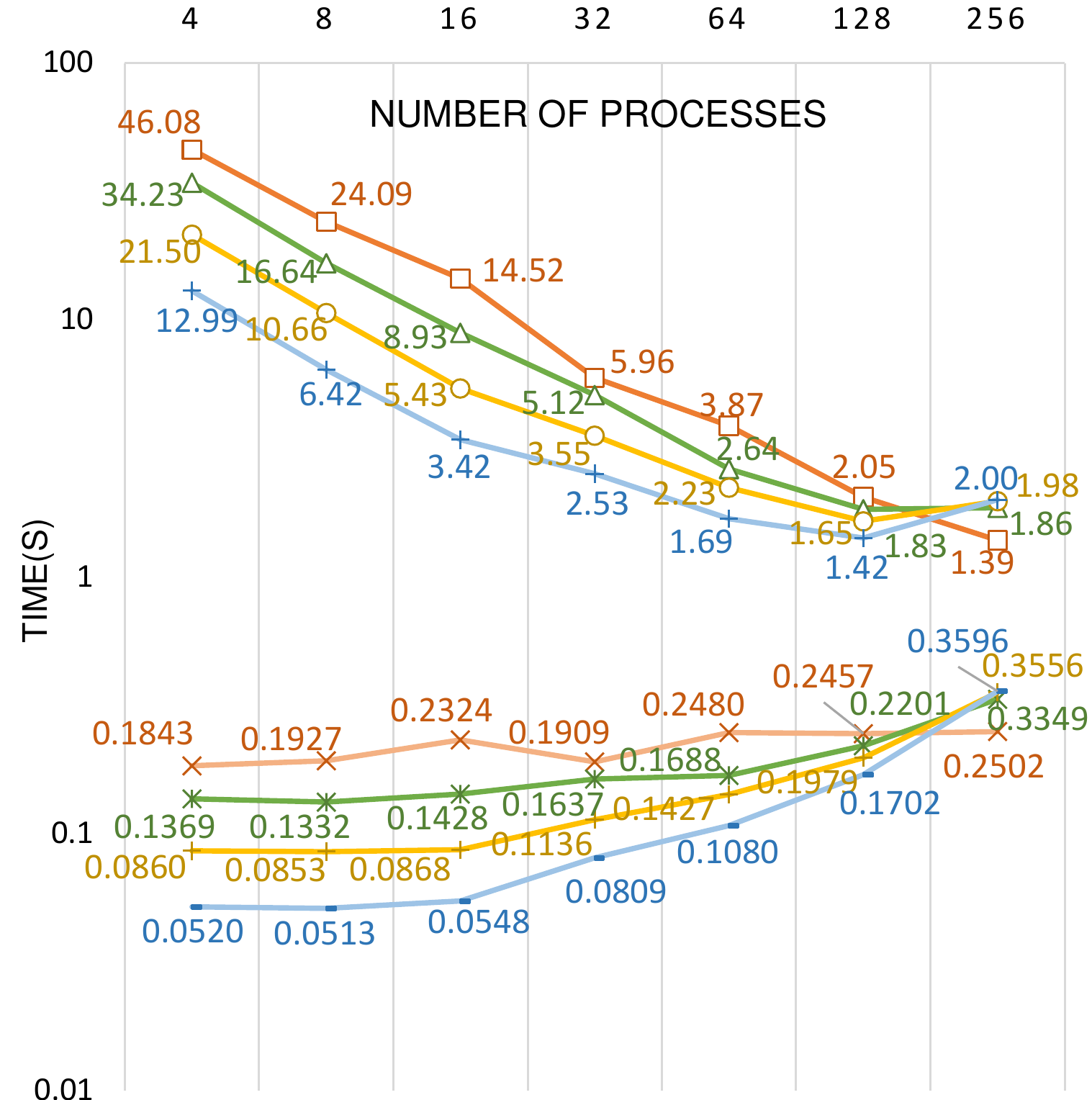}}%
			\vfill
			\subfigure[A$_{1}$]{
				\includegraphics[width=0.29\textwidth,trim=215 65 190 90,clip]{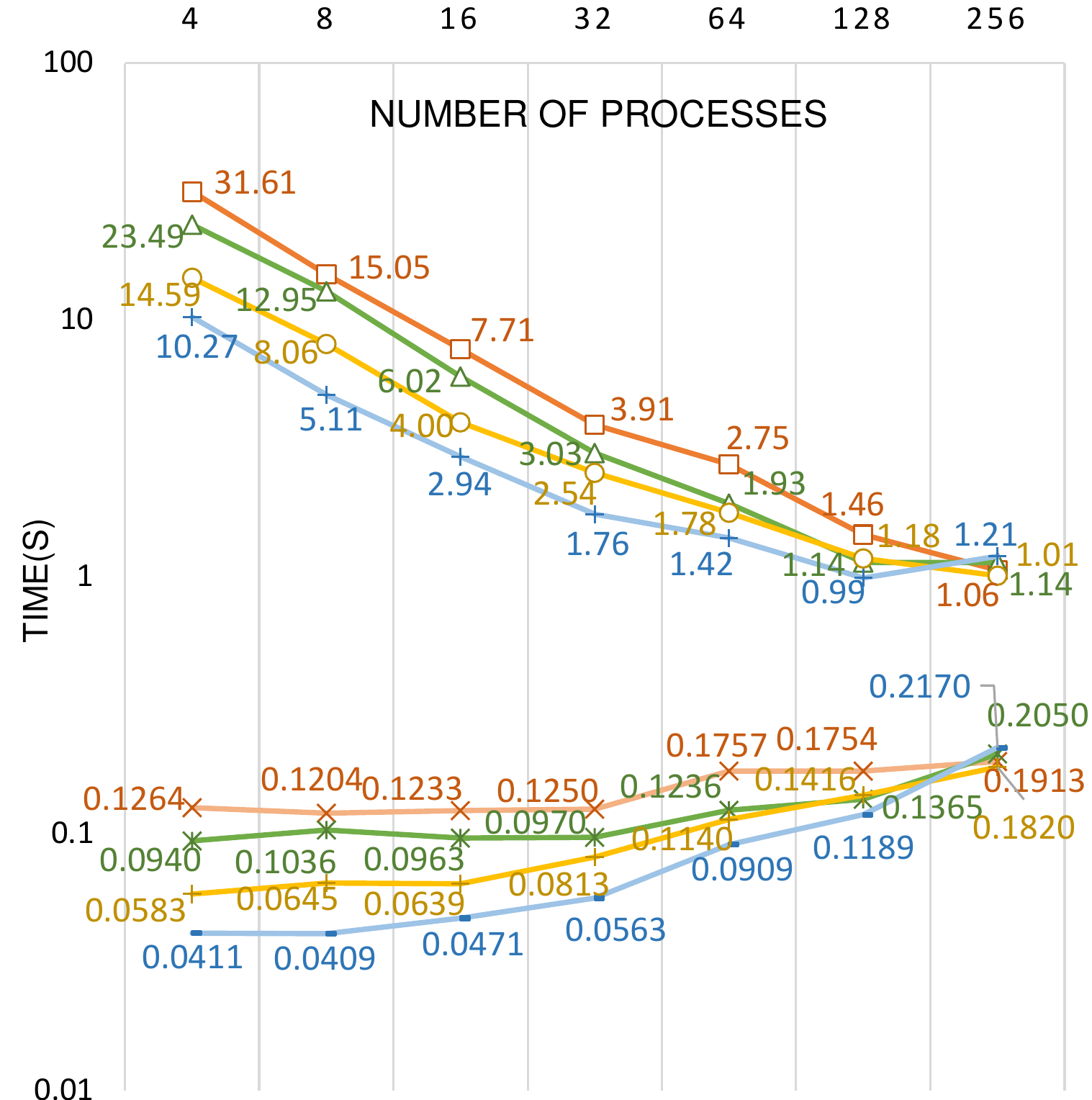}}%
			\subfigure[A$_{2}$]{
				\includegraphics[width=0.29\textwidth,trim=215 65 190 90,clip]{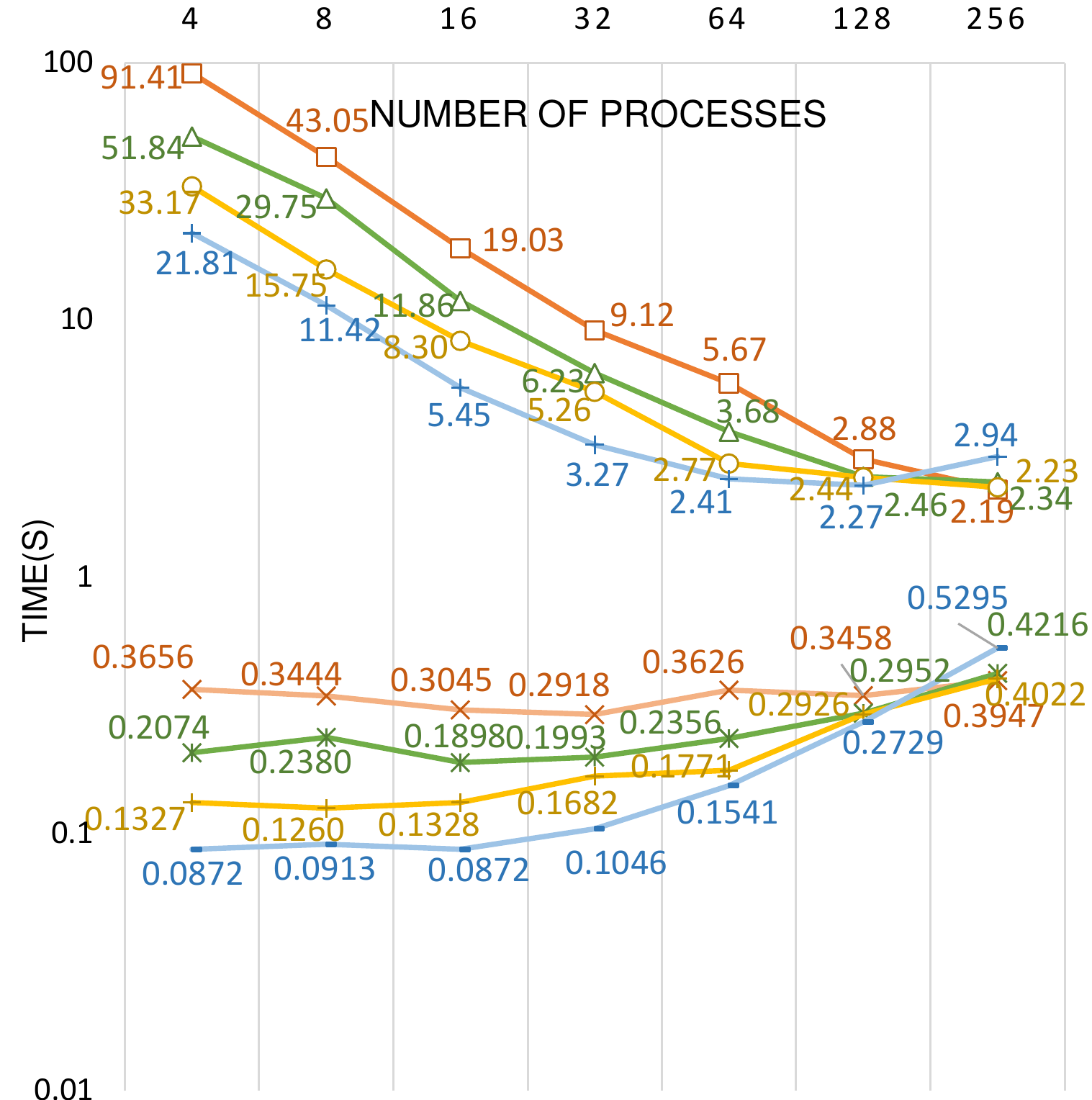}}%
			\subfigure{
				\includegraphics[width=0.29\textwidth,trim=300 10 250 400,clip]{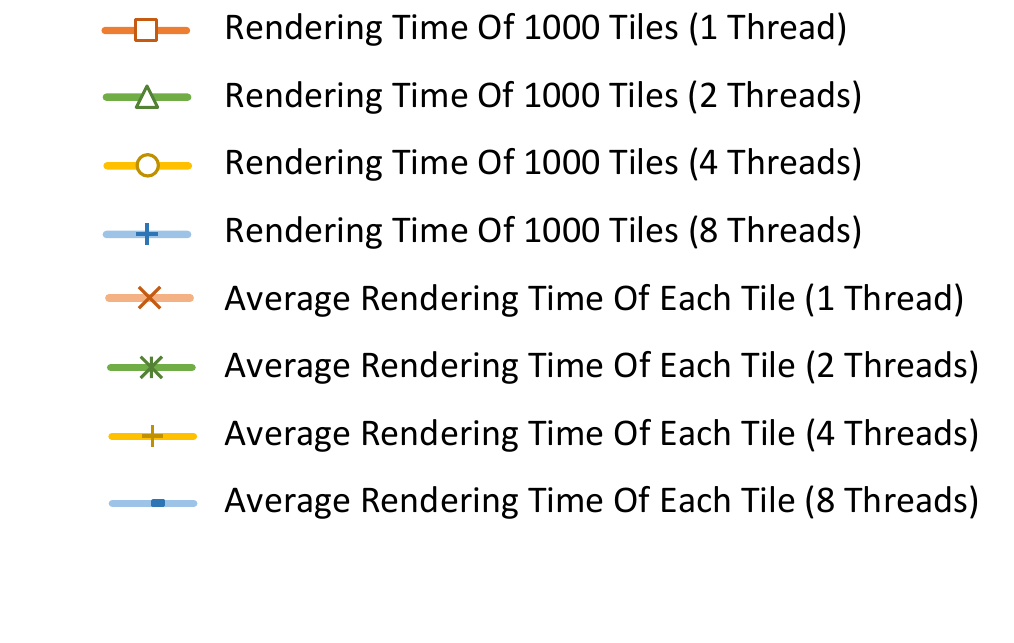}}%
			\vspace{-0.3cm}			
			\caption{Parallel performance of HiVision with different numbers of MPI processes and OpenMP threads.}%
			\label{f5_5}
		\end{center}
	\end{figure}

	To evaluate the parallel scalability, HiVision is respectively tested to run on 4, 8, 16, 32, 64, 128 and 256 MPI processes with 1, 2, 4 and 8 OpenMP threads in each process. For each pair of MPI processes and OpenMP threads, we generate 1000 random tile requests of different zoom levels on each dataset. The experimental results are plotted in Figure~\ref{f5_5}.

	We analyze the rendering time of 1000 tiles. HiVision achieves high performance of parallel acceleration when the process number is below 32, which is approximate to linearity; and the performance of parallel acceleration decreases as the process number is over 32, especially while running with 8 OpenMP threads in each process. It is because the increase of process numbers intensifies the resource competition, and the competition is even more intense while running with multiple threads. For example, see Figure~\ref{f5_5} (d), the rendering time of 1000 tiles with 8 threads increases as the process numbers increase from 128 to 256. Then, we analyze the average rendering time of each tile line with different OpenMP threads. As shown in the figures, multi-thread parallel processing reduces the rendering time of a tile when the resource competition is not intense. Surprisingly, for L$_{1}$, L$_{2}$ and P$_{1}$ which have the smaller scale, running with 2 thread produces weaker performance than 1 thread even if the resource competition is not intense, it is because that the initialization cost of multiple threads is higher compared with the parallel acceleration. Based on the experimental results and the analysis, a conclusion can be drawn about the deployments of HiVision in the given hardware environment: 1) if the number of request tiles is high, 256 processes $\times$ 1 thread is suggested; 2) if the number of request tiles is low, 32 processes $\times$ 8 threads is suggested, as this setting has low response time of each tile.
	
	\section{Online Demo}\label{section 6}

	An online demonstration of HiVision is provided\footnote{https://github.com/MemoryMmy/HiVision}. The 10-million-scale datasets (see Table~\ref{t6_2}) used in the demonstration are provided by map service providers. As the datasets are not open published, the raw datasets are encrypted by adding offsets. To note that, a current demonstration is deployed on a stand-alone server with 4 cores CPU and 32 GB Memory (see Table~\ref{t6_1}), which is accessible for an up-to-date personal computer. Even so, as illustrated in the demonstration, it is still possible to provide an interactive visualization of 10-million-scale datasets in HiVsion.
	
	\begin{table}[!h]
		\caption{Environment of the online demo.}
		\centering
		{
			\begin{tabular}{ll}
				\hline
				\textbf{Item}&\textbf{Description}\\
				\hline
				\ CPU&4cores, Intel(R) Xeon(R) E5-2680@2.50GHz\\
				\ Memory&32 GB\\
				\ Operating System& Centos 7\\
				\hline
			\end{tabular}	
		}
		\label{t6_1}
	\end{table}

	\begin{table}[!h]
		\caption{Datasets of China.}
		\centering
		{
			\begin{tabular}{@{}llll}
				\hline 
				\textbf{Dataset}&\textbf{Type}& \textbf{Records}& \textbf{Size}\\
				\hline 
				China roads&Linestring&21,898,508 &163,171,928 segments\\
				China points&Point&20,258,450 &20,258,450 points\\
				China farmland&Polygon&10,520,644 &133,830,561 edges\\
				\hline 
			\end{tabular}	
		}
		\label{t6_2}
	\end{table}
	
	\begin{figure}[htbp]
		\begin{center}
			\makeatletter
			\def
			\@captype{figure}
			\makeatother
			\subfigure[China POI]{
				\includegraphics[width=0.99\textwidth,trim=0 0 0 0,clip]{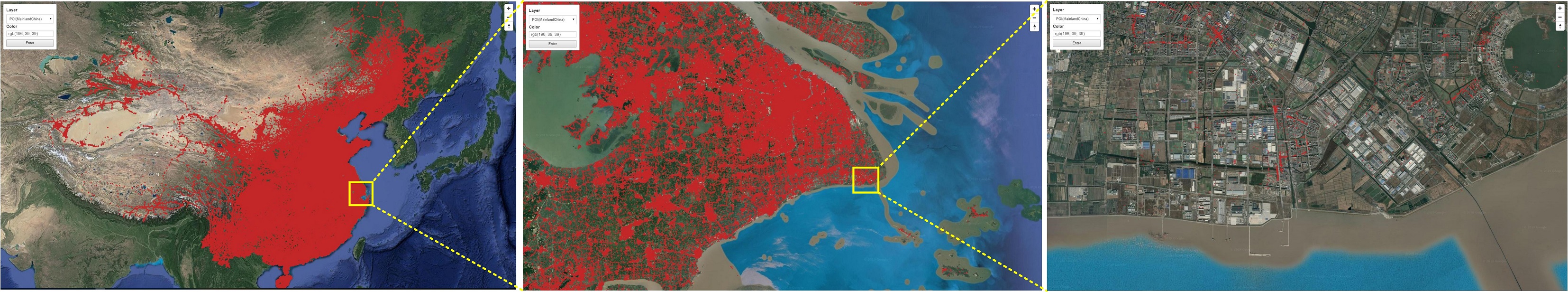}}%
			\vfill
			\subfigure[China roads]{
				\includegraphics[width=0.99\textwidth,trim=0 0 0 0,clip]{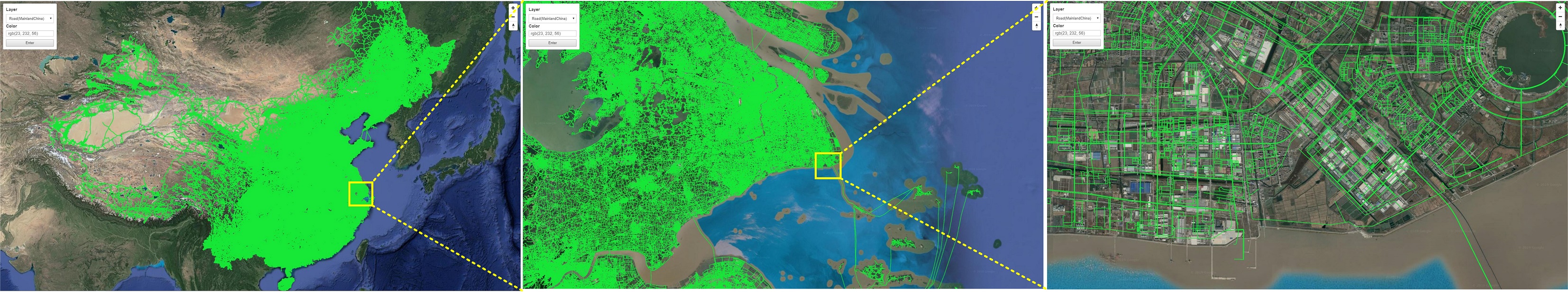}}%
			\vfill
			\subfigure[China farmlands]{
				\includegraphics[width=0.99\textwidth,trim=0 0 0 0,clip]{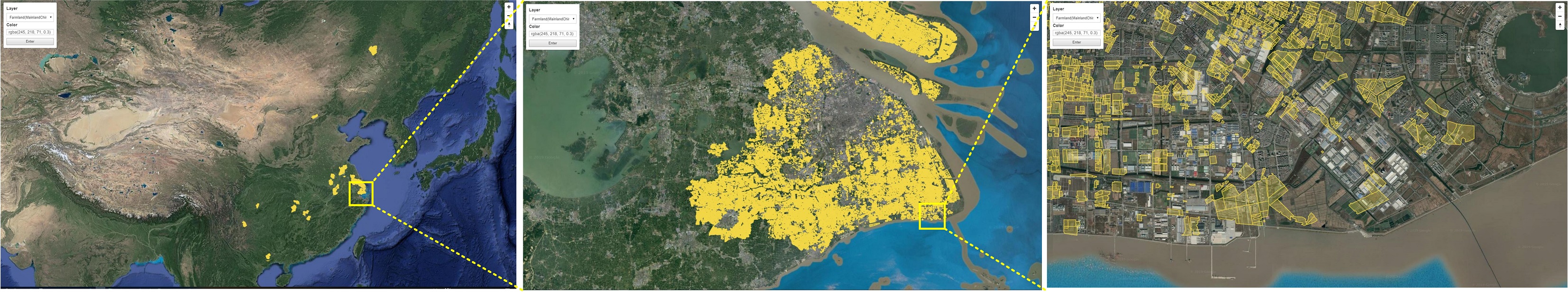}}%
			\vfill
			\subfigure[Patterns filling for polygon objects]{
				\includegraphics[width=0.99\textwidth,trim=0 0 0 0,clip]{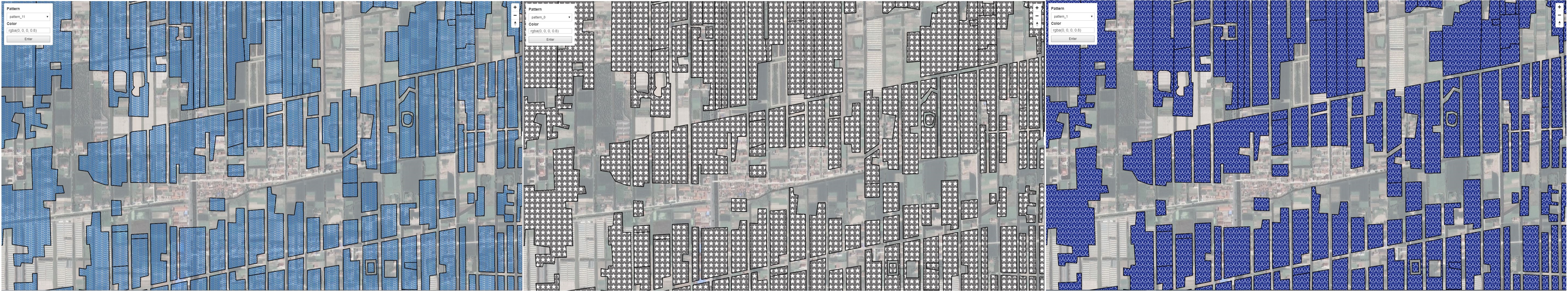}}%
			\caption{Visualized results of the online demo.}
			\label{f6_1}			
		\end{center}
	\end{figure}

	\section{Conclusions and Future Work}\label{section 7}
	
	In this paper, we present a display-driven visualization model, HiVision, for interactive exploration of large-scale spatial vector data. Different from traditional methods, in HiVision, the computing units are pixels rather than spatial objects to achieve the goal of being less sensitive to data volumes. Different experiments are designed and conducted to evaluate various system performance: experiment 1 shows that, compared with traditional data-driven methods, our approach produces higher performance with better visual effects; experiment 2 demonstrates the ability of HiVision to provide an interactive exploration of large-scale spatial vector data; in experiment 3, we analyze the impact of the request rate in HiVision and demonstrate that higher performance can be achieved in practical applications compared with the results in experiments; experiment 4 tests the parallel scalability of HiVision, and the results show that HiVision achieves high performance of parallel acceleration while the resource competition is not intense. Moreover, an online demonstration of HiVision is provided on the Web, which verifies that HiVision is capable of handling 10-million scale spatial data even deployed on a personal computer. Our future work will focus on extending HiVision to support more complex visualization styles and applying HiVision to the field of map cartography.
	
	\section*{Sourcecode availability}
	The source code of HiVision, including test data and user manuals, is available for download from Github.
	
	(https://github.com/MemoryMmy/HiVision)

	\bibliographystyle{cas-model2-names}
	\bibliography{HiVision}
	
\end{document}